\documentclass[11pt, a4paper]{article}
\usepackage{amssymb}
\usepackage{amsfonts}
\usepackage{amsmath}
\usepackage{bm}
\usepackage{indentfirst}
\usepackage{color}
\usepackage{graphicx}
\usepackage{enumerate}
\usepackage{subfigure}
\usepackage{multicol}
\usepackage{multirow}
\usepackage{float}
\usepackage{caption}
\usepackage{color}
\usepackage{setspace}
\usepackage{booktabs}
\usepackage{CJKutf8}
\usepackage{CJKpunct}
\usepackage{geometry}
\usepackage{changepage}
\usepackage{titlesec}
\usepackage{tcolorbox}
\usepackage{mathpazo}
\usepackage{threeparttable}
\usepackage{authblk}

\begin{document}
	
	\baselineskip=20pt
	
	
	\renewcommand{\baselinestretch}{1.18}	
	
	\title{\huge{\textrm{Identification and inference of outcome conditioned partial effects of general interventions}}}
	\author[a]{Zhengyu Zhang}
	\author[a]{Zequn Jin		\thanks{Address correspondence to Zequn Jin, Shanghai University of Finance and Economics, Shanghai, China;
			\textit{E-mail address}: polaris@163.sufe.edu.cn.}$^,$}
	\author[b]{Lihua Lin}
	\affil[a]{School of Economics, Shanghai University of Finance and Economics, China}
	\affil[b]{School of Economics, Jiangxi University of Finance and Economics, China}
	\maketitle
\renewcommand{\baselinestretch}{1.3}

\begin{abstract}\normalsize
	
	This paper proposes a new class of distributional causal quantities, referred to as the \textit{outcome conditioned partial policy effects} (OCPPEs), to measure the \textit{average} effect of a general counterfactual  intervention of a target covariate on the individuals in different quantile ranges of the outcome distribution.

The OCPPE approach is valuable in several aspects: (i) Unlike the unconditional quantile partial effect (UQPE) that is not $\sqrt{n}$-estimable, an OCPPE is $\sqrt{n}$-estimable.  Analysts can use it to capture heterogeneity across the unconditional distribution of $Y$ as well as obtain accurate estimation of the aggregated effect at the upper and lower tails of $Y$. (ii) The semiparametric efficiency bound for an OCPPE is explicitly derived. (iii) We propose an efficient debiased estimator for OCPPE, and provide feasible uniform inference procedures for the OCPPE process.
(iv) The efficient doubly robust score for an OCPPE can be used to optimize infinitesimal nudges to a
continuous treatment by maximizing a quantile specific Empirical Welfare function.  We  illustrate the method by analyzing how anti-smoking  policies impact low percentiles of live infants'
birthweights.

\end{abstract}

	\baselineskip=20pt
	
	\thispagestyle{empty}
	
	\renewcommand{\baselinestretch}{1.2}
	\newpage
	
	\section{Introduction}

One of the main objectives of policy analysis is to evaluate the effect of a counterfactual change in
some covariates on the unconditional distribution of an outcome variable of interest. Several methods have been proposed in the literature (Firpo et al. (2009), Rothe (2012), Sasaki et al. (2022),
Mart\'{i}nez-Iriarte et al. (2024)). A leading casual quantity studied by them is called the unconditional quantile partial effect (UQPE). Assume that the covariates and the outcome variable are related via a structural function $Y=m(D,X,U)$, where $D$ is the target variable\footnote{In parallel with the literature, the target covariate, denoted as $D$, is the variable a policy maker aims to change. $D$ is also called a treatment variable.  A policy maker hopes to change $D$ in order to achieve a desired effect on $Y$. }, $X$ is a vector of observed control variables and $U$ includes all unobservable covariates.  Suppose that every member of the population would experience an exogenous change in $D$, which is characterized by $D_{\delta}=\mathcal{G}_{\delta}(D)$ for some smooth function $\mathcal{G}_{\delta}(\cdot)$. For example, $\mathcal{G}_{\delta}(D)=D+\delta$ means
increasing each person's $D$ by a fixed amount; $\mathcal{G}_{\delta}(D)=D(1+\delta)$ means increasing $D$ by a fixed proportion. Other policy relevant interventions include
\[
\mathcal{G}_{\delta}(D,\sigma)=(D-\mu)\times(1+\sigma\delta)+\mu, \eqno(1.1)
\]
with $\mu=ED$ and $\sigma\in [-1,1]$ a free parameter. With a positive (or negative) $\sigma$, intervention (1.1) raises (or shrinks) the dispersion (inequality) of $D$ while keeps its mean unchanged.

With the above induced
change in $D$, the outcome variable becomes $Y_{\delta}=m\left(\mathcal{G}_{\delta}(D),X,U\right)$. Let $Q_{A}(\tau)$  denote the $\tau$-th quantile of  a generic continuous random variable $A$. The $\tau$-th UQPE of $D$ on $Y$ associated with a counterfactual change $\mathcal{G}_{\delta}(\cdot)$
is
\[
\mbox{UQPE}(\tau,\mathcal{G})=\lim_{\delta\rightarrow 0}\frac{Q_{Y_\delta}(\tau)-Q_{Y}(\tau)}{\delta}.\eqno(1.2)
\]
Firpo et al. (2009) develop the \textit{recentered influence function} (RIF) regression to study what corresponds to a location
shift. Rothe (2012) and Mart\'{i}nez-Iriarte et al. (2024) generalize Firpo et al. (2009) to cases with general intervention. Sasaki et al. (2022)
study the high-dimensional setting.

This paper proposes a new class of counterfactual causal quantities, referred to as the \textit{outcome conditioned} partial policy effects (OCPPEs). An OCPPE measures the
\textit{average} effect of an infinitesimal change in the value or the marginal distribution of $D$ on the individuals in different quantile ranges of the outcome distribution.

To introduce an OCPPE,  let us again consider the same structural model $Y=m(D,X,U)$. Suppose that $Y$ has a continuous
support $\mathcal{S}_{Y}\subset \mathbb{R}$.  A counterfactual change in $D$ to  $D_{\delta}=\mathcal{G}_{\delta}(D)$ induces a counterfactual outcome $Y_{\delta}=m\left(\mathcal{G}_{\delta}(D),X,U\right)=m\left(D_{\delta},X,U\right)$.
For some $0<\tau_1<\tau_2<1$, the outcome conditioned average partial policy effect of $D$ on $Y$ associated with  $\mathcal{G}$ is
\begin{eqnarray*}
	\theta(\tau_1,\tau_2,\mathcal{G})&=&E\left[\lim_{\delta\rightarrow 0}\frac{Y_{\delta}-Y}{\delta}\bigg|Y\in\left(Q_Y(\tau_1),Q_Y(\tau_2)\right)\right]\\
	&=&E\left[\lim_{\delta\rightarrow 0}\frac{m\left(\mathcal{G}_{\delta}(D),X,U\right)-m\left(D,X,U\right)}{\delta}\bigg|Y\in\left(Q_Y(\tau_1),Q_Y(\tau_2)\right)\right].
\end{eqnarray*}

There are several differences
between $\mbox{UQPE}(\tau,\mathcal{G})$ and  $\theta(\tau_1,\tau_2,\mathcal{G})$, some of which make OCPPE more attractive relative to UQPE.
First, UQPE, by measuring the causal difference between the unconditional quantiles of the observable $Y$ and the unobservable $Y_{\delta}$, does not possess a mean effect interpretation. Applied researchers prefer average partial effect to quantile partial effect, because
the formers are easier to interpret. Our OCPPE is unique in terms of combining features of both average and quantile partial effect; it measures the average causal effect of $\mathcal{G}_{\delta}(\cdot)$
on individuals in a given quantile range of $Y$.

An OCPPE is indexed by a pair of quantile indices
$(\tau_1,\tau_2 )\in(0,1)\times(0,1)$, as opposed to a single point $\tau\in(0,1)$ in UQPE. This subtle distinction has important implications: it ensures
that an OCPPE  has many desirable properties relative to the UQPE. \textit{First}, it is well known in the literature that an UQPE is not $\sqrt{n}$-estimable. However, an OCPPE is $\sqrt{n}$-estimable.
In practice, there are two ways to utilize OCPPE. One can use OCPPE to explore heterogeneity, by estimating a series of OCPPEs with $(\tau_1,\tau_2)\in\{(0.01,0.02), (0.02,0.03),\cdots, (0.98,0.99)\}$. On the other hand, OCPPE delivers fairly accurate estimation of aggregated tail
information on treatment effects. For example, in analyzing how anti-smoking policies impact infants' birth weight, one may extremely care about the impact on
infants at the extremal low tail of birthweight distribution. To this end, one may estimate OCPPEs at extremal percentile ranges such as $(\tau_1,\tau_2)\in\{(0.01,0.02), (0.01,0.03),\cdots, (0.01,0.10)\}$.
Moreover, fix $0<\bar{\delta}<1$.
We can predict the effect of a fixed anti-smoking policy, denoted by $\mathcal{G}_{\bar{\delta}}(D)$, on low birthweight infants by the mean value theorem:
\begin{eqnarray*}
	E\left(Y_{\bar{\delta}}\big|Y\in\left(Q_Y(\tau_1),Q_Y(\tau_2)\right)\right)\approx\bar{\delta}\times\theta(\tau_1,\tau_2,\mathcal{G})+E\left(Y\big|Y\in\left(Q_Y(\tau_1),Q_Y(\tau_2)\right)\right).
\end{eqnarray*}
Because both $\theta(\tau_1,\tau_2,\mathcal{G})$ and $E\left(Y\big|Y\in\left(Q_Y(\tau_1),Q_Y(\tau_2)\right)\right)$ are $\sqrt{n}$-estimable, the righthand side of ``$\approx$'' is also $\sqrt{n}$-estimable.

The \textit{second} advantage of OCPPE is that it has finite semiparametric efficiency bound with general interventions, as shown by this paper (Thoerem 4.2, Theorem 5.1), while the UQPE has no  efficiency bound. \textit{Third}, there has been no uniform inference method for the UQPE process available in the existing literature. The prior studies (e.g. Mart\'{i}nez-Iriarte et al. (2024), Sasaki et al. (2022),
Alejo et al. (2024) ) at most established inference of UQPE for \textit{fixed} $\tau\in(0,1)$ or for a \textit{fixed} policy intervention. In this paper, we prove that $\theta(\tau_1,\tau_2,\mathcal{G})$ converges to a Gaussian process uniformly in $\mathcal{G}_{\delta}(\cdot)$ that belongs to a compact function space.
Such uniform Gaussianity can be used to test whether a given policy intervention is ``optimal'' in the sense of generating the largest OCPPE for some $(\tau_1,\tau_2)\subset (0,1)\times (0,1)$ relative to a class of alternative policies.
To our best knowledge,  there is no such ``policy optimality'' test in the literature on \textit{unconditional policy effect}. Our work is
the first to develop uniform inference procedure of causal quantities measuring distributional impact of counterfactual changes in a continuous target variable.\footnote{Rothe (2012) briefly discusses the estimation of partial distribution policy effects and did not develop formal inference theory. Mart\'{i}nez-Iriarte et al. (2024) establish the pointwise limiting distribution of the UQPE estimator for a given $\tau\in(0,1)$. Sasaki
	et al. (2022) do not obtain uniform Gaussianity of their
	high dimensional UQPE estimator, whereas we establish  uniform Gaussianity
	of the OCPPE process in a high dimensional setting.}

To be concrete, we provide two empirical examples to motivate the use of OCPPE.

\noindent \textbf{Example 1 }. (Optimal job training program for low wage earners). Sasaki et al. (2022) use
UQPE to analyze the effect on wages of counterfactual increase in the days of participation in a job
training program.  In this example, $D$ is the duration in days of participation in Job Corps, $Y$ is the hourly wages.
Policymakers may care about whether the program (and its various counterfactual exercises) benefits
low wage earners more than high wages earners. OCPPE can answer this question directly, by estimating OCPPEs
with $(\tau_1,\tau_2)$ being low/middle/high ranges of $Y$.

In addition to considering the effect of extending the duration of the program ($D_{\delta}=D+\delta$),  one may consider the effect of a class of interventions:
\begin{eqnarray*}
	\mathcal{G}_{\delta}(D,\sigma)=(D-ED)\times(1+\sigma\delta)+ED,
\end{eqnarray*}
indexed by $\sigma\in[-1,1]$, which change the inequality of $D$ while keeping the mean of $D$ (total cost of the program) unchanged. A further question maybe is: among these interventions indexed by $\sigma$,
which one is most beneficial to ``low wage earners''? Such question can be formally answered by performing an optimal policy test. Choose a low quantile range, say, $(\tau_1,\tau_2)=(0,0.25)$,
and fix some $\sigma^{*}$ (determined by the estimation result).
The null hypothesis is:
\begin{eqnarray*}
	\mathbf{H_{0}}:~~\theta\left(\tau_1,\tau_2,\mathcal{G}_{\delta}\big(\cdot,\sigma^{*}\big)\right)\geq \theta\left(\tau_1,\tau_2,\mathcal{G}_{\delta}\big(\cdot,\sigma\big)\right),  ~\mbox{for~ any}~ \sigma\in[-1,1].
\end{eqnarray*}
The theory developed in this paper can be used to perform the above test.

\noindent\textbf{Example 2}. (Smoking and low percentiles of live infants'
birthweights). This topic was previously studied by Rothe (2010), Chernozhukov and Fern\'{a}ndez-Val (2011), and Mart\'{i}nez-Iriarte et al. (2024).
In this example, $Y$ is infant birthweight in grams, $D$ is the number of cigarettes smoked daily during
pregnancy.  Our interest is whether an anti-smoking intervention
can reduce the incidence of low-birthweight infants, which is usually defined by infants weight at birth falling below 2500 grams. This problem can be analyzed by estimating an OCPPE such as $\theta(0,\tau_2,\mathcal{G})$, with $\tau_{2}=F_{Y}(2500)$.
In the empirical application, we consider two hypothetical interventions that induce smoking cutoff: (i) $\mathcal{G}_{\delta}=D-\delta$ and (ii) $\mathcal{G}_{\delta}=\frac{D}{1+\delta}$. We find that  neither intervention has
positive impact on extremal quantiles of birth weight, whereas the effect is significantly positive on
typical birth weights. These results show that low birthweight infants may not effectively benefit from the rule of ``treating everyone''.  In Section 6, we show a simple empirical welfare-maximizing treatment rule that conditioning on a small number of pretreatment variables can produce positive welfare gain for low birthweight infants.

\textit{The paper makes three contributions to the Literature}. First, we propose OCPPEs as novel
and $\sqrt{n}$-estimable causal quantities to evaluate the effect of a counterfactual change in
a target covariate that is heterogeneous across the unconditional
distribution of $Y$. We show an OCPPE is identifiable for a general class of intervention functions (Assumption 2.1) under the unconfoundedness.
In addition to changing the values of $D$ directly, we also provide doubly robust identification of OCPPE with a manipulation of the marginal CDF of $D$ (in Section 5.1).

Second, we propose a novel debiased machine
learning (DML) estimator for an OCPPE, compatible with high-dimensional settings, and develop comprehensive asymptotic theories for it. We
show the estimator is
$\sqrt{n}$-consistent and asymptotically normal uniformly over $\mathcal{G}_{\delta}(\cdot)$ in a compact function space.
We prove uniform validity of the multiplier bootstrap procedure for inference of an OCPPE process. One of the most relevant applications of such uniformity  is to test whether a given counterfactual intervention, relative to a prespecified class of alternative interventions,
can generate the largest policy effect for the individuals in a given quantile range of $Y$.
Our work is
the first to develop uniform limiting  theories for causal quantities measuring distributional impacts of counterfactual changes in covariates.

Third, in the process of studying the OCPPE, we
obtain some new auxiliary results, which are of independent interest. The OCPPE provides the first example in the literature (on estimating causal and structural effect via high-dimensional regressions) in that the
Riesz representer (denoted as $L(D,X;\mathcal{G})$ in this paper) relies on an intervention function $\mathcal{G}_{\delta}(\cdot)$. \footnote{According to Newey (1994), Chernozhukov et al. (2022 a,b), a Riesz representer is an unknown nuisance function that appears in the Neyman orthogonal moment function of the causal quantity, in addition
	to the regressions in the raw moment function. }
To address this new feature, we generalize Chernozhukov et al. (2022)'s Lasso minimum distance estimator of the representer, and prove the estimator has a desirable convergence rate uniformly over $\mathcal{G}_{\delta}(\cdot)$.

We show an OCPPE with a general intervention has finite semiparametric efficiency bound, and the proposed DML estimator achieves this bound. Similar results have not been obtained in the literature on unconditional policy effect with general interventions (Rothe 2012,  Mart\'{i}nez-Iriarte et al. (2024)).
We also show how the doubly robust score for an OCPPE estimand can be related to learning optimal treatment assignment rules
within the framework of Empirical Welfare Maximization (in Section 5.2) in the spirit of Kitagawa and Tetenov (2018), Athey and Wager (2021). To our knowledge, our work is the first to consider learning
optimal infinitesimal intervention by maximizing a quantile specific utility function.

\textit{Relationship to the Literature}. This paper is related to the literature on UQPE and those studying unconditional effects of a policy. Rothe (2012) considers unconditional policy effect of a general intervention of the marginal distribution of a covariate.
Mart\'{i}nez-Iriarte et al. (2024) extend Firpo et al. (2009) to general
counterfactual policy changes, allowing for any smooth and invertible intervention of the target covariates.\footnote{Like  Mart\'{i}nez-Iriarte et al. (2024), OCPPE can be extended to explore the effect of
	simultaneous changes in two or more covariates. }
Sasaki et al. (2022)
study the estimation of UQPE corresponding to a location shift in a high-dimensional setting.  Alejo et al. (2024) develop a semi-parametric procedure for estimation of UQPE using conditional quantile regressions.

As a main departure from this line of the literature, this paper introduces the OCPPE to characterize heterogeneous counterfactual policy effects.  The conception of OCPPE extends Jin et al. (2024), who study a causal quantity called \textit{outcome conditioned average structural derivative} (OASD), in several directions. OASD is a special case of OCPPE with $\mathcal{G}_{\delta}(D)=D+\delta$. Identification of OCPPE under the unconfoundedneses generalizes Jin et al. (2024) to what corresponds to a general counterfactual intervention. Estimation and inference of OCPPE is more general than OASD in that we prove $\theta(\tau_1,\tau_2,\mathcal{G})$ converges to a Gaussian process not only uniformly in $(\tau_1,\tau_2)$, but also uniformly in $\mathcal{G}_{\delta}(\cdot)$. The latter uniformity is necessary for testing whether a
given policy intervention is superior to other alternative
policies.

This paper is also related to the literature on the estimation and inference of causal quantities based on orthogonal scores (Newey (1994), Chernozhukov et al. (2022, 2022a, 2022b)).
Chernozhukov et al. (2022a, 2022b)
develop a lasso minimum-distance learner of the Riesz representer. Our DML estimator exploits this knowledge and generalizes their theory to the situation with the Riesz representer relying on an intervention function.

This study is also connected to a growing literature on personalized treatment rules, including
Manski (2004), Kitagawa and Tetenov (2018), and Athey and Wager (2021). The efficient doubly robust score for an OCPPE can be used to optimize infinitesimal nudges to a
continuous treatment within the framework of Empirical Welfare Maximization proposed by Kitagawa and Tetenov (2018), Athey and Wager (2021). We describe the OCPPE related policy learning problem in Section 5.2 and sketch
an algorithm for choosing whom to treat.

The remainder of this paper is organized as follows: Section 2
presents the set-up of OCPPE and establishes its doubly robust identification under the unconfoundedness. Sections 3 develops a DML estimator for an OCPPE.
Section 4 establishes the asymptotic properties of the
DML estimator, and show the DML estimator is efficient. Section 5 discusses several important extensions of OCPPE. The previous sections focus on counterfactual interventions that change the value of $D$ directly. Section 5.1 provides the double robust identification of an OCPPE by changing the marginal CDF of $D$.  Section 5.2 connects the efficient doubly robust score for an OCPPE to the empirical welfare maximization framework to determine personalized assignment of intervention based on individual
characteristics.  Section 6 presents an empirical example. Section 7 concludes.  The Appendix collects proofs and all other figures.
	
	\noindent \textit{Notations}.
	
	Our study is based on independent and identically distributed (i.i.d.) data, $\{W_{i}\}_{i=1}^{n}$, which are defined on the probability space, $\left(\mathcal{W},\mathcal{A}_{\mathcal{W}},P\right)$. We denote by $\mathbb{P}_{n}$ the empirical probability measure that assigns probability $n^{-1}$ to each $W_{i}\in\{W_{i}\}_{i=1}^{n}$. $\mathbb{E}_{n}$ denotes the expectation with respect to the empirical measure, and $\mathbb{G}_{n}$ denotes the empirical process, that is,
	\[
	\mathbb{G}_{n}B(W)=\frac{1}{\sqrt{n}}\sum_{i=1}^{n}\bigg[B(W_{i})-EB(W)\bigg].
	\]
	indexed by a measurable class of functions, $\mathcal{B}:\mathcal{W}\mapsto\mathbb{R}$. In what follows, we use $\Vert\cdot\Vert_{P,q}$ to denote the $L^{q}(P)$ norm. $\left\Vert{A}\right\Vert_{\infty}=\max_{i,j}|a_{ij}|$, $\left\Vert{A}\right\Vert_{1}=\sum_{i,j}|a_{ij}|$, and $\Vert{A}\Vert_{0}$ equals the number of nonzero components of $A$ for a matrix, $A=[a_{ij}]$.
	
	\section{OCPPE and Its Identification}
	
As mentioned before, we assume that $Y=m(D,X,U)$, where $Y$ has a continuous
support $\mathcal{S}_{Y}\subset \mathbb{R}$.  A counterfactual change in $D$ to  $D_{\delta}=\mathcal{G}_{\delta}(D)$ induces a counterfactual outcome $Y_{\delta}=m\left(\mathcal{G}_{\delta}(D),X,U\right)=m\left(D_{\delta},X,U\right)$.
For some $0<\tau_1<\tau_2<1$, the OCPPE of $D$ on $Y$ associated with  $\mathcal{G}$ is
\begin{eqnarray*}
	\theta(\tau_1,\tau_2,\mathcal{G})=E\left[\lim_{\delta\rightarrow 0}\frac{m\left(\mathcal{G}_{\delta}(D),X,U\right)-m\left(D,X,U\right)}{\delta}\bigg|Y\in\left(Q_Y(\tau_1),Q_Y(\tau_2)\right)\right].
\end{eqnarray*}
Note that to ensure $\theta(\tau_1,\tau_2,\mathcal{G})$ has a causal interpretation, we hold the joint distribution of $(D,X,U)$ and $m(\cdot,\cdot,\cdot)$ constant in generating $Y$ and $Y_{\delta}$.  To attain point identification of $\theta(\tau_1,\tau_2,\mathcal{G})$, we need the following assumptions.

\noindent \textbf{Assumption 2.1.} (i) There exists  $\overline{\delta}>0$, such that for any $\delta\in [-\overline{\delta},\overline{\delta}]$, $\mathcal{G}_{\delta}(d)$ is strictly increasing in $d$; (ii) $\mathcal{G}_{0}(d)=d$; (iii) $\mathcal{G}_{\delta}(d)$ is continuously differentiable with respect to $(d,\delta)$.

Assumption 2.1 imposes some restrictions on intervention function $\mathcal{G}_{\delta}(d)$.  Interventions satisfying Assumption 2.1 include (i) (location shift) $\mathcal{G}_{\delta}(D)=D+\delta$,
(ii) (scale transformation) $\mathcal{G}_{\delta}(D)=D(1+\delta)$; (iii) (general location-scale transformation) $\mathcal{G}_{\delta}(D;
\mu,\sigma_1,\sigma_2 )=(D-\mu)(1+\sigma_{1}\delta)+\mu+\sigma_{2}\delta$; and (iv) (marginal perturbation in the direction of $g_{0}(\cdot)$)  $\mathcal{G}_{\delta}(D)=D+\delta\left(g_{0}(D)-D\right)$.\footnote{For intervention (iv),  $g_{0}(D)$ is the target that a policy maker desires $D$ to become in the long run.  Intervention (iv) changes each person's $D$ by a small step $\delta$ from the starting point $D$   towards the prespecified target $g_{0}(D)$. In this case, Assumption 2.1 is satisfied if there exists some $c>0$ such that
    $\sup_{d\in\mathbb{R}}\left|\frac{\partial g_{0}(d)}{\partial d}\right|\leq c<\infty$.}

\noindent\textbf{Assumption 2.2.}  $U$ is independent of $D$ conditional on $X$.

Assumption 2.2 is the unconfoundedness condition that is standard in the literature . It is substantially weaker than assuming $U$ is jointly independent of $(D,X)$.

	\noindent\textbf{Proposition 2.1.} \textit{If Assumptions 2.1--2.2 and other regularity conditions (Assumptions A.1-A.3 listed in Appendix A) hold, then
	\[
	\theta(\tau_1,\tau_2,\mathcal{G})=\frac{-1}{\tau_2-\tau_1}E\left[\vartheta\left(D;\mathcal{G}\right)\int_{Q_Y(\tau_1)}^{Q_Y(\tau_2)}\partial_DF_Y(y|D,X)dy\right], \eqno{(2.1)}
	\]
	where $\vartheta\left(d;\mathcal{G}\right)=\dfrac{\partial }{\partial \delta}\mathcal{G}_\delta(d)\Big|_{\delta=0}$ and $\partial_{D}$ denotes the first order derivative with respect to $D$.
}

\noindent\textbf{Remark 2.1.}  When $\mathcal{G}_{\delta}$ is a location shift or $\mathcal{G}_{\delta}(D)=D+\delta$,
\begin{eqnarray*}
	\theta(\tau_1,\tau_2,\mathcal{G})=\frac{-1}{\tau_2-\tau_1}E\left[\int_{Q_Y(\tau_1)}^{Q_Y(\tau_2)}\partial_DF_Y(y|D,X)dy\right],
\end{eqnarray*}
which is showed by Jin et al. (2024).  Proposition 2.1 extends Jin et al. (2024) to a general counterfactual intervention of $D$. Moreover, in Section 5.1, we provide identification results of an OCPPE when the policymaker aims to manipulate the marginal distribution of $D$ (Rothe (2012)) instead of changing the value of $D$ directly.

Although we can directly estimate $\theta(\tau_1,\tau_2,\mathcal{G})$ using  formula (2.1), such estimator exhibits significant limitations, such as lacking $\sqrt{n}$-consistency and being severely biased when $F_Y(y|D,X)$ is estimated by neural nets, random forests, Lasso, boosting, and other
high-dimensional methods. Moreover, it is not semi-parametrically efficient.  Below, we provide another identification result based on an
orthogonal score, which is necessary for estimation with high
dimensional controls.
Let $W=(Y,D,X)$, and let $\eta=\eta(W;\tau_1,\tau_2,\mathcal{G})$ encompass the nuisance parameters that are potentially infinite-dimensional:
\[
\eta(W;\cdot)=\Big(Q_{Y}(\cdot),F_Y(\cdot|D,X),L(D,X;\cdot),f_{Y}(\cdot),E\Big[\vartheta(D;\cdot)\cdot\partial_DF_Y\left(Q_Y(\cdot)\big|D,X\right)\Big]\Big),
\]
where
\[
L(D,X;\mathcal{G})=\frac{\partial_D\Big(\vartheta\left(D;\mathcal{G}\right)f(D,X)\Big)}{f(D,X)}
\]
is the Riesz representer for an OCPPE.
We show the efficient orthogonal moment for an OCPPE is
\[\begin{aligned}
	&\psi\Big(W,\theta,\eta;\tau_1,\tau_2,\mathcal{G}\Big)\\
	=&\frac{-1}{\tau_2-\tau_1}\vartheta\left(D;\mathcal{G}\right)\int_{Q_Y(\tau_1)}^{Q_Y(\tau_2)}\partial_{D}F_Y(y|D,X)dy-\theta\\
	&+\frac{1}{\tau_2-\tau_1}\frac{\partial_D\Big(\vartheta\left(D;\mathcal{G}\right)f(D,X)\Big)}{f(D,X)}\left(\int_{Q_Y(\tau_1)}^{Q_Y(\tau_2)}F_Y(y|D,X)dy-\int_{Q_Y(\tau_1)}^{Q_Y(\tau_2)}1\{Y\leq y\}dy\right)\\
	&-\frac{1}{\tau_2-\tau_1}\frac{E\Big[\vartheta\left(D;\mathcal{G}\right)\cdot\partial_{D}F_Y\left(Q_Y(\tau_1)\big|D,X\right)\Big]}{f_Y\Big(Q_Y(\tau_1)\Big)}\Big(1\{Y\leq Q_Y(\tau_1)\}-\tau_1\Big)\\
	&+\frac{1}{\tau_2-\tau_1}\frac{E\Big[\vartheta\left(D;\mathcal{G}\right)\cdot\partial_{D}F_Y\left(Q_Y(\tau_2)\big|D,X\right)\Big]}{f_Y\Big(Q_Y(\tau_2)\Big)}\Big(1\{Y\leq Q_Y(\tau_2)\}-\tau_2\Big). \\
\end{aligned}\eqno{(2.2)}\]
	
	\noindent\textbf{Proposition 2.2.} \textit{Under the same assumptions as those in Proposition 2.1, we have that
		\begin{enumerate}[(i)]
			\item 
			\[
			E\Big[\psi\Big(W,\theta\left(\tau_1,\tau_2,\mathcal{G}\right),\eta;\tau_1,\tau_2,\mathcal{G}\Big)\Big]=0.
			\]
			\item $\psi\Big(W, \theta, \eta; \tau_1,\tau_2,\mathcal{G}\Big)$ satisfies the Neyman orthogonal property
			\begin{eqnarray*}
				\frac{\partial E\Big[\psi\Big(W,\theta\left(\tau_1,\tau_2,\mathcal{G}\right),\eta+r(\widetilde{\eta}-\eta);\tau_1,\tau_2,\mathcal{G}\Big)\Big]}{\partial r}\Bigg|_{r=0}=0.
			\end{eqnarray*}
			\item The orthogonal score $\psi$ is doubly robust, such that for any $\widetilde{\eta}=\left(\widetilde{\eta}_{1},\widetilde{\eta}_{2},\widetilde{\eta}_3,\widetilde{\eta}_4,\widetilde{\eta}_5\right)$ belonging to the space of nuisance parameters,
			\[
			\begin{aligned}
				E\Big[\psi\Big(W,\theta,{\eta};\tau_1,\tau_2,\mathcal{G}\Big)\Big]&=E\Big[\psi\Big(W,\theta,{\eta}_{1},\widetilde{\eta}_{2},{\eta}_{3},\widetilde{\eta}_4,\widetilde{\eta}_5;\tau_1,\tau_2,\mathcal{G}\Big)\Big]\\
				&=E\Big[\psi\Big(W,\theta,{\eta}_{1},{\eta}_{2},\widetilde{\eta}_3,\widetilde{\eta}_4,\widetilde{\eta}_5;\tau_1,\tau_2,\mathcal{G}\Big)\Big].
			\end{aligned}
			\]
	\end{enumerate}}
	
The result in (i) means $\theta(\tau_1,\tau_2,\mathcal{G})$ is identified by the moment condition
\begin{eqnarray*}
	E\psi\Big(W,\theta\left(\tau_1,\tau_2,\mathcal{G}\right),\eta;\tau_1,\tau_2,\mathcal{G}\Big)=0.
\end{eqnarray*}
The result in (ii) means $\psi(\cdot)$ exhibits some local robustness, that is, it is insensitive to minor perturbation of the nuisance function $\eta(W;\cdot)$ around its true value. This local robustness or orthogonality property is crucial for reducing model selection and/or regularization biases which
are common for machine learning first steps, and obtaining $\sqrt{n}$-consistency of an OCPPE when the dimension of $X$ is high. The double robustness in (iii) says $\theta(\tau_1,\tau_2,\mathcal{G})$ remains identifiable if $Q_{Y}(\cdot)$ is consistently estimated, and either $F_{Y}(\cdot|D,X)$ or $L(D,X;\cdot)$ (but not both) is correctly specified.

\noindent\textbf{Remark 2.2.} The orthogonal moment function $\psi(\cdot)$ consists of four terms. The first line is the sample analog for Proposition 2.1.
The second line is the influence function adjustment for the estimation of nuisance parameter $F_{Y}(\cdot|D,X)$. The third and fourth lines are for $Q_{Y}(\tau_1)$ and $Q_{Y}(\tau_2)$, respectively. The adjustment
for $Q_{Y}(\tau_1)$ and $Q_{Y}(\tau_2)$ is unnecessary in terms of debiased estimation and
inference. However, these adjustment terms are necessary to
achieve the semiparametric efficiency bound.

\noindent\textbf{Remark 2.3.} Like UQPE (Mart\'{i}nez-Iriarte et al. (2024)), our OCPPE framework can accommodate situations where interventions depend on covariates $X$. To illustrate this, let us consider Example 2 again.  Suppose that the authority decides to levy a tax on the cigarettes consumption.  The taxation would reduce $D$ to $D/(1+\delta)$, with $\delta>0$ the tax rate.
The literature has shown that mothers who paid no prenatal care visit are more likely to give birth to low weight infants (Abrevaya 2001,  Chernozhukov and Fern\'{a}ndez-Val (2011)).
Let $X_1$ be a dummy variable, with $X_{1}=1$ indicating a mother paid no prenatal visit.
A taxation intervention that takes into account the birth weight differential between mothers with $X_{1}=1$ and $X_{1}=0$ is
\begin{eqnarray*}
	D_{\delta}=\mathcal{G}_{\delta}(D,X_{1})=\frac{D}{1+a_{1}\delta}1\{X_{1}=1\}+\frac{D}{1+a_{2}\delta}1\{X_{1}=0\}.
\end{eqnarray*}
with $a_{1}>a_{2}\geq 1$. Such intervention reflects more stringent tax burden imposed on mothers who paid no prenatal medical care visit during pregnancy. It is straightforward to show that if the intervention is determined by individual characteristics, or $D_{\delta}=\mathcal{G}_{\delta}(D,X)$, Propositions 2.1-2.2 remain valid with $\vartheta\left(D;\mathcal{G}\right)$ replaced by $\vartheta\left(D,X;\mathcal{G}\right)$, with
\begin{eqnarray*}
	\vartheta\left(d,x; \mathcal{G}\right)=\dfrac{\partial }{\partial \delta}\mathcal{G}_\delta(d,x)\Big|_{\delta=0}.
\end{eqnarray*}

	\section{Estimation Procedure}
	We propose an debiased machine learning (DML) procedure for estimating OCPPEs with high dimensional covariates. The procedure is easily implemented, and the DML estimator is semiparametrically efficient.
	We outline the estimation procedure as follows:
	\begin{enumerate}[(i)]
		\item  Estimate the unconditional quantile $Q_Y(\cdot)$ and density $f_Y(\cdot)$ by conventional nonparametric method.
		\item Estimate the CDF $F_{Y}\left(y|d,x\right)$, its integral over $[Q_Y(\tau_1),Q_Y(\tau_2)]$ and corresponding derivative via high-dimensional nonparametric methods with model selection.
		\item Estimate the Riesz representer  $L(d,x;\mathcal{G})=\dfrac{\partial_{d}\Big(\vartheta\left(d;\mathcal{G}\right)f(d,x)\Big)}{f(d,x)}$ by the (automatic) Lasso minimum distance method.
		\item Estimate $\theta\left(\tau_1,\tau_2,\mathcal{G}\right)$ based on the orthogonal score via the plug-in rule.
	\end{enumerate}
	We now describe the estimation procedure in detail.
	
	\textbf{Step 1.} (\textit{Estimate unconditional quaniles}) Given a pair of $(\tau_{1},\tau_{2})$, the unconditional quantiles $Q_Y(\tau_1)$ and $Q_Y(\tau_2)$ are estimated by
	\[
	\begin{aligned}
		\widehat{Q}_Y(\tau_1)=&\arg\min_{q_1}\frac{1}{n}\sum_{i=1}^{n}\rho_{\tau_1}(Y_i-q_1), \\
		\widehat{Q}_Y(\tau_2)=&\arg\min_{q_2}\frac{1}{n}\sum_{i=1}^{n}\rho_{\tau_1}(Y_i-q_2), \\
	\end{aligned}\eqno{(3.1)}
	\]
	where $\rho_{\tau}(u)=(\tau-1\{u<0\})u$ is the check function.
	
	(\textit{Estimate unconditional densities}) Given $\left(\widehat{Q}_Y(\tau_1), \widehat{Q}_Y(\tau_2)\right)$, the unconditional densities $f_Y\left(Q_Y(\tau_1)\right)$ and $f_Y\left(Q_Y(\tau_2)\right)$ are estimated by
	\[
	\begin{aligned}
		&\widehat{f}_Y\left(\widehat{Q}_Y(\tau_1)\right)=\frac{1}{nh_Y}\sum_{i=1}^{n}\mathbb{K}\left(\frac{Y_i-\widehat{Q}_Y(\tau_1)}{h_Y}\right),\\
		&\widehat{f}_Y\left(\widehat{Q}_Y(\tau_2)\right)=\frac{1}{nh_Y}\sum_{i=1}^{n}\mathbb{K}\left(\frac{Y_i-\widehat{Q}_Y(\tau_2)}{h_Y}\right),
	\end{aligned}\eqno{(3.2)}
	\]
	where $\mathbb{K}(\cdot)$ is the univariate kernel function and $h_{Y}$ is the corresponding bandwidth.

	\textbf{Step 2.} (\textit{Estimate CDF}) Consider the approximately sparse distribution regression model for $F_Y(y|d,x)$:
	\begin{eqnarray*}
		F_Y(y|d,x)=\Lambda\left(b(d,x)'\beta(y)\right)+(\text{approximation error}),
	\end{eqnarray*}
	where $\Lambda(\cdot)$ is a known link function, $b(d,x)$ is a $p_{b}$-dimensional vector of basis functions and $\beta(y)$ is a vector of unknown parameters.
	We estimate $\beta(y)$ by the Lasso penalized distribution regression:
	\[
	\widetilde{\beta}(y)=\arg\min_{\beta}-\frac{1}{n}\sum_{i=1}^{n}\log\left(\Lambda\left(b(D_{i},X_{i})'\beta\right)^{1\{Y_i\leq{y}\}}\left(1-\Lambda\left(b(D_{i},X_{i})'\beta\right)\right)^{1\{Y_i>{y}\}}\right)+\frac{\lambda_{\beta}}{n}\left\Vert\widehat{\Psi}_y\beta\right\Vert_1, \eqno{(3.3)}
	\]
	where $\lambda_{\beta}$ denotes the penalty level to guarantee good theoretical properties of the lasso estimator, and $\widehat{\Psi}_y=\text{diag}\left(\widehat{\psi}_{y,1},\cdots,
	\widehat{\psi}_{y,p_{b}}\right)$ denotes the diagonal matrix of penalty loadings.
	According to Belloni et al. (2017) and Sasaki et al. (2022), we set the penalty level $\lambda_{\beta}$ as
	\begin{eqnarray*}
		\lambda_{\beta}=1.1\sqrt{n}\Phi^{-1}\left(1-\frac{0.1/\log(n)}{2p_{b} n}\right),
	\end{eqnarray*}
with $\Phi(\cdot)$ the CDF of the standard normal distribution.

	Penalty loadings $\widehat{\Psi}_y$ can be constructed  by Algorithm 6.1 in Belloni et al. (2017). Let $\text{supp}\left(\beta\right)$ denote the labels of the components in $\beta$ with nonzero values. The Post-Lasso estimator, $\widehat{\beta}(y)$, is a solution to
	\begin{eqnarray*}
		\widehat{\beta}(y)&=&\arg\min_{\beta}-\frac{1}{n}\sum_{i=1}^{n}\log\left(\Lambda\left(b(D_{i},X_{i})'\beta\right)^{1\{Y_i\leq{y}\}}\left(1-\Lambda\left(b(D_{i},X_{i})'\beta\right)\right)^{1\{Y_i>{y}\}}\right)\\
		&&s.t.\ \ \beta_j=0,j\notin\text{supp}\left(\widetilde{\beta}(y)\right).
	\end{eqnarray*}
	Then $F_Y(y|d,x)$ can be estimated by
	\begin{eqnarray*}
		\widehat{F}_Y(y|d,x)=\Lambda\left(b(d,x)'\widehat{\beta}(y)\right).
	\end{eqnarray*}
	
	(\textit{Estimate derivative of CDF}) Let
	\[
	DF(d,x;y)=\partial_dF_Y(y|d,x).
	\]
	Similar to Sasaki et al. (2022), we directly estimate $DF(d,x;y)=\partial_dF_Y(y|d,x)$ as follows
	\begin{eqnarray*}
		\widehat{DF}(d,x;y)=\partial_d\widehat{F}_Y(y|d,x).
	\end{eqnarray*}

	(\textit{Estimate integral of CDF and its derivative}) By definition of integration,
	\[\begin{aligned}
		&IF(d,x;\tau_{1},\tau_{2}):=\int_{Q_{Y}(\tau_{1})}^{Q_{Y}(\tau_{2})}F_{Y}(y|d,x)dy=\lim_{J\to\infty}\sum_{j=1}^{J}F_{Y}\Big(Q_{Y}(\tau_{1})+j\Delta{y}|d,x\Big)\Delta{y}, \\
		&IDF(d,x;\tau_{1},\tau_{2}):=\int_{Q_{Y}(\tau_{1})}^{Q_{Y}(\tau_{2})}DF(y,d,x)dy=\lim_{J\to\infty}\sum_{j=1}^{J}DF\Big(Q_{Y}(\tau_{1})+j\Delta{y},d,x\Big)\Delta{y}, \\
	\end{aligned}\]
	where $\Delta y=\left(Q_Y(\tau_2)-Q_Y(\tau_1)\right)/J$. Thus, the corresponding estimators can be constructed as
	\[
	\widehat{IF}(d,x;\tau_{1},\tau_{2})=\sum_{j=1}^{J}\widehat{F}_{Y}\Big(\widehat{Q}_{Y}(\tau_{1})+j\widehat{\Delta{y}}|d,x\Big)\widehat{\Delta{y}}=\sum_{j=1}^{J}\Lambda\left(b^{\prime}(d,x)\widehat{\beta}\left(\widehat{Q}_{Y}(\tau_{1})+j\widehat{\Delta{y}}\right)\right)\widehat{\Delta{y}}, \eqno{(3.4)}
	\]
	and
	\[
	\widehat{IDF}(d,x;\tau_{1},\tau_{2})=\sum_{j=1}^{J}\widehat{DF}\Big(\widehat{Q}_{Y}(\tau_{1})+j\widehat{\Delta{y}},d,x\Big)\widehat{\Delta{y}}, \eqno{(3.5)}
	\]
	where $\widehat{\Delta{y}}=\Big(\widehat{Q}_{Y}(\tau_{2})-\widehat{Q}_{Y}(\tau_{1})\Big)\Big/J$.

	\textbf{Step 3.} (\textit{Estimate  Riesz Representer} $L(d,x;\mathcal{G})$)  Suppose that
	\begin{eqnarray*}
		L(d,x;\mathcal{G})=h(d,x)'\gamma(\mathcal{G})+(\text{approximation error})
	\end{eqnarray*}
	holds, where $h(d,x)$ is a $p_{h}$-dimensional vector of basis functions and $\gamma(\mathcal{G})$ is a vector of unknown parameters. According to locally robust property described in Proposition 2.2(ii), for any real function $\delta(d,x)$, we have
	\begin{eqnarray*}
		E\left[\partial_D\delta(D,X)\vartheta\left(D;\mathcal{G}\right)+\frac{\partial_{D}\Big(\vartheta\left(D;\mathcal{G}\right)f(D,X)\Big)}{f(D,X)}\delta(D,X)\right]=0.
	\end{eqnarray*}
	Let $\delta(d,x)$ be one element of $\{h_{k}(d,x), k=1,\dots,p_{h}\}$ each time. We obtain a $p_{h}\times1$ vector of moment conditions as
	\begin{eqnarray*}
		E\left[\partial_{D}h(D,X)\vartheta\left(D;\mathcal{G}\right)+\frac{\partial_{D}\Big(\vartheta\left(D;\mathcal{G}\right)f(D,X)\Big)}{f(D,X)}h(D,X)\right]=0.
	\end{eqnarray*}
	 $L(d,x;\mathcal{G})$ can be estimated by
	\begin{eqnarray*}
		\widehat{L}(d,x;\mathcal{G})=h(d,x)'\widehat{\gamma}(\mathcal{G}),
	\end{eqnarray*}
	where $\widehat{\gamma}(\mathcal{G})$ is the Lasso estimator:
			\begin{eqnarray*}
		\widehat{\gamma}(\mathcal{G})=\arg\min_{\gamma}-2\widehat{M}(\mathcal{G})'\gamma+\gamma'\widehat{G}\gamma+2\lambda_{\gamma}\|\gamma\|_1,
	\end{eqnarray*}
		with $\lambda_{\gamma}>0$ a positive scalar to control the degree of penalty, and
		\[
		\widehat{M}(\mathcal{G})=-\frac{1}{n}\sum_{i=1}^{n}\partial_{D_i}h(D_i,X_i)\vartheta\left(D_i;\mathcal{G}\right),\quad\widehat{G}=\frac{1}{n}\sum_{i=1}^{n}h(D_i,X_i)h(D_i,X_i)'.
		\]
\noindent\textbf{Remark 3.1.} The above Steps 1-3 largely extend those of Jin et al. (2024) to account for a general intervention function $\mathcal{G}_{\delta}(\cdot)$. The chief difference between Jin et al. (2024) and the current paper is the penalty level
$\lambda_{\gamma}$ needs to be set to control selection errors uniformly
over $\mathcal{G}_{\delta}(\cdot)$ in a compact function space. To do so, we set
\[
	\lambda_{\gamma}=A\sqrt{\frac{\log(p_{h}\vee{n})}{n}} \eqno{(3.6)}
	\]
with $A$ a constant. As demonstrated by the subsequent asymptotic analysis (Section 4.2.2), $\lambda_{\gamma}$ set above achieves the fastest possible mean square convergence rate for Lasso minimum distance estimator $\widehat{L}(d,x;\mathcal{G})$.
To determine $A$, we can follow Sasaki et al. (2022) to set $A=\log\left(\log(n)\right)$. 
	 Note that when the Riesz representer does not rely on $\mathcal{G}_{\delta}(\cdot)$ or the function space containing $\mathcal{G}_{\delta}(\cdot)$ is a singleton,   Chernozhukov et al. (2022), Sasaki et al. (2022) and Jin et al. (2024) all set
$\lambda_{\gamma}=A\sqrt{\log{p_{h}}/n}$, converging to zero faster than Eq. (3.6).

	\textbf{Step 4.} (\textit{Estimate OCPPE})
	Plugging the estimators obtained by the previous steps into the orthogonal score gives the estimator of an OCPPE:
	\begin{eqnarray*}
		\widehat{\theta}(\tau_1,\tau_2,\mathcal{G})&=&\frac{1}{\tau_2-\tau_1}\frac{1}{n}\sum_{i=1}^{n}\Bigg[-\vartheta\left(D_i;\mathcal{G}\right)\widehat{IDF}(D_i,X_i;\tau_1,\tau_2)\\
		&&-\widehat{L}(D_i,X_i;\mathcal{G})\left(\widehat{IF}(D_i,X_i;\tau_1,\tau_2)-\int_{\widehat{Q}_Y(\tau_1)}^{\widehat{Q}_Y(\tau_2)}1\{Y_i<y\}dy\right)\\
		&&-\left(\frac{1}{n}\sum_{j=1}^{n}\frac{\vartheta\left(D_j;\mathcal{G}\right)\widehat{DF}\left(D_j,X_j;\widehat{Q}_Y(\tau_1)\right)}{\widehat{f}_Y\left(\widehat{Q}_Y(\tau_1)\right)}\right)\left(1\left\{Y_i\leq \widehat{Q}_Y(\tau_1)\right\}-\tau_1\right)\\
		&&+\left(\frac{1}{n}\sum_{j=1}^{n}\frac{\vartheta\left(D_j;\mathcal{G}\right)\widehat{DF}\left(D_j,X_j;\widehat{Q}_Y(\tau_2)\right)}{\widehat{f}_Y\left(\widehat{Q}_Y(\tau_2)\right)}\right)\left(1\left\{Y_i\leq \widehat{Q}_Y(\tau_2)\right\}-\tau_2\right)\Bigg].
	\end{eqnarray*}

\section{Asymptotic Properties}
	
	In this section, we establish the asymptotic properties of $\widehat{\theta}(\tau_1,\tau_2,\mathcal{G})$. To facilitate application of the uniformity over $\mathcal{G}_{\delta}(\cdot)$, we assume $\mathcal{G}_{\delta}(\cdot)$'s are indexed  by a  $p_{\sigma}$-dimension real vector $\sigma\in\mathbb{R}^{p_{\sigma}}$, such that $\mathcal{G}_{\delta}(\cdot)\equiv\mathcal{G}_{\delta}(\cdot,\sigma)$, $\sigma\in\mathcal{S}$, with $\mathcal{S}$ a compact set of $\mathbb{R}^{p_{\sigma}}$. Although parameterizing the function space containing $\mathcal{G}_{\delta}(\cdot)$  by a finite vector $\sigma\in\mathcal{S}$ seems restrictive, it is flexible enough to characterize most counterfactual interventions considered in empirical studies. For example, let $D$ be the consumption level of a commodity. To characterize that the level of consumption drops more with heavier taxation,
one may set $\mathcal{G}_{\delta}(D,\sigma)=D/(1+\sigma\delta)$, where $\sigma>0$ quantifies the intensity of tax burden. A general location-scale transformation of $D$ can be expressed as  $\mathcal{G}_{\delta}(D,\sigma)=(D-\sigma_1)\times(1+\sigma_{2}\delta)+\sigma_{1}+\sigma_{3}\delta$, with $\sigma=(\sigma_{1},\sigma_{2},\sigma_{3})$. More generally, let $\left\{g_{0}(D,\sigma):\sigma\in\mathcal{S}\right\}$ be a sequence of target variables that the policymaker desires $D$ to become, with $g_{0}(D,\sigma)$ parameterized by a finite real vector $\sigma$.   Then $\mathcal{G}_{\delta}(D,\sigma)=D+\delta\left(g_{0}(D,\sigma)-D\right)$ represents a class of interventions that induce a perturbation of $D$ towards the target $g_{0}(D,\sigma)$.
	
	  Let $u=(\tau_{1},\tau_{2},\sigma)$. Because $\mathcal{G}$ is fully represented by $\sigma$, we omit the dependence on $\mathcal{G}$, e.g., abbreviating $\vartheta(D;\mathcal{G}(\sigma))$ to $\vartheta(D;\sigma)$, $\theta(\tau_1,\tau_2,\mathcal{G}(\sigma))$ to $\theta(\tau_1,\tau_2,\sigma)=\theta(u)$, and so forth. Section 4 is organized as follows:
Section 4.1.1 proves that the DML estimator is uniformly Gaussian in $u$ under some high-level conditions. We show  the DML estimator achieves semi-parametric efficiency in Section 4.1.2 and establish the uniform validity of the multiplier bootstrap method used for constructing uniform confidence bands in Section 4.1.3. We end Section 4.1 by providing several applications of the uniformity results. Section 4.2 provides sufficient low-level conditions for the high-level conditions  introduced in Section 4.1.1 to hold.   We also discuss on how accounting for selection errors uniformly over $\sigma\in \mathcal{S}$ affect the convergence rate of  $L\left(d,x,\sigma\right)$.

	\subsection{Main Results}
	
        Let $\epsilon_{n}\to{0^{+}}$ be a fixed sequence of numbers approaching zero at a speed at most polynomial in $n$. Let  $C$, $H$, $\overline{c}$ and $\underline{c}$ be positive constants.
	
	\subsubsection{Uniform Gaussianity of  DML Estimator}
	
	\noindent\textbf{Assumption 4.1.} The random element $W$ takes values in a compact measure space $(\mathcal{W},\mathcal{A}_{\mathcal{W}})$ and its law is determined by a probability measure $P$. The observed data $\{W_{i}\}_{i=1}^{n}$ consist of $n$ i.i.d. copies of a random element $W=(Y,D,X)\in\mathcal{W}\subset\mathbb{R}^{2+d_{X}}$.
	
	\noindent\textbf{Assumption 4.2.} Let $u=(\tau_{1},\tau_{2},\sigma)\in\mathcal{U}\subset(0,1)^{2}\times\mathcal{S}$ be the index of target parameter $\theta$. $\mathcal{U}$ is a totally bounded metric space equipped with a semi-metric $d_{\mathcal{U}}$.\footnote{Our OCPPEs are defined on $(\tau_{1},\tau_{2},\sigma)\in (0,1) \times (0,1)\times\mathcal{S}$ with $\tau_1<\tau_2$. Let $c_{0}, c_{1}>0$ and $c_{2}<1$ be three constants. The metric space $\mathcal{U}$ in our paper can be defined as $\mathcal{U}=\left\{(\tau_{1},\tau_{2},\sigma):c_{1}\leq{\tau_{1}+c_{0}}\leq{\tau_{2}}\leq{c_{2}},\sigma\in\mathcal{S}\right\}$.} Denote $W(u)$ as a measurable transform of $W$ and $u$. Specifically, we have
	\[
	W(u)\in\left\{\displaystyle\int_{Q_{Y}(\tau_{1})}^{Q_{Y}(\tau_{2})}1\{Y\leq{y}\}dy,1\Big\{Q_{Y}(\tau_{1})\leq{Y}\leq{Q_{Y}(\tau_{2})}\Big\},\vartheta\left(D;\sigma\right)\right\}
	\]
	in this paper. The map $u\mapsto{W(u)}$ obeys the following uniform continuity property:
	\[
	\lim\limits_{\epsilon\to{0^{+}}} \sup_{d_{\mathcal{U}}(u,\bar{u})\leq\epsilon}\left\Vert W(u)-W(\bar{u})\right\Vert_{P,2}=0, \quad E\sup_{u\in\mathcal{U}}|W(u)|^{2+c}<\infty,
	\]
	where the supremum in the first expression is taken over $u,\bar{u}\in\mathcal{U}$.
	
	Assumption 4.2 defines a valid metric space for OCPPEs and restricts the continuity and boundedness of $W$. According to Assumption 4.2, there exists a positive constant, $H<1/2$, that ensures  $\mathcal{U}\subset[H,1-H]\times[H,1-H]\times\mathcal{S}$. Denote the space $\mathcal{H}=\{Q_{Y}(\tau): H\leq\tau\leq{1-H}\}$.
	
	\noindent\textbf{Assumption 4.3.} Assume the functions $F_{Y}(y|d,x)$ and $L(d,x;\sigma)$ can be approximated by
	\[
	F_{Y}(y|d,x)=\Lambda\bigg(b(d,x)^{\prime}\beta(y)\bigg)+r_{F}(d,x,y),
	\]
	and
	\[
	L(d,x;\sigma)=h(d,x)^{\prime}\gamma(\sigma)+r_{L}(d,x,\sigma),
	\]
	where $r_{F}(d,x,y)$ and $r_{L}(d,x,\sigma)$ are the approximation errors. Then uniformly over $y\in\mathcal{H}$ and $u\in\mathcal{U}$,
	
	\begin{enumerate}[(i)]
		\item 
		\begin{enumerate}[(a)]
			\item The sparsity conditions $\Vert\beta(y)\Vert_{0}\leq{s_{\beta}}$ and $\Vert\gamma(\sigma)\Vert_{0}\leq{s_{\gamma}}$ hold.
			\item The approximation errors satisfy $\Vert \partial_{D}r_{F}\Vert_{P,2}=o\left(n^{-1/4}\right)$, $\Vert \partial_{D}r_{F}\Vert_{P,\infty}=o\left(1\right)$ and $\Vert r_{L}\Vert_{P,2}=o\left(n^{-1/4}\right)$, $\Vert r_{L}\Vert_{P,\infty}=o(1)$.
			\item Let $\left\Vert\left\Vert\partial^{l}_{D} b(D,X)\right\Vert_{\infty}\right\Vert_{P,\infty}\leq{K_{nb}}$ and $\left\Vert\left\Vert h(D,X)\right\Vert_{\infty}\right\Vert_{P,\infty}\leq{K_{nh}}$ for $l\in\{0,1\}$.\footnote{The definition of $\partial_{D}^{0}$ denotes the primitive function, for example, $\partial_{D}^{0}b(D,X)=b(D,X)$.} The sparsity indices $s_{\beta}$, $s_{\gamma}$ and the numbers of terms $p_{b}$, $p_{h}$ in the vectors $b(d,x)$, $h(d,x)$ obeying $K_{nb}^{2}s_{\beta}^{2}\log^{2}(p_{b}\vee n)/n=o(1)$ and $K_{nh}^{2}s_{\gamma}^{2}\log^{2}(p_{h}\vee n)/n=o(1)$.
			\item $\mathbb{K}(\cdot)$ is a symmetric, continuous and bounded kernel. Let the corresponding bandwidth $h_{Y}=Cn^{-\nu}$. Then $\nu$ should satisfy $1/8<\nu<1/2$.
		\end{enumerate}
		\item 
		\begin{enumerate}[(a)]
			\item There are estimators $\widehat{\beta}(y)$ and $\widehat{\gamma}(\sigma)$ such that for $l\in\{0,1\}$, the estimation errors satisfy 
			\[\begin{aligned}
				&\left\Vert \partial^{l}_{D}b(D,X)^{\prime}\left(\widehat{\beta}(y)-\beta(y)\right)\right\Vert_{\mathbb{P}_{n},2}=o_{p}\left(n^{-1/4}\right), \quad K_{nb}\left\Vert \widehat{\beta}(y)-\beta(y)\right\Vert_{1}=o_{p}\left(1\right), \\
				&\left\Vert h(D,X)^{\prime}\left(\widehat{\gamma}(\sigma)-\gamma(\sigma)\right)\right\Vert_{\mathbb{P}_{n},2}=o_{p}\left(n^{-1/4}\right),\quad K_{nh}\left\Vert \widehat{\gamma}(\sigma)-\gamma(\sigma)\right\Vert_{1}=o_{p}(1). \\
			\end{aligned}\] 
			\item With probability approaching to 1,
			the estimators are sparse such that $\left\Vert\widehat{\beta}(y)\right\Vert_{0}\leq Cs_{\beta}$ and $\Vert\widehat{\gamma}(\sigma)\Vert_{0}\leq Cs_{\gamma}$.
		\end{enumerate}
		\item 
		\begin{enumerate}[(a)]
			\item The empirical and population norms induced by the Gram matrix formed by $\{b_{j}(d,x)\}_{j=1}^{p_{b}}$ and $\{h_{j}(d,x)\}_{j=1}^{p_{h}}$ are equivalent on sparse subsets, such as for $l\in\{0,1\}$,
			\[
			\sup_{\Vert\upsilon\Vert_{0}\leq{s_\beta\cdot\log{n}}}\left|\frac{\Vert \partial_{D}^{l}b(D,X)^{\prime}\upsilon\Vert_{\mathbb{P}_{n},2}}{\Vert \partial_{D}^{l}b(D,X)^{\prime}\upsilon\Vert_{P,2}}-1\right|\leq{\epsilon_{n}} \quad\text{and}\quad \sup_{\Vert\upsilon\Vert_{0}\leq{s_\gamma\cdot\log{n}}}\left|\frac{\Vert h(D,X)^{\prime}\upsilon\Vert_{\mathbb{P}_{n},2}}{\Vert h(D,X)^{\prime}\upsilon\Vert_{P,2}}-1\right|\leq{\epsilon_{n}}.
			\]
			\item $F_{Y}(y|d,x)$ and $\partial_{d}F_{Y}(y|d,x)$ are continuously differentiable and bounded with respect to $y$.  $\Lambda(\cdot)$ is twice continuously differentiable and bounded away from zero and infinity. $L(d,x;\cdot)$ is continuously differentiable and bounded with respect to $(d,x)$. $f_Y(\cdot)$ is continuously differentiable and bounded away from zero and infinity. 
			\item The boundedness conditions hold: $\Vert W(u)\Vert_{P,\infty}\leq{C}$, $\left\Vert\frac{\partial_{D}f(Y|D,X)}{f(Y|D,X)}\right\Vert_{P,2}\leq{C}$ .
		\end{enumerate}
	\end{enumerate}
	
Assumption 4.3 imposes high-level conditions which
encode both the approximate sparsity of the models as well as reasonable behavior
of the sparse estimators of $F_{Y}\left(y|d,x\right)$ and $L\left(d,x,\sigma\right)$. These conditions extend Jin et al. (2024) where $\mathcal{S}$ is a singleton.  
Primitive conditions for Lasso estimators to satisfy various bounds in Assumption 4.3(ii) are provided in Section 4.2.

	\noindent\textbf{Theorem 4.1.} \textit{Suppose that the assumptions in Proposition 2.2 and Assumptions 4.1-4.3 hold, then 
		\[
		\widehat{Z}_{n}(u)=\sqrt{n}\left(\widehat{\theta}(u)-\theta(u)\right)=Z_{n}(u)+o_{p}(1) \quad in \quad \mathbb{D}=\ell^{\infty}\left(\mathcal{U}\right)
		\]
		where $Z_{n}(u)=\mathbb{G}_{n}\psi(W,\theta(u),\eta;u)$. The process $\widehat{Z}_{n}(u)$ is asymptotically Gaussian, namely
		\[
		\widehat{Z}_{n}(u)\leadsto Z(u) \quad in \quad \mathbb{D}=\ell^{\infty}\left(\mathcal{U}\right)
		\]
		where $Z(u)=\mathbb{G}\psi(W,\theta(u),\eta;u)$ with $\mathbb{G}$ denoting Gaussian process and with $Z(u)$ having bounded, uniformly continuous paths:
		\[
		E\sup_{u\in\mathcal{U}}\Vert{Z}(u)\Vert<\infty, \quad \lim_{\epsilon\to{0}^{+}}E\sup_{d_{\mathcal{U}}\left(u,\widetilde{u}\right)\leq{\epsilon}}\Vert{Z}(u)-{Z}\left(\widetilde{u}\right)\Vert=0.
		\]}
	
	\subsubsection{Efficiency of DML Estimator }

	In the causal inference literature,  Hahn (1998) calculated the semi-parametric efficiency bounds for the ATE and the ATE on the treated. Firpo (2007) calculated the efficiency bounds for unconditional QTE and QTE on the treated. Fr\"{o}lich and Melly (2013) derived the efficiency bound for unconditional local QTE for compliers. An interesting question is whether there exists finite semi-parametric efficiency bound for unconditional partial effects of a general counterfactual policy.
Such a question has not been answered by Rothe (2012) and Mart\'{i}nez-Iriarte et al. (2024). The next theorem shows that the doubly robust score derived in Proposition 2.2 is efficient and the DML estimator achieves this bound.

	\vspace{5pt}
	
	\noindent\textbf{Theorem 4.2.} \textit{Suppose the Assumptions in Theorem 4.1 hold, then for any $u\in\mathcal{U}$, the semi-parametric efficiency bound of $\theta(u)$ is $E\left[\psi\big(W,\theta(u),\eta;u\big)\right]^2$.}

	\subsubsection{Multiplier Bootstrap}
	
	In practice, inference based on directly estimating the asymptotic variance of the limit process can be overly complicated. In such cases, bootstrap methods can effectively be applied to construct the confidence bands. Let $\{\xi_i\}_{i=1}^{n}$ be a random sample drawn from a distribution with zero mean and unit variance. We then define the estimated multiplier process for $Z(u)$ as follows:
	\[
	\widehat{Z}_{n}^{*}(u)=\sqrt{n}\left(\widehat{\theta}^{*}(u)-\widehat{\theta}(u)\right)=\frac{1}{\sqrt{n}}\sum_{i=1}^{n}\xi_{i}\psi\bigg(W_{i},\widehat{\theta},\widehat{\eta};u\bigg).
	\]
	The main result of this section shows that the bootstrap law for the process, $\widehat{Z}_{n}^{*}(u)$, provides a valid approximation to the large-sample law for $\widehat{Z}_{n}(u)$. 
	
	We develop such validity by imposing the following regular assumption:
	
	\noindent\textbf{Assumption 4.4.} A random element, $\xi$, with values in a measure space, $(\Omega,\mathcal{A}_{\Omega})$ that is independent of $(\mathcal{W},\mathcal{A}_{\mathcal{W}})$, and a law determined by a probability measure, $P_{\xi}$, with zero mean and unit variance. The observed data, $\{\xi_{i}\}_{i=1}^{n}$, comprise $n$ i.i.d. copies of a random element, $\xi$.
	
	We introduce some useful notations to describe the following results. We define the conditional weak convergence of the bootstrap law in probability, denoted by $\widetilde{Z}_{n}(u)\leadsto_{B}\widetilde{Z}(u)$ in $\ell^{\infty}\left(\mathcal{U}\right)$, by
	\[
	\sup_{T\in{BL_{1}\left(\ell^{\infty}\left(\mathcal{U}\right)\right)}}\left|E_{\xi|P}T\left(\widetilde{Z}_{n}(u)\right)-ET\left(\widetilde{Z}(u)\right)\right|=o_{p}(1),
	\]
	where $BL_{1}\left(\mathbb{D}\right)$ denotes the space of functions mapping $\mathbb{D}$ to [0,1] with Lipschitz norm at most 1, and $E_{\xi|P}$ denote the expectation over the multiplier weights $\{\xi_{i}\}_{i=1}^{n}$ holding the data $\{W_{i}\}_{i=1}^{n}$ fixed.
		\vspace{5pt}
	
	\noindent\textbf{Theorem 4.3.} \textit{Suppose that the assumptions in Theorem 4.1 and Assumption 4.4 hold, the bootstrap law consistently approximates the large sample law $Z(u)$ of $Z_{n}(u)$, namely}
	\[
	\widehat{Z}_{n}^{*}(u)\leadsto_{B} Z(u) \quad in \quad \mathbb{D}=\ell^{\infty}\left(\mathcal{U}\right).
	\]

	One application of the uniform multiplier bootstrap procedure is to perform various tests such as treatment nullity and homogeneity for a given counterfactual intervention.
	
	\noindent\textit{Example 4.1.} We test the treatment homogeneity across quantile ranges of $Y$ for a given counterfactual intervention $\mathcal{G}$ indexed by $\sigma_{0}\in\mathcal{S}$.
\begin{eqnarray*}
	\mathbf{H_{10}}: \theta\left(\tau_{1},\tau_{2},\sigma_{0}\right)=\theta\left(\tau_{1}',\tau_{2}',\sigma_{0}\right) \  \ for \ \ any \  \  a<\tau_{1}<\tau_{2}<1-a,\ a<\tau_{1}'<\tau_{2}'<1-a,
	\end{eqnarray*}
	with $0<a<1/2$ a fixed constant. To test this hypothesis, we use
	\begin{eqnarray*}
	\sup_{(\tau_{1},\tau_{2})\subset [a,1-a]}\sqrt{n}\left|\widehat{\theta}\left(\tau_{1},\tau_{2},\sigma_{0}\right)-\dfrac{2}{(1-2a)^{2}}\displaystyle\int_{a<t_{1}<
t_{2}<1-a}\widehat{\theta}\left(t_{1},t_{2},\sigma_{0}\right)dt_{1}dt_{2}\right|
	\end{eqnarray*}
	 as the test statistic, and use

\begin{eqnarray*}
	\sup_{(\tau_{1},\tau_{2})\subset [a,1-a]}\sqrt{n}\left|\widehat{Z}_{n}^{*}\left(\tau_{1},\tau_{2},\sigma_{0}\right)-\dfrac{2}{(1-2a)^{2}}\displaystyle\int_{a<t_{1}<
t_{2}<1-a}\widehat{Z}_{n}^{*}\left(t_1,t_2,\sigma_{0}\right)dt_{1}dt_{2}\right|
	\end{eqnarray*}
	 to simulate its asymptotic distribution.

	Another application of the uniform multiplier bootstrap is to test treatment homogeneity and policy optimality uniformly in a class of policy interventions $\sigma\in\mathcal{S}$.

	\noindent\textit{Example 4.2.} A policymaker may be interested in whether a class of interventions indexed by $\sigma\in\mathcal{S}$ has similar impacts on a given target subgroup indexed by $(\tau_1,\tau_2)$:
	\[
	\mathbf{H_{20}}: \theta\left(\tau_{1},\tau_{2},\sigma\right)=\theta\left(\tau_{1},\tau_{2},\sigma'\right) \  \ for \ \ any \  \ \sigma,\sigma'\in\mathcal{S}.
	\]
Let $\#\mathcal{S}$ denote the area of space $\mathcal{S}$.	 To test this hypothesis, we construct 
\begin{eqnarray*}
\sup_{\sigma\in\mathcal{S}}\sqrt{n}\left|\widehat{\theta}\left(\tau_{1},\tau_{2},\sigma\right)-\dfrac{1}{\#\mathcal{S}}\displaystyle\int_{\mathcal{S}}\widehat{\theta}\left(\tau_{1},\tau_{2},{\sigma}'\right)d{\sigma}'\right|
 \end{eqnarray*}
as the test statistic, and use
\begin{eqnarray*} 
 \sup_{\sigma\in\mathcal{S}}\sqrt{n}\left|\widehat{Z}_{n}^{*}\left(\tau_{1},\tau_{2},\sigma\right)-\dfrac{1}{\#\mathcal{S}}\displaystyle\int_{\mathcal{S}}\widehat{Z}_{n}^{*}\left(\tau_{1},\tau_{2},{\sigma}'\right)d{\sigma}'\right|
\end{eqnarray*}  
to simulate its asymptotic distribution. If the test statistic value exceeds the simulated critical value, we reject the null hypothesis; otherwise, we fail to reject it.
	
	\noindent\textit{Example 4.3.} For a target subgroup $(\tau_1,\tau_2)$, a policymaker wants to learn whether a given policy intervention, denoted by $\sigma^{*}$, is uniformly superior to others indexed by $\sigma\in\mathcal{S}$:
	\begin{eqnarray*}
	\mathbf{H_{30}}: \theta\left(\tau_{1},\tau_{2},\sigma^{*}\right)\geq\theta\left(\tau_{1},\tau_{2},\sigma\right) \  \ for \ \ any \  \ \sigma\in\mathcal{S}.
	\end{eqnarray*}
	To test this hypothesis, we reformulate the null hypothesis  as
	\begin{eqnarray*}
	\mathbf{H_{30}}: \max\Big\{\theta\left(\tau_{1},\tau_{2},\sigma\right)-{\theta\left(\tau_{1},\tau_{2},\sigma^{*}\right)},0\Big\}=0 \  \ for \ \ any \  \ \sigma\in\mathcal{S}.
	\end{eqnarray*}
	We construct the test statistic  
\begin{eqnarray*}
\sup_{\sigma\in\mathcal{S}}\sqrt{n}\max\Big\{\widehat{\theta}\left(\tau_{1},\tau_{2},\sigma\right)-\widehat{\theta}\left(\tau_{1},\tau_{2},\sigma^{*}\right),0\Big\},
\end{eqnarray*} 
and use 
\begin{eqnarray*}
\sup_{\sigma\in\mathcal{S}}\sqrt{n}\max\Big\{\widehat{Z}_{n}^{*}\left(\tau_{1},\tau_{2},\sigma\right)-\widehat{Z}_{n}^{*}\left(\tau_{1},\tau_{2},\sigma^{*}\right),0\Big\}
\end{eqnarray*}
 to simulate its asymptotic distribution.

	\subsection{Primitive Conditions for Uniform Convergence of Nuisance Functions}
	
	This subsection provides primitive conditions for Lasso estimators of $F_{Y}(y|d,x)$ and Lasso minimum distance estimator of $L(d,x,\sigma)$
to satisfy the bounds specified in Assumption 4.3(ii).

	\subsubsection{Lasso Estimator of $F_{Y}(y|d,x)$}
	
	We now list sufficient conditions for the (Post-)Lasso estimator $\widehat{\beta}(y)$ (described in Step 2, Section 3) to satisfy the following bounds specified in Assumption 4.3(ii):
	\[
	\left\Vert \partial^{l}_{D}b(D,X)^{\prime}\left(\widehat{\beta}(y)-\beta(y)\right)\right\Vert_{\mathbb{P}_{n},2}=o_{p}\left(n^{-1/4}\right), \quad K_{nb}\left\Vert \widehat{\beta}(y)-\beta(y)\right\Vert_{1}=o_{p}\left(1\right)
	\]
	and $\left\Vert \widehat{\beta}(y)\right\Vert_{0}\leq{Cs}_{\beta}$ uniformly over $y\in\mathcal{H}$. 
	
	\noindent\textbf{Assumption 4.5.} For some generic positive constants $\overline{c}$ and $\underline{c}$ (which may vary case by case), 
	\begin{enumerate}[(i)]
		\item $\underline{c}\leq\|b_{j}(D,X)\|_{P,2}\leq\overline{c}$ for any $j\in\{1,2,\dots,p_{b}\}$; 
		\item $\partial_{d}F_{Y}(y|d,x)$ is bounded by $\overline{c}$ uniformly over $(y,d,x)$; 
		\item $\partial_{d}b(d,x)'\beta(y)$ is bounded by $\overline{c}$ uniformly over $(y,d,x)$.
	\end{enumerate}
	
	\noindent\textbf{Assumption 4.6.} Assume the function $F_{Y}(y|d,x)$ can be approximated by
	\[
	F_{Y}(y|d,x)=\Lambda\bigg(b(d,x)^{\prime}\beta(y)\bigg)+r_{F}(d,x,y),
	\]
	where $r_{F}(d,x,y)$ is the approximation error. Then uniformly over $y\in\mathcal{H}$, 
	
	\begin{enumerate}[(i)]
		\item the sparsity condition $\Vert\beta(y)\Vert_{0}\leq{s_{\beta}}$ holds; 
		\item the approximation error satisfies $\Vert \partial_{D}r_{F}\Vert_{P,2}=O\left(\sqrt{s_{\beta}\log(p_{b}\vee{n})/n}\right)$ and $\Vert \partial_{D}r_{F}\Vert_{P,\infty}=O\left(\sqrt{K_{nb}^{2}s_{\beta}^{2}\log(p_{b}\vee{n})/n}\right)$; further, $K_{nb}^{2}s_{\beta}^{2}\log^{2}(p_{b}\vee{n})/n=o(1)$;
		\item the penalty level $\lambda_{\beta}$ is chosen as $\lambda_{\beta}=C\sqrt{n}\Phi^{-1}\Big(1-\delta_{n}/(2p_{b}n)\Big)$ for some constant $C>1$ and $\delta_{n}=o(1)$.
	\end{enumerate}
	
	\noindent\textbf{Assumption 4.7.} For some generic positive constants $\overline{c}$ and $\underline{c}$, with probability approaching to 1 and $l\in\{0,1\}$,
	\[
	\underline{c}\leq\inf_{1\leq\Vert{v}\Vert_{0}\leq{s_{\beta}\cdot\log{n}}}\frac{\Big\Vert{\partial^{l}_{D}b(D,X)'v}\Big\Vert_{\mathbb{P}_{n},2}}{\Vert{v}\Vert}\leq\sup_{1\leq\Vert{v}\Vert_{0}\leq{s_{\beta}\cdot\log{n}}}\frac{\Big\Vert{\partial^{l}_{D}b(D,X)'v}\Big\Vert_{\mathbb{P}_{n},2}}{\Vert{v}\Vert}\leq\overline{c}.
	\]
	
	Assumptions 4.5-4.7 are common in the literature on logistic regressions with an $\ell_{1}$ penalty. See, for instance, Belloni et al. (2017) and Sasaki et al. (2022). We note that Assumption 4.6(ii) implies the condition on $\partial_{d}r_{F}$ stated in Assumption 4.3(i,b) once $K_{nb}^{2}s_{\beta}^{2}\log^{2}(p_{b}\vee n)/n=o(1)$ holds. Based on these low-level conditions, we can obtain sharper uniform convergence rate of $\widehat{\beta}(y)$, as stated by the following theorem.
	\vspace{5pt}
	
	\noindent\textbf{Theorem 4.4.} \textit{If Assumptions 4.1-4.2, 4.3(iii) together with 4.5-4.7 hold, then for $l\in\{0,1\}$,
		\[
		\sup_{y\in\mathcal{H}}\left\Vert \partial^{l}_{D}b(D,X)^{\prime}\left(\widehat{\beta}(y)-\beta(y)\right)\right\Vert_{\mathbb{P}_{n},2}=O_{p}\left(\sqrt{\frac{s_{\beta}\log(p_{b}\vee{n})}{n}}\right)
		\]
		and
		\[
		K_{nb}\sup_{y\in\mathcal{H}}\left\Vert \widehat{\beta}(y)-\beta(y)\right\Vert_{1}=O_{p}\left(\sqrt{\frac{K_{nb}^{2}s_{\beta}^{2}\log(p_{b}\vee{n})}{n}}\right).
		\]
		Furthermore, $\widehat{\beta}(y)$ is uniformly sparse, such that with probability approaching to 1,
		\[
		\sup_{y\in\mathcal{H}}\left\|\widehat{\beta}(y)\right\|_{0}\leq{C}s_{\beta}.
		\] 
	}
	
		\noindent\textbf{Remark 4.1.} Under the condition $K_{nb}^{2}s_{\beta}^{2}\log^{2}(p_{b}\vee n)/n=o(1)$, the uniform convergence rate established in Theorem 4.4 yields the desired results $\sqrt{s_{\beta}\log(p_{b}\vee{n})/n}=o\left(n^{-1/4}\right)$ and $\sqrt{K_{nb}^{2}s_{\beta}^{2}\log(p_{b}\vee{n})/n}=o\left(1\right)$, which satisfies Assumption 4.3(ii).
	
	\subsubsection{Lasso Minimum Distance Learner of $L(d,x,\sigma)$}
	
	Assumption 4.3(ii) requires the Lasso minimum distance learner should converge faster than a usual rate, such as
	\[
	\left\Vert h(D,X)^{\prime}\left(\widehat{\gamma}(\sigma)-\gamma(\sigma)\right)\right\Vert_{\mathbb{P}_{n},2}=o_{p}\left(n^{-1/4}\right),\quad K_{nh}\left\Vert \widehat{\gamma}(\sigma)-\gamma(\sigma)\right\Vert_{1}=o_{p}(1)
	\]
	and $\left\|\widehat{\gamma}(\sigma)\right\|_{0}\leq{C}s_{\gamma}$ uniformly over $\sigma\in\mathcal{S}$. Below we provide sufficient conditions to achieve these bounds.
	
	\noindent\textbf{Assumption 4.8.} There exists a constant $C$ such that for $l\in\{0,1\}$, $\left\|\partial_{D}^{l}h(D,X)\right\|_{\infty}\leq{C}$ with probability approaching to 1.
	
	\noindent\textbf{Assumption 4.9.} Assume the function $L(d,x;\sigma)$ can be approximated by
	\[
	L(d,x;\sigma)=h(d,x)^{\prime}\gamma(\sigma)+r_{L}(d,x,\sigma),
	\]
	where $r_{L}(d,x,\sigma)$ is the approximation error. Then uniformly over $\sigma\in\mathcal{S}$,
	
	\begin{enumerate}[(i)]
		\item the sparsity condition $\Vert\gamma(\sigma)\Vert_{0}\leq{s_{\gamma}}$ holds, where $s_{\gamma}=O\left(\left(\log(p_{h}\vee{n})/n\right)^{-1/(1+2\xi)}\right)$ for $\xi\geq1/2$; 
		\item the approximation error satisfies  $\Vert r_{L}\Vert_{P,2}=O\left(s_{\gamma}^{-\xi}\right)$ and $\Vert r_{L}\Vert_{P,\infty}=O\left(K_{nh}s_{\gamma}^{1/2-\xi}\right)$; further, $K_{nh}^{2}s_{\gamma}^{2}\log^{2}(p_{h}\vee n)/n=o(1)$;
		\item the penalty level $\lambda_{\gamma}$ is chosen as $\lambda_{\gamma}=\kappa_{n}\sqrt{\log(p_{h}\vee{n})/n}$ with $\kappa_{n}=O\left(\log(\log{n})\right)$.
	\end{enumerate}

	\noindent\textbf{Assumption 4.10.} For some positive constants $\overline{c}$ and $\underline{c}$, with probability approaching to 1,
	\[
	\underline{c}\leq\inf_{1\leq\Vert{v}\Vert_{0}\leq{s_{\gamma}\cdot\log{n}}}\frac{\Big\Vert{h(D,X)'v}\Big\Vert_{\mathbb{P}_{n},2}}{\Vert{v}\Vert}\leq\sup_{1\leq\Vert{v}\Vert_{0}\leq{s_{\gamma}\cdot\log{n}}}\frac{\Big\Vert{h(D,X)'v}\Big\Vert_{\mathbb{P}_{n},2}}{\Vert{v}\Vert}\leq\overline{c}.
	\]
	
	Assumptions 4.8-4.10 follow Chernozhukov et al. (2022) and Sasaki et al. (2022), besides that $\log{p_{h}}$ is replaced with $\log(p_{h}\vee{n})$, to ensure uniform bounds for Lasso minimum distance learners. We also note that Assumption 4.9(i)-(ii) implies the condition on $r_{L}$ stated in Assumption 4.3(i,b) once $K_{nh}^{2}s_{\gamma}^{2}\log^{2}(p_{h}\vee n)/n=o(1)$ holds. 
	\vspace{5pt}
	
	\noindent\textbf{Theorem 4.5.} \textit{If Assumptions 4.1-4.2, 4.3(iii) together with 4.8-4.10 hold, then 
		\[
		\sup_{\sigma\in\mathcal{S}}\left\Vert h(D,X)^{\prime}\left(\widehat{\gamma}(\sigma)-{\gamma}(\sigma)\right)\right\Vert_{\mathbb{P}_{n},2}=O_{p}\left(\kappa_{n}\left(\frac{\log(p_{h}\vee{n})}{n}\right)^{\frac{\xi}{1+2\xi}}\right)
		\]
		and
		\[
		K_{nh}\sup_{\sigma\in\mathcal{S}}\left\Vert \widehat{\gamma}(\sigma)-{\gamma}(\sigma)\right\Vert_{1}=O_{p}\left(\kappa_{n}K_{nh}\left(\frac{\log(p_{h}\vee{n})}{n}\right)^{\frac{2\xi-1}{2(1+2\xi)}}\right).
		\]
		Furthermore, $\widehat{\gamma}(\sigma)$ is uniformly sparse, such that with probability approaching to 1,
		\[
		\sup_{\sigma\in\mathcal{S}}\left\|\widehat{\gamma}(\sigma)\right\|_{0}\leq{C}s_{\gamma}.
		\]
	}
	
\noindent\textbf{Remark 4.2.} Theorem 4.5 provides the uniform convergence rate for Lasso minimum distance learners. To our best knowledge, this result is new in the literature on
direct estimation of the Riesz representer for a broad class of casual quantities in a high-dimensional setting.
 The conventional (pointwise) convergence rate for an automatic estimator is\footnote{Notice that the empirical and population norms are asymptotically equivalent under Assumption 4.3(iii).}
	\[
	\left\Vert h(D,X)^{\prime}\left(\widehat{\gamma}(\sigma)-{\gamma}(\sigma)\right)\right\Vert_{\mathbb{P}_{n},2}=O_{p}\left(\kappa_{n}\left(\frac{\log p_{h}}{n}\right)^{\frac{\xi}{1+2\xi}}\right)
	\]
	for any given $\sigma\in\mathcal{S}$. See, for example, Theorem 1 in Chernozhukov et al. (2022), Lemma D.2 in Chernozhukov et al. (2022a) or Theorem A.2 in Sasaki et al. (2022). We show that to ensure uniformity in $\sigma\in\mathcal{S}$, the term $\log p_{h}$ should be replaced by  $\log(p_{h}\vee{n})$. The intuition is that $\widehat{M}(\sigma)$  converges pointwise to $M(\sigma)$ at the rate of $\sqrt{\log{p_{h}}/n}$. See, for example, Assumption 6 in Chernozhukov et al. (2022), Assumption D.2 in Chernozhukov et al. (2022a) or Assumption 3.2 in Sasaki et al. (2022). However, the uniform convergence rate of $\widehat{M}(\sigma)$ with respect to $\sigma\in\mathcal{S}$ should be corrected to $\sqrt{\log(p_{h}\vee{n})/n}$, yielding a slower convergence rate of $L(d,x,\sigma)$.
	
	\noindent\textbf{Remark 4.3.} Under the conditions $s_{\gamma}=O\left(\left(\log(p_{h}\vee{n})/n\right)^{-1/(1+2\xi)}\right)$ and $K_{nh}^{2}s_{\gamma}^{2}\log^{2}(p_{h}\vee n)/n=o(1)$, the uniform convergence rate in Theorem 4.5 yields the following desired results $\kappa_{n}\left(\log(p_{h}\vee{n})/n\right)^{\xi/(1+2\xi)}=o\left(n^{-1/4}\right)$ and $\kappa_{n}K_{nh}\left(\log(p_{h}\vee{n})/n\right)^{(2\xi-1)/(2+4 \xi)}=o\left(1\right)$ with $\xi>1/2$, which satisfies Assumption 4.3(ii).

\section{Extensions}
	
	In this section, we consider several important extensions of  OCPPE. In Section 5.1, we provide doubly robust identification of OCPPE when the policymaker changes the marginal CDF of the target variable $D$ instead of changing its value directly. Section 5.2 connects the derived efficient doubly robust score for an OCPPE (in Proposition 2.2) to optimizing infinitesimal
nudges to a continuous treatment by maximizing a quantile specific empirical welfare function.

	\subsection{OCPPE with Distributional Perturbation}

	In Section 2, we define an OCPPE as the causal effect of counterfactually changing the value of $D$ to $D_{\delta}=\mathcal{G}_{\delta}(D)$. Alternatively, an OCPPE can be defined in terms of
counterfactually changing the unconditional distribution of one covariate while holding everything else constant, in the spirit of Rothe (2012). Let $F_{D}(\cdot)$ be the marginal CDF of $D$. Note that since $D$ is continuously distributed, $F_{D}(\cdot)$ is strictly increasing. There exists a unique random variable $R_{D}\sim^{d}\mbox{Uniform}(0,1)$ which is a one-to-one transformation of $D$, such that $D=F_{D}^{-1}\left(R_{D}\right)$, with $F^{-1}_{D}(\cdot)$ the quantile function of $D$ and $R_{D}$ being the rank of $D$. With this formulation, $Y=m(D,X,U)$ can be equivalently expressed as $Y=m\left(F_{D}^{-1}\left(R_{D}\right),X,U\right)$.
	
	Let $\Psi_{\delta}(\cdot)$ be an element of a continuum of CDFs indexed by $\delta\in\mathbb{R}$ such that $\Psi_{\delta}(\cdot)\rightarrow F_{D}(\cdot)$ as $\delta\rightarrow 0$. For example, $\Psi_{\delta}(\cdot)=F_{D}(\cdot)+\delta\left(G_{0}(\cdot)-F_{D}(\cdot)\right)$, with $G_{0}(\cdot)$ representing the target CDF that a policymaker desires $F_{D}(\cdot)$ to be.
A counterfactual change in the marginal CDF from $F_{D}(\cdot)$ to $\Psi_{\delta}(\cdot)$ induces a counterfactual outcome $Y_{\delta}=m\left(\Psi_{\delta}^{-1}\left(R_{D}\right),X,U\right)$. The outcome conditioned partial policy effect of $D$ on $Y$ associated with a \textit{distributional} intervention characterized by $\Psi_{\delta}(\cdot)$ is
	\begin{eqnarray*}
		\theta^{D}(\tau_1,\tau_2,\Psi)&=&E\left[\lim_{\delta\rightarrow 0}\frac{m\left(\Psi_{\delta}^{-1}\left(R_{D}\right),X,U\right)-m\left(F_{D}^{-1}\left(R_{D}\right),X,U\right)}{\delta}\bigg|Y\in\left(Q_Y(\tau_1),Q_Y(\tau_2)\right)\right]\\
&=&E\left[\lim_{\delta\rightarrow 0}\frac{m\left(\Psi_{\delta}^{-1}\circ F_{D}\left(D\right),X,U\right)-m\left(D,X,U\right)}{\delta}\bigg|Y\in\left(Q_Y(\tau_1),Q_Y(\tau_2)\right)\right].
	\end{eqnarray*}

	\vspace{5pt}
	
	\noindent\textbf{Proposition 5.1.} \textit{Suppose that $\Psi_{\delta}(\cdot)=F_{D}(\cdot)+\delta\left(G_{0}(\cdot)-F_{D}(\cdot)\right)$ with $G_{0}$ the CDF of a continuous random variable.  
Under Assumption 2.2 and other regularity conditions (listed in Appendix A), then
		\[
		\theta^{D}\left(\tau_1,\tau_2,\Psi\right)=\frac{-1}{\tau_2-\tau_1}E\left[\frac{F_{D}(D)-G_{0}(D)}{f_{D}(D)}\int_{Q_Y(\tau_1)}^{Q_Y(\tau_2)}\partial_DF_Y(y|D,X)dy\right].
		\]
	}
	
Similar to the discussion in Section 2 (below Remark 2.1), the estimator of $\theta^{D}\left(\tau_1,\tau_2,\Psi\right)$ based on the formula in Proposition 5.1 is neither robust when $X$ is high dimensional nor efficient. 
	We can show that the doubly robust and orthogonal score for $\theta^{D}\left(\tau_1,\tau_2,\Psi\right)$ is
	\[\begin{aligned}
		&\psi^{D}\left(W,\theta,\eta^{D};\tau_1,\tau_2,\Psi\right)\\
		=&\frac{-1}{\tau_2-\tau_1}\vartheta\Big(F_{D}(D);\Psi\Big)\int_{Q_Y(\tau_1)}^{Q_Y(\tau_2)}\partial_DF_Y(y|D,X)dy-\theta\\
		&+\frac{1}{\tau_2-\tau_1}\frac{\partial_D\left(\vartheta\Big(F_{D}(D);\Psi\Big)f(D,X)\right)}{f(D,X)}\left(\int_{Q_Y(\tau_1)}^{Q_Y(\tau_2)}1\{Y\leq y\}dy-\int_{Q_Y(\tau_1)}^{Q_Y(\tau_2)}F_Y(y|D,X)dy\right)\\
		&+\frac{1}{\tau_2-\tau_1}\int \alpha\left(\widetilde{d};\tau_{1},\tau_{2},\Psi\right)\Big(1\left\{D\le \widetilde{d}\right\}-F_{D}\left(\widetilde{d}\right)\Big)f_{D}\left(\widetilde{d}\right)d\widetilde{d}\\
		&-\frac{1}{\tau_2-\tau_1}\frac{E\left[\vartheta\Big(F_{D}(D);\Psi\Big)\cdot\partial_DF_Y\left(Q_Y(\tau_1)|D,X\right)\right]}{f_Y\left(Q_Y(\tau_1)\right)}\Big(1\{Y\leq Q_Y(\tau_1)\}-\tau_1\Big)\\
		&+\frac{1}{\tau_2-\tau_1}\frac{E\left[\vartheta\Big(F_{D}(D);\Psi\Big)\cdot\partial_DF_Y\left(Q_Y(\tau_2)|D,X\right)\right]}{f_Y\left(Q_Y(\tau_2)\right)}\Big(1\{Y\leq Q_Y(\tau_2)\}-\tau_2\Big), \\
	\end{aligned}\]
	in which $\vartheta\Big(F_{D}(D);\Psi\Big)=\Big(F_{D}(D)-G_{0}(D)\Big)\Big/f_{D}(D)$, the nuisance parameters
	\[\begin{aligned}
		&\eta^{D}(W;\cdot)\\
		=&\Big(Q_{Y}(\cdot),F_Y(\cdot|D,X),F_{D}(D),L(D,X;\cdot),f_{Y}(\cdot),\alpha(D;\cdot),E\Big[\vartheta\Big(F_{D}(D);\cdot\Big)\cdot\partial_DF_Y\left(Q_Y(\cdot)\big|D,X\right)\Big]\Big), \\
	\end{aligned}\]
		where 
	\[
	L(D,X;\Psi)=\frac{\partial_D\left(\vartheta\Big(F_{D}(D);\Psi\Big)f(D,X)\right)}{f(D,X)},
	\]
	 and  
	\[\begin{aligned}
		\alpha(D;\tau_{1},\tau_{2},\Psi)=&-\frac{1}{f_{D}(D)}\partial_{D}\left(\int_{Q_Y(\tau_1)}^{Q_Y(\tau_2)}E\left[\partial_DF_Y(y|D,X)\Big|D\right]dy\cdot\vartheta\Big(F_{D}(D);\Psi\Big)\right)\\
		&-\frac{1}{f_{D}(D)}\int_{Q_Y(\tau_1)}^{Q_Y(\tau_2)}E\left[\partial_DF_Y(y|D,X)\Big|D\right]dy. \\
	\end{aligned}\]

		\noindent\textbf{Theorem 5.1.} \textit{Under the same assumptions as those in Proposition 5.1, we have 
			\begin{enumerate}[(i)]
				\item 
				\[
				E\Big[\psi^{D}\left(W,\theta^{D}(\tau_1,\tau_2,\Psi),\eta^{D};\tau_1,\tau_2,\Psi\right)\Big]=0.
				\]
				\item $\psi^{D}\left(W,\theta^{D}(\tau_1,\tau_2,\Psi),\eta^{D};\tau_1,\tau_2,\Psi\right)$ satisfies the Neyman orthogonal property
				\begin{eqnarray*}
					\frac{\partial E\Big[\psi^{D}\Big(W,\theta^{D}\left(\tau_1,\tau_2,\Psi\right),\eta^{D}+r\left(\widetilde{\eta}^{D}-\eta^{D}\right);\tau_1,\tau_2,\Psi\Big)\Big]}{\partial r}\Big|_{r=0}=0.
				\end{eqnarray*}
				\item The orthogonal score $\psi^{D}$ is doubly robust, such that for any $\widetilde{\eta}^{D}=\left(\widetilde{\eta}_{1},\widetilde{\eta}_{2},\widetilde{\eta}_3,\widetilde{\eta}_4,\widetilde{\eta}_5,\widetilde{\eta}_6,\widetilde{\eta}_7\right)$ belonging to the space of nuisance parameters,
				\[
				\begin{aligned}
					E\Big[\psi^{D}\left(W,\theta,\eta^{D};\tau_1,\tau_2,\Psi\right)\Big]&=E\Big[\psi^{D}\Big(W,\theta,{\eta}_{1},\widetilde{\eta}_{2},\widetilde{\eta}_{3},{\eta}_4,\widetilde{\eta}_5,{\eta}_6,\widetilde{\eta}_7;\tau_1,\tau_2,\Psi\Big)\Big]\\
					&=E\Big[\psi^{D}\Big(W,\theta,{\eta}_{1},{\eta}_{2},{\eta}_3,\widetilde{\eta}_4,\widetilde{\eta}_5,\widetilde{\eta}_6,\widetilde{\eta}_7;\tau_1,\tau_2,\Psi\Big)\Big].
				\end{aligned}
				\]
				\item The semi-parametric efficiency bound of  $\theta^{D}(\tau_1,\tau_2,\Psi)$ is  $E\left[\psi^{D}\left(W,\theta^{D}(\tau_1,\tau_2,\Psi),\eta^{D};\tau_1,\tau_2,\Psi\right)\right]^2$.
		\end{enumerate}}

\subsection{Optimize Infinitesimal Nudge to Continuous Treatment by OCPPE}

From the perspective of Rubin's potential outcome framework, the OCPPE
\begin{eqnarray*}
	\theta(\tau_1,\tau_2,\mathcal{G})=E\left[\lim_{\delta\rightarrow 0}\frac{m\left(\mathcal{G}_{\delta}(D),X,U\right)-m\left(D,X,U\right)}{\delta}\bigg|Y\in\left(Q_Y(\tau_1),Q_Y(\tau_2)\right)\right].
\end{eqnarray*}
can be interpreted as the average \textit{welfare gain} for the individuals in quantile range $\left(\tau_1,\tau_2\right)$ of the outcome distribution (the target subpopulation), that would be realized if all individuals were mandated to set their treatment level to $\mathcal{G}_{\delta}\left(D_{i}\right)$, compared to the regime
in which all individuals maintain their preexisting treatment level $D_{i}$. In practice, when implementing a policy change entails a cost and the
policymaker faces a budget or capacity constraint that limits the proportion of individuals
who would experience the intervention, then how we optimally allocate the eligibility of intervention in order to produce
the highest welfare to the target subpopulation?

This problem can be formulated within the Empirical Welfare Maximization (EWM) framework (Kitagawa and Tetenov (2018), Athey and Wager (2021)), and can be described as follows.
Given $(\tau_1,\tau_2)\subset (0,1)$ indicating the target subgroup a policymaker cares about and the policy change $\mathcal{G}_{\delta}(\cdot)$ desired by the policymaker,
our goal is to learn a policy $\pi\in\Pi$ that maps a subject's features $X_{i}\in \mathcal{S}_{X}$ to a binary decision:
\[
\pi\left(\cdot,\tau_1,\tau_2,\mathcal{G}\right):  \mathcal{S}_{X}\rightarrow \{0,1\}.\eqno(5.2.1)
\]
$\Pi$ is a function space that encodes problem specific
constraints pertaining to budget, functional form, fairness, and so on. Since $(\tau_1,\tau_2)$ and $\mathcal{G}_{\delta}(\cdot)$ are regarded fixed throughout solving the policy learning problem,
we suppress the dependence of $\pi\left(\cdot,\tau_1,\tau_2,\mathcal{G}\right)$ on $(\tau_1,\tau_2)$ and $\mathcal{G}_{\delta}(\cdot)$ to simplify the notations.
When treatment variable $D$ is continuous, following Athey and Wager (2021), we define the \textit{outcome conditioned}
utility of deploying a binary nudge policy $\pi(\cdot)$ as $V\left(\tau_1,\tau_2,\pi,\mathcal{G}\right)=$
\[
E\left[\lim_{\delta\to 0}\frac{\pi(X)m\left(\mathcal{G}_{\delta}(D),X,U\right)+\left(1-\pi(X)\right)m\left(D,X,U\right)-m\left(D,X,U\right)}{\delta}\bigg|Y\in\left(Q_Y(\tau_1),Q_Y(\tau_2)\right)\right].\eqno(5.2.2)
\]
The (infeasible) optimal policy assignment rule is 
\[
\pi^*\left(\tau_1,\tau_2,\mathcal{G}\right)=\arg\max_{\pi\in\Pi}V\left(\tau_1,\tau_2,\pi,\mathcal{G}\right).\eqno(5.2.3)
\]

When $\mathcal{G}_{\delta}(D)=D+\delta$ and the quantile range is unit interval $(0,1)$,  $V\left(\tau_1,\tau_2,\pi,\mathcal{G}\right)$ becomes the
utility of  an infinitesimal location shift intervention for the entire population:
\begin{eqnarray*}
	V(\pi)=E\left[\frac{\partial m\left(D+\delta \pi(X),X,U\right)}{\partial \delta}\bigg|_{\delta=0}\right],
\end{eqnarray*}
which is Equation (8) in Athey and Wager (2021).

Compared with the literature, in particular the formulation that optimizes infinitesimal nudges to continuous treatments,
the OCPPE-related policy learning (OCPPE-PL) problem defined by (5.2.1)-(5.2.3) is novel in two aspects. First, as previously discussed, OCPPE is able to capture heterogeneous policy effects across the unconditional distribution of $Y$. Thus OCPPE-PL allows us to explore optimal treatment allocation policies that target distributional welfare, and the resulting assignment rule $\pi\left(\cdot,\tau_1,\tau_2,\mathcal{G}\right)$ can differ across different quantile ranges of $Y$. For example, if the policymaker aims to maximize the welfare of low wage earners or infants with low birthweight,  the OCPPE-PL framework may be useful by setting $(\tau_1,\tau_2)$ at extremal quantiles.
Second, OCPPE-PL accommodates solving personalized nudges to continuous treatment with the desired intervention characterized by $\mathcal{G}_{\delta}(\cdot)$, not limited to the location shift of $D$ only.

The key idea of Athey and Wager (2021) is that whenever one can estimate the average utility (treatment effects) of treating
everyone (like OCPPE, ATE, average partial effect and other mean causal quantities) using an estimator like
\begin{eqnarray*}
	\widehat{\theta}=\frac{1}{n}\sum_{i=1}^{n}\widehat{\psi}_{i},
\end{eqnarray*}
where $\theta$ is the causal quantity, $\psi_{i}$ is a doubly robust score constructed based on Chernozhukov et al. (2022), we can use these scores to learn $\pi$ by solving
\[
\widehat{\pi}=\arg\max\left\{\frac{1}{n}\sum_{i=1}^{n}\left(2\pi(X_{i})-1\right)\widehat{\psi}_{i}:\pi\in\Pi\right\}.
\]

Since we have obtained the efficient doubly robust score for an OCPPE in Proposition 2.2, and proposed a feasible efficient estimator of OCPPE in Section 3, the OCPPE-PL problem can be solved as follows:

\noindent\textbf{Step 1.} Divide the data into $K$ even-sized folds.

\noindent\textbf{Step 2.} Choose a fold $k=1,\cdots,K$. Use the data from other $K-1$ folds to estimate the orthogonal score $\psi$ in Equation (2.2), 
by applying identical procedures outlined in Section 3, except that data from $k$-th fold are excluded. Denote the estimated score as
$\widehat{\psi}_{-k}$.

\noindent\textbf{Step 3.} Estimate the optimal policy assignment rule $\pi^{*}$ by
\[
\widehat{\pi}_{k}=\arg\max\left\{\sum_{i\in{n_{k}}}\Big(2\pi(X_i)-1\Big)\widehat{\psi}_{i,-k}:\pi\in\Pi\right\},
\]
where $n_{k}$ denotes the set of observations in $k$-th fold.

\noindent\textbf{Step 4.} Repeat Step 2 and 3 $K$ times. Use the average of the resulting estimators as the final estimate of the optimal policy assignment rule.

Let $R\left(\pi,\tau_1,\tau_2,\mathcal{G}\right)=\mbox{max}\{V\left(\pi^{\prime},\tau_1,\tau_2,\mathcal{G}\right),\pi^{\prime}\in \Pi\}-V\left(\pi,\tau_1,\tau_2,\mathcal{G}\right)$.
Following the proof strategy of Athey and Wager (2021), under similar regularity conditions, the resulting policies $\widehat{\pi}$
have regret $R\left(\widehat{\pi},\tau_1,\tau_2,\mathcal{G}\right)$ bounded on the order of
$\sqrt{\mbox{VC}\left(\Pi\right)/n}$ with high probability, with $\mbox{VC}(\Pi)$  the Vapnik Chervonenkis dimension
of the class $\Pi$ and $n$ the sample size.

\section{An Empirical Example}

In this section, we utilize the OCPPE to analyze how anti-smoking policies impact low percentiles of infants'
birth weight.
Abrevaya (2001), Koenker and Hallock (2001) and Chernozhukov and Fern\'{a}ndez-Val (2011) utilize \textit{quantile regression} (QR) to quantify the effects of demographics and maternal behavior during pregnancy at various quantiles of the birth weight distribution.
They all find smoking has a negative impact on the distribution of birth weights.
In terms of capturing heterogeneity, the OCPPE differs from QR in two major aspects. QR coefficients measure the heterogeneous effect of a marginal increase in cigarette consumption
on birth weight \textit{conditioning on a large number of demographic controls}, whereas OCPPE answers a more straightforward question: what is the mean effect of reducing cigarette consumption on low/medium/high birth weight infants, irrespective of a mother's characteristics.  Second, OCPPE is flexible enough to analyze the effect of various counterfactual changes in cigarette consumption distribution in addition to changing the mean of cigarette consumption only.

Mart\'{i}nez-Iriarte et al. (2024) estimate the unconditional effects of
smoking on birth weight using the UQPE approach. Our empirical exercise furthers their analysis in three aspects. First,  our DML estimator of an OCPPE is compatible with high dimensional controls.  We differ from
Mart\'{i}nez-Iriarte et al. (2024) by using the high-dimensional methods developed in this paper to allow
ourselves to consider a broader set of controls than has previously been considered. We consider two counterfactual interventions: (i) reducing the number of cigarette by a fixed amount and (ii) by a fixed proportion. Our general finding is that these policies increase infants'
birth weight over almost the entire range of quantiles considered except for extremal low quantiles. Second, as predicted by the theory, OCPPE delivers efficient and $\sqrt{n}$-consistent estimation of aggregated tail
information on treatment effects. For the current application, a policymaker may extremely care about the impact on
infants at the extremal low tail of birth weight distribution. Motivated by this, we estimate OCPPEs at a series of extremal percentile ranges of birth weight using the DML method. The estimation results confirm the previous finding that anti-smoking policies have
no positive impact on low birth weight infants.

Third, prior analysis shows that low birth weight infants hardly benefit by an anti-smoking policy that treats everyone uniformly without accounting for personal characteristics. We then make a preliminary attempt at whether conditioning intervention assignment on a small number of covariates can achieve higher welfare gain for the target subpopulation (infants with birth weight lower than 2500 grams).  We consider conditioning treatment assignment on two dummy covariates: (i) whether a mother paid a prenatal visit and (ii) whether a mother's BMI is higher than the medium of all mothers' BMI in the sample. Preliminary quantile regressions indicate that both variables significantly explain low quantiles of birth weight.
We do not use infant's gender, mother's age or education as the conditioning variable. Though treatment effects may vary with these characteristics,
policy makers usually cannot use them to determine treatment assignment. We find the optimal intervention that achieves the highest welfare gain for low birth weight infants are highly determined by mother's prenatal visit during pregnancy. Assigning intervention to mothers who paid no visit at all can raise the baseline welfare gain (by assigning intervention to every mother in the sample) by 80\%.

\subsection{Data description}

We use the National Vital Statistics System for the year 2018.
Previous study by Mart\'{i}nez-Iriarte et al. (2024) uses the same data set.
The outcome variable $Y$ is birth weight in grams, and the target variable $D$ is the average number of cigarettes smoked daily during pregnancy.
We focus on the sample of black mothers who are smokers, aged between 18 and 45, gave birth to a live infant. The sample consists of 10995
observations. Our decision to focus the analysis on black mothers is similar to Chernozhukov and Fern\'{a}ndez-Val (2011), that there is a considerable proportion (about 19.7\%) of infants weight at birth falling below 2500 grams for black mothers.
Figure C.1 in Appendix C shows the density of birth weights for black and white mothers.
The descriptive statistics for $Y$, $D$  and other covariates are presented in Table 6.1.
\begin{table}[H]\centering
	\begin{threeparttable}
			\begin{tabular}{lccccccc}
					\multicolumn{8}{l}{\small{\textbf{Table 6.1: Descriptive Statistics}}}\\
					\toprule
					& Mean  & Std.dev & Min   & Q25   & Median & Q75   & Max \\
					\midrule
					Birth Weights & 2925  & 578.002 & 227   & 2608  & 2960  & 3289  & 5585 \\
					Cigarettes & 7.732 & 6.845 & 3.333 & 5.333 & 7.332 & 10    & 90 \\
					Married & 0.206 & 0.404 & 0     & 0     & 0     & 0     & 1 \\
					Mother's Age		  & 28.446 & 5.290 & 18    & 24    & 28    & 32    & 45 \\
					Mother's Education & 1.800 & 1.141 & 0     & 0     & 2     & 3     & 3 \\
					Father's Age  & 32.041 & 8.111 & 15    & 26    & 31    & 37    & 83 \\
					Father's Education & 0.685 & 0.826 & 0     & 0     & 0     & 1     & 2 \\
					Prenatal Times & 9.737 & 5.096 & 0     & 7     & 10    & 12    & 74 \\
				Prenatal Second& 0.267 & 0.443 & 0     & 0     & 0     & 1     & 1 \\
					Prenatal Third & 0.072 & 0.258 & 0     & 0     & 0     & 0     & 1 \\
					Mother's BMI  & 28.806 & 7.872 & 13.1  & 22.7  & 27.4  & 33.5  & 65.9 \\
					Mother's Height & 64.550 & 2.890 & 51    & 63    & 64    & 66    & 78 \\
					Mother's Weight gain & 26.587 & 16.873 & 0     & 15    & 25    & 37    & 91 \\
					WIC   & 0.635 & 0.482 & 0     & 0     & 1     & 1     & 1 \\
					Gestation & 37.988 & 2.756 & 22    & 37    & 38    & 39    & 47 \\
					Boy   & 0.509 & 0.500 & 0     & 0     & 1     & 1     & 1 \\
					\bottomrule
				\end{tabular}%
				\begin{tablenotes}
					\footnotesize
					\item Notes:
		``Married'', an indicator of whether the mother was married or not; ``Mother's Education'', a categorical variable taking a value of 0 if the mother had less than a high-school education, 1 if she  completed high-school education, 2 if she obtained some college education, and 3 if she  graduated from college; ``Father's Education'' define the information on father's education;  ``Prenatal Times'', the times of prenatal visits; ``Prenatal Second'' and ``Prenatal Third'', indicators of whether mother's first prenatal visit was in the second, and the third trimester;
		``Mother's BMI'', mother's pre-pregnancy body mass index;
		 ``Mother's Height'', mother's height in inches; ``Mother's Weight Gain'', weight gain during the pregnancy in pounds;
		``WIC'', an indicator of whether mother get WIC (The Special Supplementary Nutrition Program for Women, Infants, and Children) food for herself during the pregnancy; ``Gestation'', gestation in weeks; ``Boy'', an indicator of infant's gender.
				\end{tablenotes}
		\end{threeparttable}
	\label{tab:addlabel}%
\end{table}%

\subsection{Heterogeneous effect of anti-smoking policies on birth weights}

We examine the effect of two anti-smoking policies on birth weight distribution. The first one reduces $D$ by a fixed amount ($D_{\delta}=D-\delta$), the second reduces $D$ by a fixed proportion ($D_{\delta}=D/(1+\delta)$).
We estimate OCPPEs on nine quantile ranges:
\begin{eqnarray*}
	(\tau_1,\tau_2)\in\left\{(0.05,0.15),(0.15,0.25),\cdots,(0.85,0.95)\right\}.
\end{eqnarray*}
We consider three different specifications of controls $b(D,X)$.
The first specification uses the original forms of all covariates listed in Table 6.1  (Basic Specification).
The second specification arguments Basic Specification by incorporating second polynomials of all covariates except the dummy ones, and all two-way interactions among the treatment and control variables (Quadratic Plus Interactions).
The last specification further incorporates the third polynomials of all covariates except the dummy ones, and all three-way interactions among the control covariates (Cube Plus Interactions).
The dimensions of the set of controls are thus 14, 128, and 424 variables for Basic Specification, Quadratic Plus Interactions, and Cube Plus Interactions, respectively.
For the method that does not use model selection, we use 14, 127, and 413 variables, respectively, as we remove terms that are perfectly collinear.

For each specification of controls, we compute two estimators.  Both estimators largely follow the steps outlined in Section 3.
The estimation results are summarized in Table 6.2 for $D_{\delta}=D-\delta$ and in Table 6.3 for $D_{\delta}=D/(1+\delta)$.
As for Table 6.2, the third column indicates whether  variable selection is used or not in estimation. When it indicates ``No'',  it means we estimate nonparametric nuisances without penalty terms, without using any variable selection. When it indicates ``Yes'', we use the lasso estimator to select variables.

\begin{table}[H]\centering\small
	\begin{threeparttable}
			\begin{tabular}{lccccccc}
					\multicolumn{8}{l}{\small{\textbf{Table 6.2: OCPPE for $\mathcal{G}_\delta=D-\delta$}}}\\
					\toprule
					\multicolumn{1}{c}{Specification} & Dimension & Selection & \multicolumn{5}{c}{OCPPE for $\mathcal{G}_\delta=D-\delta$} \\
					&       &       & 0.05-0.15 & 0.15-0.25 & 0.25-0.35 & 0.35-0.45 & 0.45-0.55 \\
					\midrule
					Basic Specification & 14    & N     & -1.254 & .226  & .321  & .372  & 1.158* \\
					& (14)  &       & (1.370) & (1.061) & (.975) & (.745) & (.819) \\
					&       &       & \{1.357\} & \{1.045\} & \{1.011\} & \{.775\} & \{.813\} \\
					Basic Specification & 14    & Y     & -1.060 & .219  & .273  & .342  & 1.062* \\
					& (14)  &       & (1.290) & (.986) & (.903) & (.688) & (.757) \\
					&       &       & \{1.281\} & \{.972\} & \{.939\} & \{.715\} & \{.752\} \\
					Quadratic Plus Interactions & 128   & N     & -2.825** & .006  & .229  & .394  & 1.206* \\
					& (127) &       & (1.399) & (1.069) & (.958) & (.741) & (.802) \\
					&       &       & \{1.392\} & \{1.030\} & \{.999\} & \{.771\} & \{.806\} \\
					Quadratic Plus Interactions & 128   & Y     & -.998 & .310  & .363  & .397  & 1.076* \\
					& (127) &       & (1.246) & (.968) & (.885) & (.677) & (.745) \\
					&       &       & \{1.239\} & \{.949\} & \{.918\} & \{.705\} & \{.742\} \\
					Cube Plus Interactions & 424   & N     & -4.857 & -.870 & -.709 & .040  & .956 \\
					& (415) &       & (1.186) & (.946) & (.880) & (.692) & (.751) \\
					&       &       & \{1.173\} & \{.923\} & \{.932\} & \{.710\} & \{.766\} \\
					Cube Plus Interactions & 424   & Y     & -1.054 & .280  & .396  & .426  & 1.170* \\
					& (415) &       & (1.248) & (.965) & (.884) & (.672) & (.738) \\
					&       &       & \{1.234\} & \{.945\} & \{.920\} & \{.700\} & \{.736\} \\
					\midrule
					\multicolumn{1}{c}{Specification} & Dimension & Selection & \multicolumn{4}{c}{OCPPE for $\mathcal{G}_\delta=D-\delta$} &  \\
					&       &       & 0.55-0.65 & 0.65-0.75 & 0.75-0.85 & 0.85-0.95 & - \\
					\midrule
					Basic Specification & 14    & N     & 1.175* & 1.901** & 1.519** & 2.604*** & - \\
					& (14)  &       & (.748) & (.828) & (.898) & (.990) & - \\
					&       &       & \{.742\} & \{.808\} & \{.886\} & \{1.024\} & - \\
					Basic Specification & 14    & Y     & 1.064* & 1.818*** & 1.342* & 2.209*** & - \\
					& (14)  &       & (.696) & (.760) & (.829) & (.926) & - \\
					&       &       & \{.689\} & \{.741\} & \{.825\} & \{.943\} & - \\
					Quadratic Plus Interactions & 128   & N     & 1.192* & 1.962*** & 1.581** & 3.000*** & - \\
					& (127) &       & (.733) & (.807) & (.891) & (1.029) & - \\
					&       &       & \{.733\} & \{.795\} & \{.887\} & \{1.051\} & - \\
					Quadratic Plus Interactions & 128   & Y     & 1.077* & 1.948*** & 1.484** & 2.260*** & - \\
					& (127) &       & (.689) & (.758) & (.832) & (.939) & - \\
					&       &       & \{.682\} & \{.743\} & \{.827\} & \{.955\} & - \\
					Cube Plus Interactions & 424   & N     & 1.269** & 2.556*** & 2.683*** & 5.009*** & - \\
					& (415) &       & (.688) & (.760) & (.834) & (.993) & - \\
					&       &       & \{.694\} & \{.752\} & \{.842\} & \{1.021\} & - \\
					Cube Plus Interactions & 424   & Y     & 1.142** & 2.029*** & 1.634** & 2.394*** & - \\
					& (415) &       & (.691) & (.753) & (.819) & (.940) & - \\
					&       &       & \{.684\} & \{.738\} & \{.811\} & \{.961\} & - \\
					\bottomrule
				\end{tabular}%
			\begin{tablenotes}
					\footnotesize
					\item Notes:
					Analytical standard errors are provided in parentheses.
					Bootstrap standard errors based on 1000 repetitions
					with standard normal variables are provided in braces.
				Basic Specification includes the original forms of all covariates listed in Table 6.1.
				Quadratic Plus Interactions specification includes the controls in Basic Specification, second polynomials of all covariates except the dummy ones, and all two-way interactions among the treatment and control variables.
				Cube Plus Interactions specification includes the controls in Quadratic Plus Interactions, the third polynomials of all covariates except the dummy ones, and all three-way interactions among the control covariates.
				\end{tablenotes}
		\end{threeparttable}
	\label{tab:addlabel}%
\end{table}%

\begin{table}[H]\centering\small
	\begin{threeparttable}
			\begin{tabular}{lccccccc}
					\multicolumn{8}{l}{\small{\textbf{Table 6.3: OCPPE for $\mathcal{G}_\delta=D/(1+\delta)$}}}\\
					\toprule
					\multicolumn{1}{c}{Specification} & Dimension & Selection & \multicolumn{5}{c}{OCPPE for $\mathcal{G}_\delta=D/(1+\delta)$} \\
					&       &       & 0.05-0.15 & 0.15-0.25 & 0.25-0.35 & 0.35-0.45 & 0.45-0.55 \\
					\midrule
					Basic Specification & 14    & N     & -9.672 & 1.772 & 2.493 & 2.878 & 8.917* \\
					& (14)  &       & (10.588) & (8.202) & (7.540) & (5.749) & (6.326) \\
					&       &       & \{10.483\} & \{8.074\} & \{7.827\} & \{5.989\} & \{6.289\} \\
					Basic Specification & 14    & Y     & -8.310 & 1.712 & 2.141 & 2.678 & 8.322* \\
					& (14)  &       & (10.104) & (7.726) & (7.076) & (5.391) & (5.930) \\
					&       &       & \{10.039\} & \{7.617\} & \{7.357\} & \{5.604\} & \{5.891\} \\
					Quadratic Plus Interactions & 128   & N     & -17.443* & 2.328 & 3.478 & 3.557 & 9.336* \\
					& (127) &       & (10.927) & (8.288) & (7.445) & (5.735) & (6.203) \\
					&       &       & \{10.829\} & \{7.998\} & \{7.789\} & \{6.005\} & \{6.236\} \\
					Quadratic Plus Interactions & 128   & Y     & -7.604 & 2.367 & 2.710 & 3.030 & 8.436* \\
					& (127) &       & (9.833) & (7.613) & (6.955) & (5.311) & (5.839) \\
					&       &       & \{9.771\} & \{7.453\} & \{7.224\} & \{5.545\} & \{5.818\} \\
					Cube Plus Interactions & 424   & N     & -11.601 & 1.999 & -1.261 & 1.225 & 6.481 \\
					& (415) &       & (9.604) & (7.457) & (6.868) & (5.376) & (5.817) \\
					&       &       & \{9.417\} & \{7.234\} & \{7.277\} & \{5.548\} & \{5.873\} \\
					Cube Plus Interactions & 424   & Y     & -8.690 & 2.236 & 2.867 & 3.239 & 9.023* \\
					& (415) &       & (9.748) & (7.522) & (6.901) & (5.259) & (5.753) \\
					&       &       & \{9.687\} & \{7.374\} & \{7.173\} & \{5.487\} & \{5.747\} \\
					\midrule
					\multicolumn{1}{c}{Specification} & Dimension & Selection & \multicolumn{4}{c}{OCPPE for $\mathcal{G}_\delta=D/(1+\delta)$} &  \\
					&       &       & 0.55-0.65 & 0.65-0.75 & 0.75-0.85 & 0.85-0.95 & - \\
					\midrule
					Basic Specification & 14    & N     & 8.993* & 14.380*** & 11.442** & 18.900*** & - \\
					& (14)  &       & (5.778) & (6.391) & (6.928) & (7.676) & - \\
					&       &       & \{5.738\} & \{6.252\} & \{6.824\} & \{7.901\} & - \\
					Basic Specification & 14    & Y     & 8.325* & 13.883*** & 10.480* & 17.309*** & - \\
					& (14)  &       & (5.419) & (5.946) & (6.445) & (7.253) & - \\
					&       &       & \{5.369\} & \{5.814\} & \{6.373\} & \{7.389\} & - \\
					Quadratic Plus Interactions & 128   & N     & 8.765* & 14.192*** & 10.935** & 19.024*** & - \\
					& (127) &       & (5.648) & (6.172) & (6.776) & (7.834) & - \\
					&       &       & \{5.647\} & \{6.098\} & \{6.735\} & \{7.964\} & - \\
					Quadratic Plus Interactions & 128   & Y     & 8.477* & 14.783*** & 11.533** & 17.868** & - \\
					& (127) &       & (5.348) & (5.894) & (6.417) & (7.303) & - \\
					&       &       & \{5.293\} & \{5.789\} & \{6.342\} & \{7.419\} & - \\
					Cube Plus Interactions & 424   & N     & 7.862* & 15.213*** & 11.770** & 13.314** & - \\
					& (415) &       & (5.297) & (5.815) & (6.348) & (7.527) & - \\
					&       &       & \{5.366\} & \{5.746\} & \{6.439\} & \{7.666\} & - \\
					Cube Plus Interactions & 424   & Y     & 8.864** & 15.270*** & 12.213** & 18.091*** & - \\
					& (415) &       & (5.279) & (5.826) & (6.368) & (7.314) & - \\
					&       &       & \{5.232\} & \{5.723\} & \{6.285\} & \{7.442\} & - \\
					\bottomrule
				\end{tabular}%
			\begin{tablenotes}
					\footnotesize
					\item Notes:
					Analytical standard errors are provided in parentheses.
				Bootstrap standard errors based on 1000 repetitions
				with standard normal variables are provided in braces.
				Basic Specification includes the original forms of all covariates listed in Table 6.1.
				Quadratic Plus Interactions specification includes the controls in Basic Specification, second polynomials of all covariates except the dummy ones, and all two-way interactions among the treatment and control variables.
				Cube Plus Interactions specification includes the controls in Quadratic Plus Interactions, the third polynomials of all covariates except the dummy ones, and all three-way interactions among the control covariates.
				\end{tablenotes}
		\end{threeparttable}
	\label{tab:addlabel}%
\end{table}%

From Tables 6.2 and 6.3, we find that the estimators without variable selection perform as well as the ones with variable selection when the dimension of controls is small, e.g., in Basic Specification.
They both show that the effects are monotonically increasing
across quantiles.
These policies increase infant birth weights over almost the entire range of quantiles considered except for extremal low quantiles.
When the dimension of controls increases, the estimators without variable selection become unstable for low quantiles.
The two typical cases are as follows.
In the case of ``Quadratic Plus Interactions, ``the estimates without variable selection exaggerate the negative effects of anti-smoking over  the quantile range $(0.05,0.15)$ of live infants' birth weights at 10\% significant level.
Besides, in the case of ``Cube
Plus Interactions,'' it
cannot show the significant positive effect of anti-smoking over the quantile range $(0.45,0.55)$.
In  contrast, the DML estimator performs similarly to that in the setting with small dimensions of controls.
Along with the point estimates, the standard errors for the estimators without variable selection are almost always larger than those with variable selection, similar to the finding in Belloni et al. (2017).

\subsection{Effect of smoking on extremal low quantiles}

The prior heterogeneity analysis shows that reducing smoking appears to have
no significantly positive impact on extremal quantiles. Since  a policymaker may extremely care about the impact on
infants at the extremal low tail of birth weight, we estimate OCPPEs at a series of extremal percentile ranges of birth weight using the DML method.

During the estimation, we use the DML throughout with the controls taking Cube Plus Interactions Specification. Other specification of the controls give similar results.
The result is provided in Table 6.4. The estimation results confirm the previous finding that anti-smoking policies have
no positive impact on low birth weight infants, consistent with  the finding in Chernozhukov and Fern\'{a}ndez-Val (2011). The lack of statistical significance in the tails could be due to selection,
where only mothers confident of good outcomes smoke, or to smoking having little or no
causal effect on very extreme outcomes.

	\begin{table}[H]\centering
		\begin{threeparttable}
				\begin{tabular}{lcc}
						\multicolumn{3}{l}{\small{\textbf{Table 6.4: OCPPE for low birthweights for black mothers}}}\\
						\toprule
						\multirow{2}[4]{*}{$(\tau_1,\tau_2)$} & \multicolumn{2}{c}{$\theta(\tau_1,\tau_2,\mathcal{G})$} \\
						\cmidrule{2-3}          & $\mathcal{G}_\delta(D)=D-\delta$ & $\mathcal{G}_\delta(D)=D/(1+\delta)$ \\
						\midrule
						(0.01,0.04) & -2.279 & -18.846 \\
						& (2.712) & (21.128) \\
						& \{2.720\} & \{21.186\} \\
						(0.01,0.06) & -1.665 & -13.765 \\
						& (2.218) & (17.266) \\
						& \{2.221\} & \{17.301\} \\
						(0.01,0.08) & -1.524 & -12.960 \\
						& (1.930) & (15.027) \\
						& \{1.964\} & \{15.310\} \\
						(0.01,0.1) & -1.425 & -12.201 \\
						& (1.764) & (13.734) \\
						& \{1.678\} & \{13.042\} \\
						(0.01,0.12) & -1.288 & -11.004 \\
						& (1.614) & (12.592) \\
						& \{1.601\} & \{12.445\} \\
						(0.01,0.14) & -1.236 & -10.806 \\
						& (1.474) & (11.489) \\
						& \{1.466\} & \{11.457\} \\
						(0.01,0.16) & -1.150 & -9.902 \\
						& (1.370) & (10.695) \\
						& \{1.358\} & \{10.619\} \\
						(0.01,0.18) & -1.072 & -9.153 \\
						& (1.299) & (10.127) \\
						& \{1.371\} & \{10.704\} \\
						(0.01,0.2) & -1.003 & -8.149 \\
						& (1.217) & (9.488) \\
						& \{1.240\} & \{9.708\} \\
						\bottomrule
					\end{tabular}%
				\begin{tablenotes}
						\footnotesize
						\item Notes:
						Analytical standard errors are provided in parentheses.
						Bootstrap standard errors based on 1000 repetitions
						with standard normal variables are provided in braces.
					\end{tablenotes}
			\end{threeparttable}
		\label{tab:addlabel}%
	\end{table}%

	\subsection{Personalized Intervention Conditioning on Prenatal Visit and BMI}
	
	Prior analysis shows that an intervention that treats everyone uniformly without considering personal characteristics is not effective in reducing low birth weight incidence. Thus we consider conditioning the assignment rule on a few covariates, in order to achieve higher welfare gain for the target subpopulation (infants with birth weight lower than 2500 grams).
	
	A series of preliminary quantile regressions (Table C.1 in Appedix C) suggest that a mother's prenatal visit and her pre-pregancy body mass index significantly explain low quantiles of the birth weight.
	We condition treatment assignment on two dummy covariates: (i) whether a mother paid a prenatal visit and (ii) whether a mother's BMI is higher than the median of all mothers' BMI in the sample.
	We do not use infant's gender, mother's age or education as the conditioning variable. Though treatment effects may vary with these characteristics,
	policy makers usually cannot use them to determine treatment assignment.
	
	We solve the optimal assignment rule that maximizes the empirical welfare of target population, by searching over all but finite number of possible rules.
	Since the assignment rule is decided based on the value of two dummies, the policy class $\Pi$ contains $2^4=16$ elements:
	\begin{eqnarray*}
		\Pi&=&\Big\{\{0,0,0,0\},\{1,0,0,0\},\{0,1,0,0\},\{0,0,1,0\},\{0,0,0,1\},\{1,1,0,0\},\notag\\
		&&\{1,0,1,0\},\{1,0,0,1\},\{0,1,1,0\},\{0,1,0,1\},\{0,0,1,1\},\notag\\
		&&\{1,1,1,0\},\{1,1,0,1\},\{1,0,1,1\},\{0,1,1,1\},\{1,1,1,1\}\Big\}.
	\end{eqnarray*}
	Let $PN=1$ denote a mother paid at least one prenatal visit during pregnancy; $BMI=1$ denote a mother's BMI exceeds the medium.
	The first component of each element indicates the treatment status of the subsample with $PN=1,BMI=1$; the second component indicates $PN=1,BMI=0$; the third and fourth components indicate
	$PN=0,BMI=1$ and $PN=0,BMI=0$ respectively.
	
	Given $\mathcal{G}$ and $(\tau_1,\tau_2)$, the welfare gain for assignment rule $\pi(\cdot)$ is
	\[\begin{aligned}
		&V\left(\tau_1,\tau_2,\pi,\mathcal{G}\right) \\
		=&E\left[\lim_{\delta\rightarrow 0}\frac{m\left(\pi(X)\mathcal{G}_{\delta}(D)+(1-\pi(X))D,X,U\right)-m\left(D,X,U\right)}{\delta}\bigg|Y\in\left(Q_Y(\tau_1),Q_Y(\tau_2)\right)\right]. \\
	\end{aligned}\]
	In Appendix C, we demonstrate that $V\left(\tau_1,\tau_2,\pi,\mathcal{G}\right)$ is identifiable and the corresponding (infeasible) optimal policy assignment rule $\pi^{*}$ can be equivalently expressed as
	\[
	\pi^*\left(\tau_1,\tau_2,\mathcal{G}\right)=\arg\max_{\pi\in\Pi}E\left[\Big(2\pi(X)-1\Big)\cdot\frac{-\vartheta\left(D;\mathcal{G}\right)}{\tau_2-\tau_1}\int_{Q_Y(\tau_1)}^{Q_Y(\tau_2)}\partial_DF_Y(y|D,X)dy\right].
	\]
	The detailed estimation procedures are outlined as Steps 1-4 in Section 5.2. Specifically, we set $K=5$ for 5-fold cross-validation. We study the effect on extremely low birth weight quantiles from 0.01 to 0.197, given that 2500 grams corresponds to the 0.197 quantile. The estimation results for $V(0.01, 0.197, \pi, \mathcal{G})$ under the personalized treatment assignment rule $\pi$ and scale intervention $\mathcal{G}_\delta(D) = D/(1 + \delta)$ are presented in Table 6.5. For the assignment rule $\pi = (0, 0, 0, 0)$, no one receives the intervention, resulting in a welfare gain of 0, which is omitted in the report. We use "assign intervention to everyone" as the baseline result, corresponding to $\pi = (1, 1, 1, 1)$, with a baseline welfare gain of -8.462, not statistically significant at the 10\% level. Specifically, we find that interventions achieving positive effects for low birth weight infants are strongly influenced by the mother's prenatal visits during pregnancy. Among these desirable interventions, the optimal intervention yields a welfare gain of 1.671, corresponding to $\pi = (0, 0, 1, 1)$, which is statistically significant at the 10\% level.
		\begin{table}[H]\centering
			\begin{threeparttable}
					\begin{tabular}{cccccccccc}
							\multicolumn{10}{l}{\small{\textbf{Table 6.5: Estimated welfare gain for each assignment rule}}}\\
				\toprule
				\multicolumn{10}{c}{$\mathcal{G}_\delta(D)=D/(1+\delta)$} \\
				$\pi$ &       & $\pi$ &       & $\pi$ &       & $\pi$ &       & $\pi$ &  \\
				\midrule
				(1,1,1,1) & -8.462 & \textbf{(0,0,1,0)} & \textbf{.850**} & (1,0,1,0) & -4.811 & (0,1,0,1) & -4.243 & (1,1,0,1) & -9.059 \\
				& (9.455) &       & \textbf{(.416)} &       & (5.708) &       & (4.838) &       & (9.138) \\
				& \{9.553\} &       & \textbf{\{.428\}} &       & \{5.510\} &       & \{4.807\} &       & \{9.343\} \\
				(1,0,0,0) & -5.191 & \textbf{(0,0,0,1)} & \textbf{1.191**} & (1,0,0,1) & -4.677 & \textbf{(0,0,1,1)} & \textbf{1.671*} & (1,0,1,1) & -4.813 \\
				& (5.336) &       & \textbf{(.677)} &       & (5.321) &       & \textbf{(1.168)} &       & (5.694) \\
				& \{5.216\} &       & \textbf{\{.689\}} &       & \{5.520\} &       & \textbf{\{1.152\}} &       & \{5.774\} \\
				(0,1,0,0) & -5.014 & (1,1,0,0) & -10.235 & (0,1,1,0) & -4.661 & (1,1,1,0) & -9.669 & (0,1,1,1) & -3.663 \\
				& (4.396) &       & (8.941) &       & (4.646) &       & (9.272) &       & (5.162) \\
				& \{4.408\} &       & \{8.934\} &       & \{4.631\} &       & \{9.170\} &       & \{5.090\} \\
				\bottomrule
						\end{tabular}%
					\begin{tablenotes}
							\footnotesize
							\item Notes:
							Analytical standard errors are provided in parentheses.
							Bootstrap standard errors based on 1000 repetitions
							with standard normal variables are provided in braces.
							*, ** and *** respectively indicate the significance at 10, 5 and 1 percent level.
						\end{tablenotes}
				\end{threeparttable}
			\label{tab:addlabel}%
		\end{table}%

	To present the conclusions from Table 6.5 more intuitively, we illustrate these estimates graphically in Figure 6.1. The highlighted sections indicate which subpopulation receives the intervention. The height of each cuboid represents the corresponding welfare gain for low birth weight infants. Positive welfare gains are depicted in green, while negative gains are shown in blue. For example, in Figure 6.1(d), the highlighted section corresponds to mothers with $PN=0$ and $BMI=1$, indicating that the intervention is assigned only to this subsample, corresponding to $\pi=(0,0,1,0)$. Under this intervention, the green color and height of 0.85 indicate a positive welfare gain precisely equal to 0.85. Additionally, the estimation results for $V\left(0.01,0.197,\pi,\mathcal{G}\right)$ under the location intervention $\mathcal{G}_\delta(D)=D-\delta$ are shown in Table C.2, which yield similar results.

		\begin{figure}[H]
		\centering
				\includegraphics[width=2.1in]{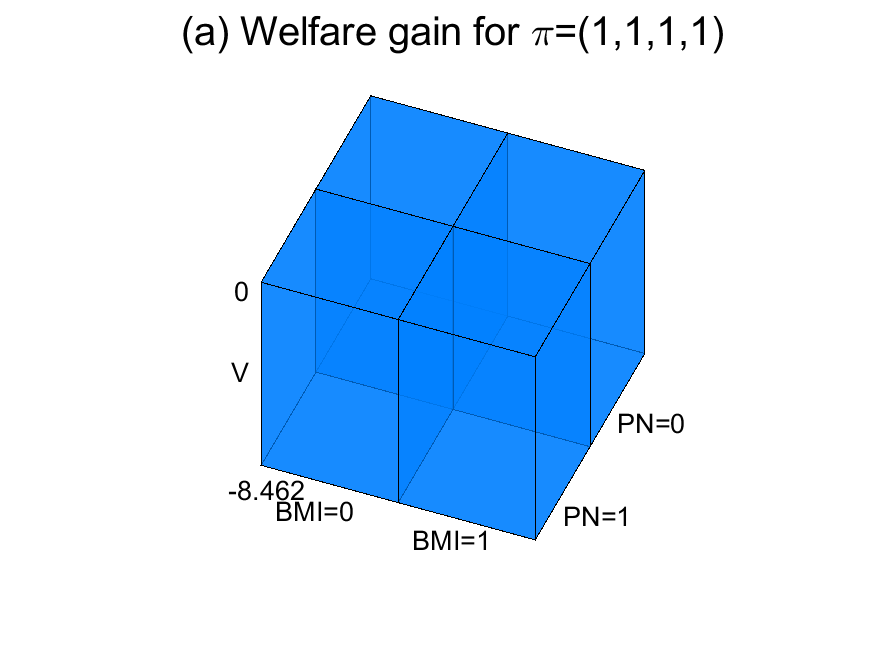}
		\includegraphics[width=2.1in]{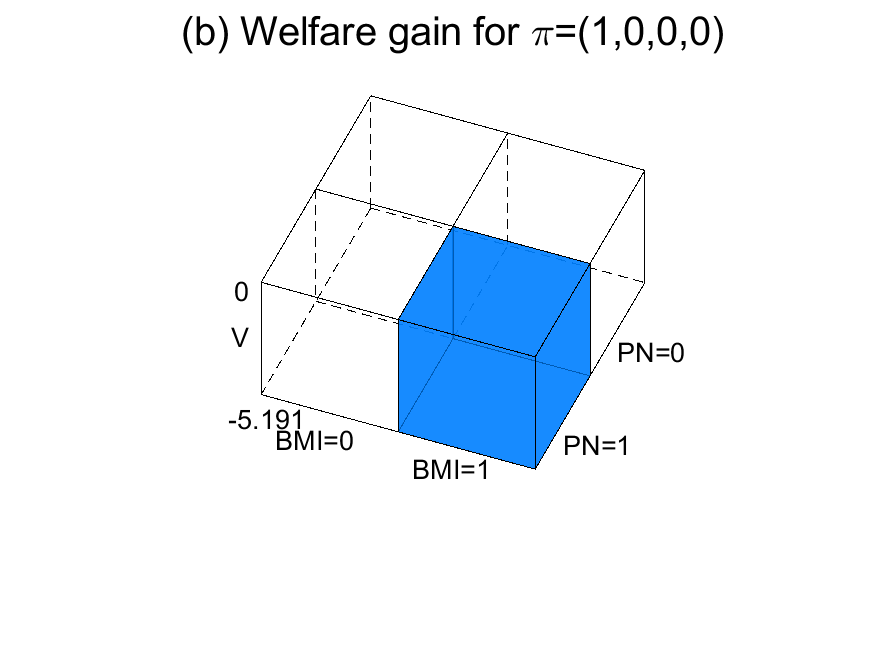}
		\includegraphics[width=2.1in]{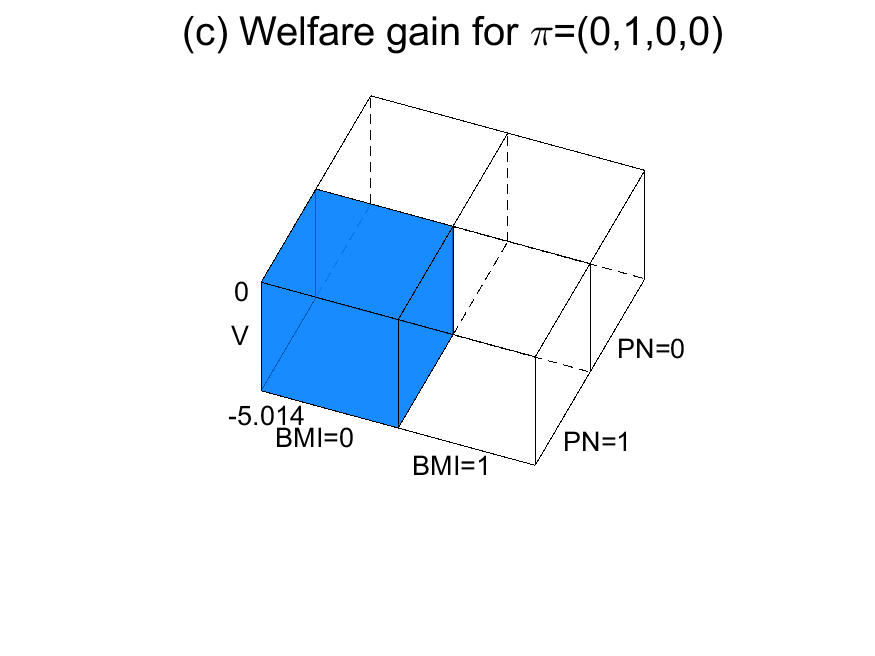}
		\includegraphics[width=2.1in]{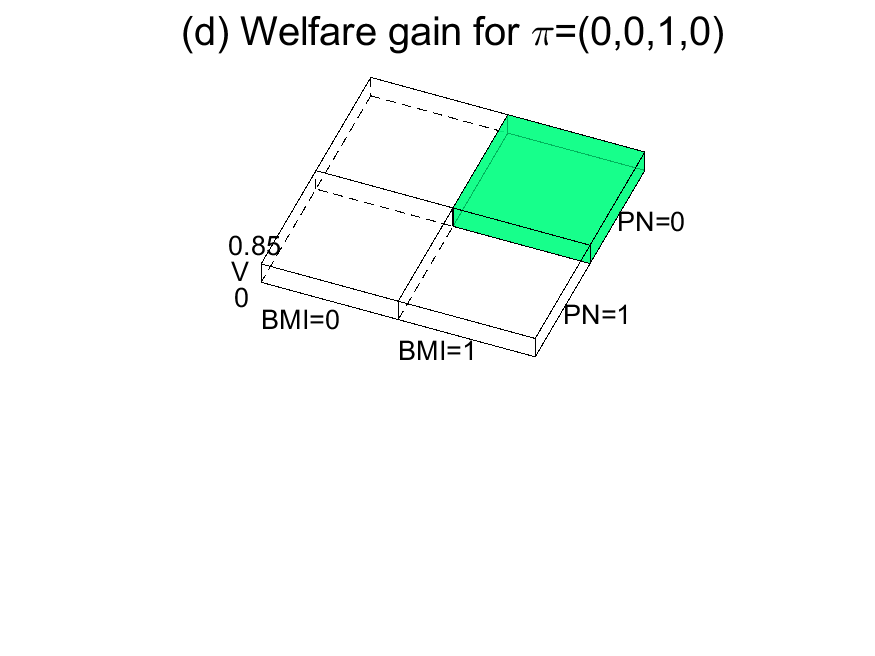}
		\includegraphics[width=2.1in]{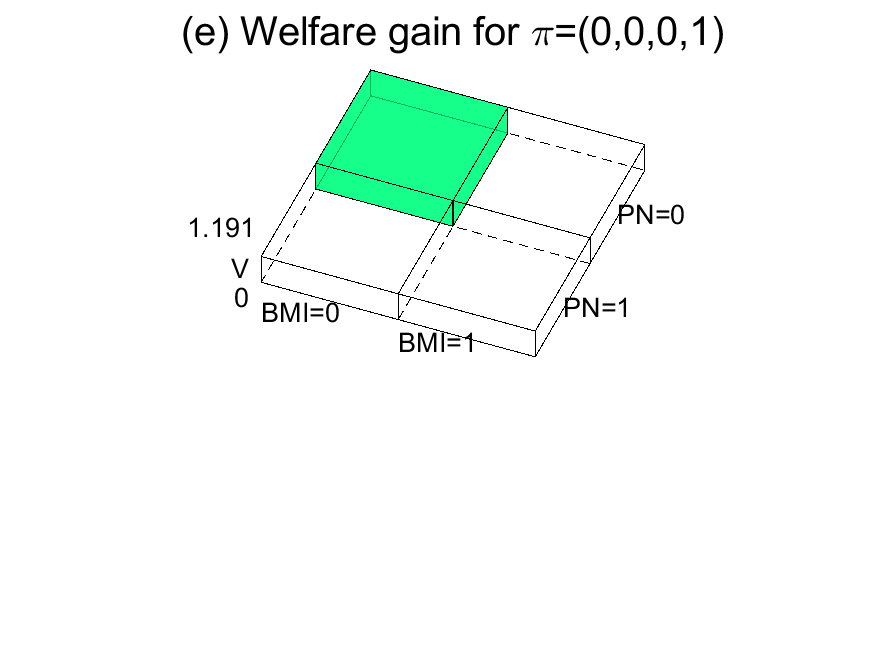}
		\includegraphics[width=2.1in]{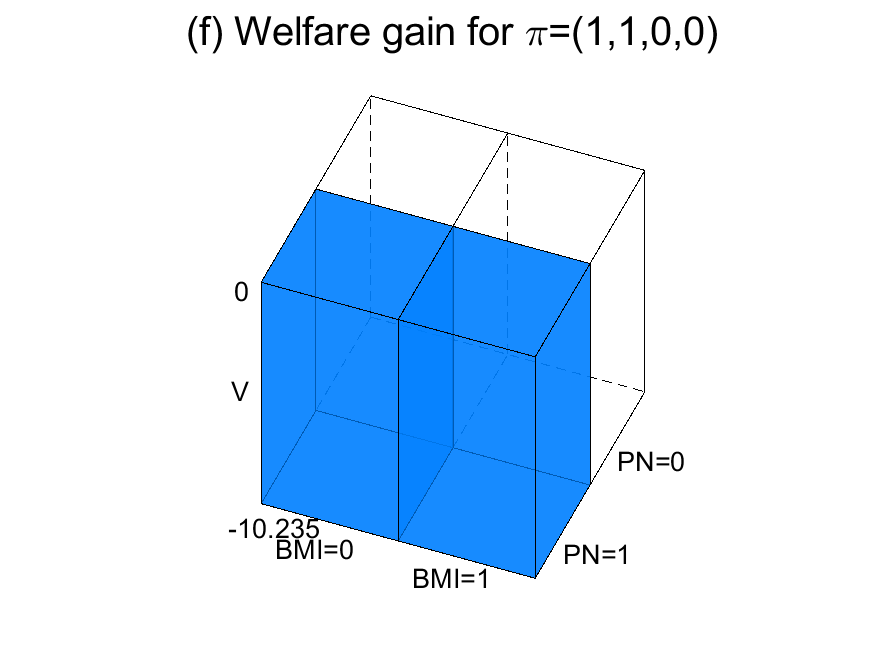}
		\includegraphics[width=2.1in]{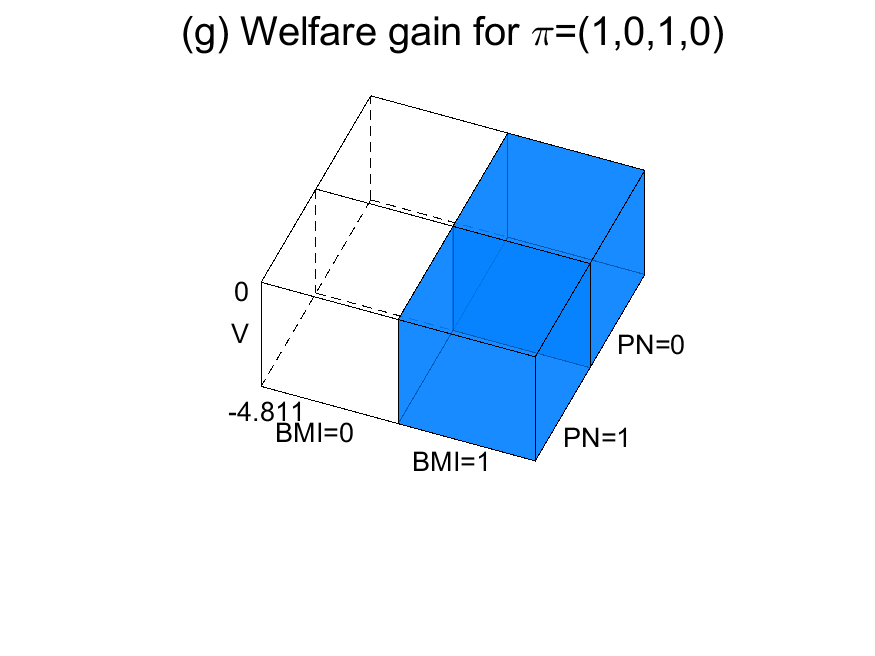}
		\includegraphics[width=2.1in]{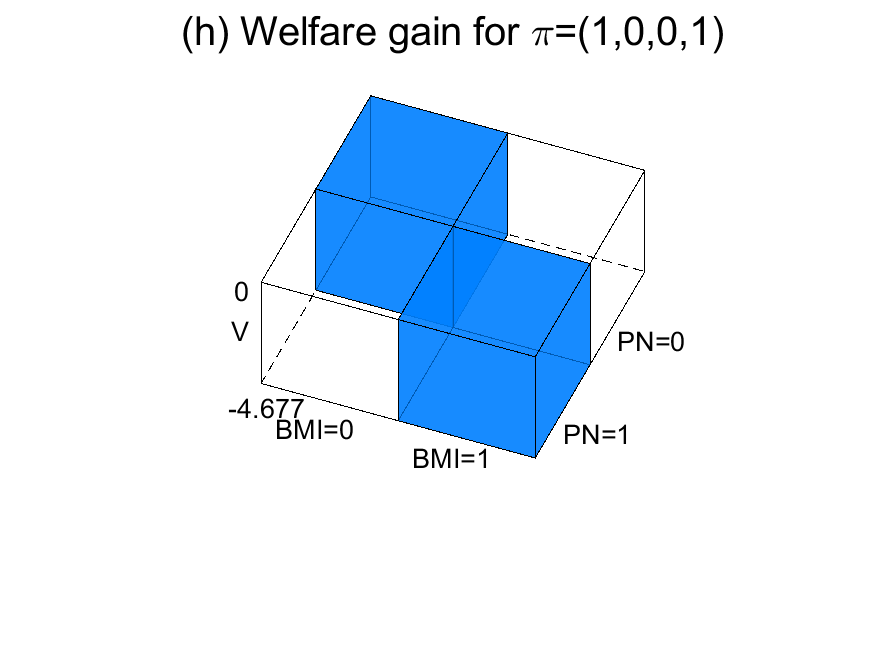}
		\includegraphics[width=2.1in]{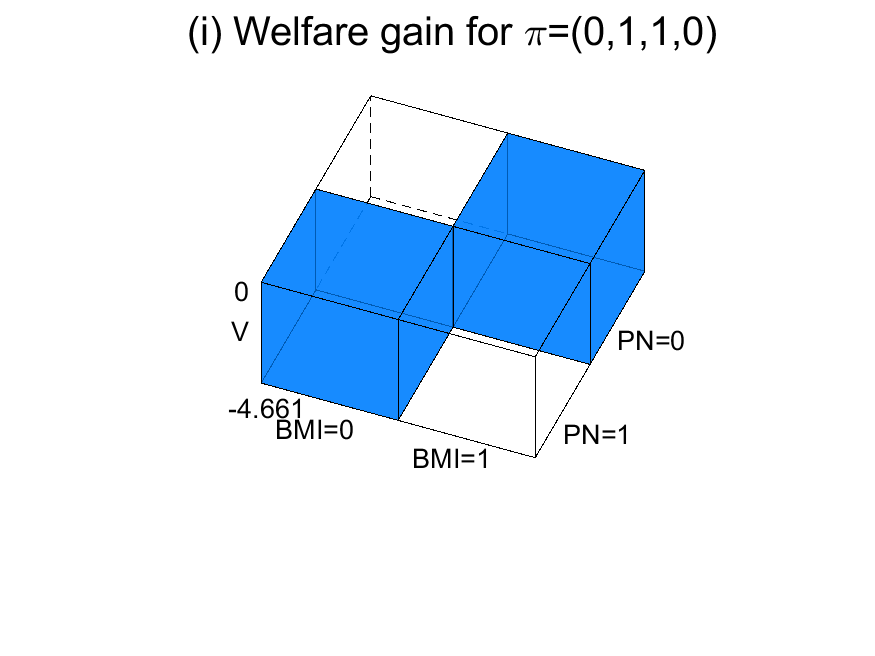}
		\includegraphics[width=2.1in]{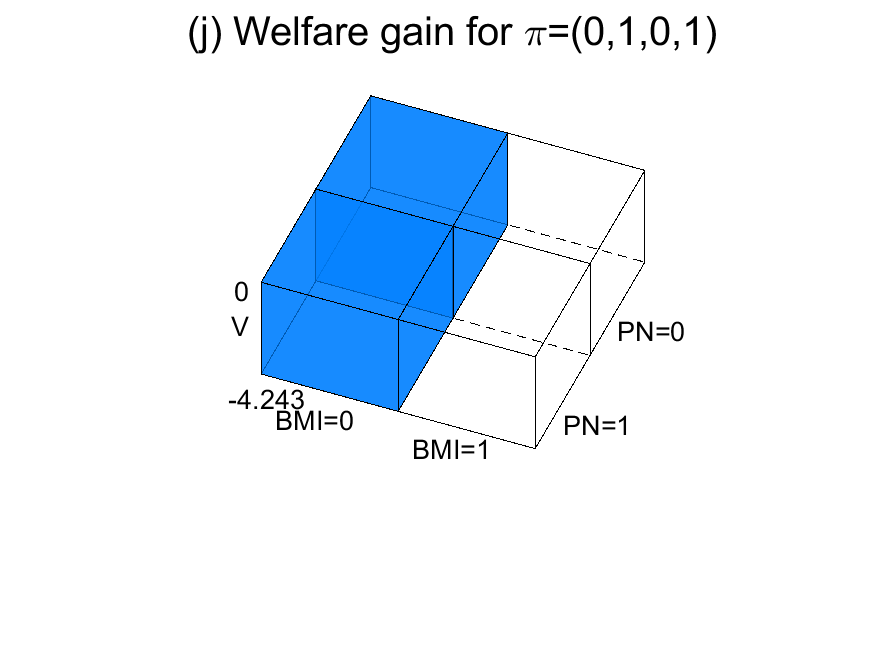}
		\includegraphics[width=2.1in]{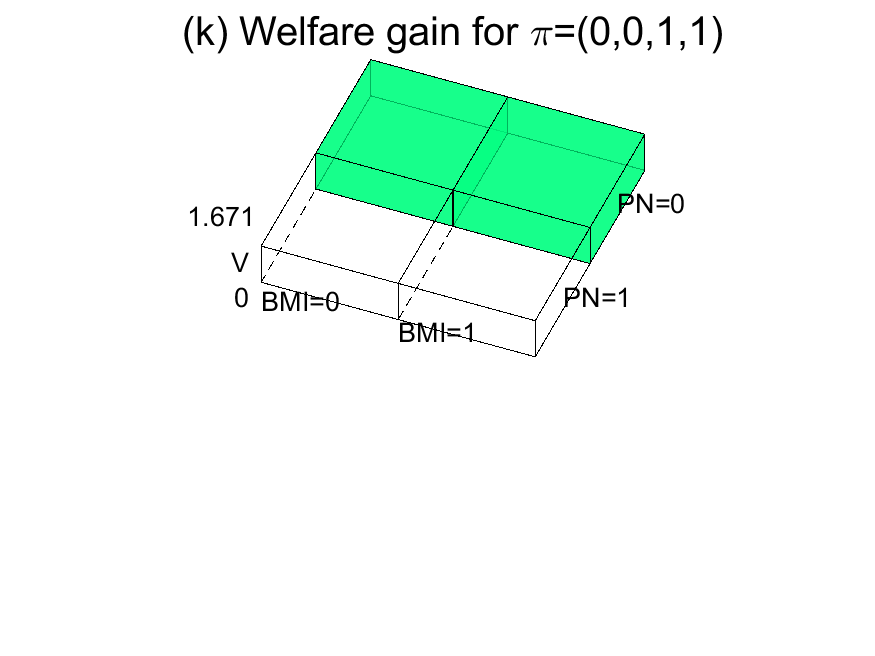}
		\includegraphics[width=2.1in]{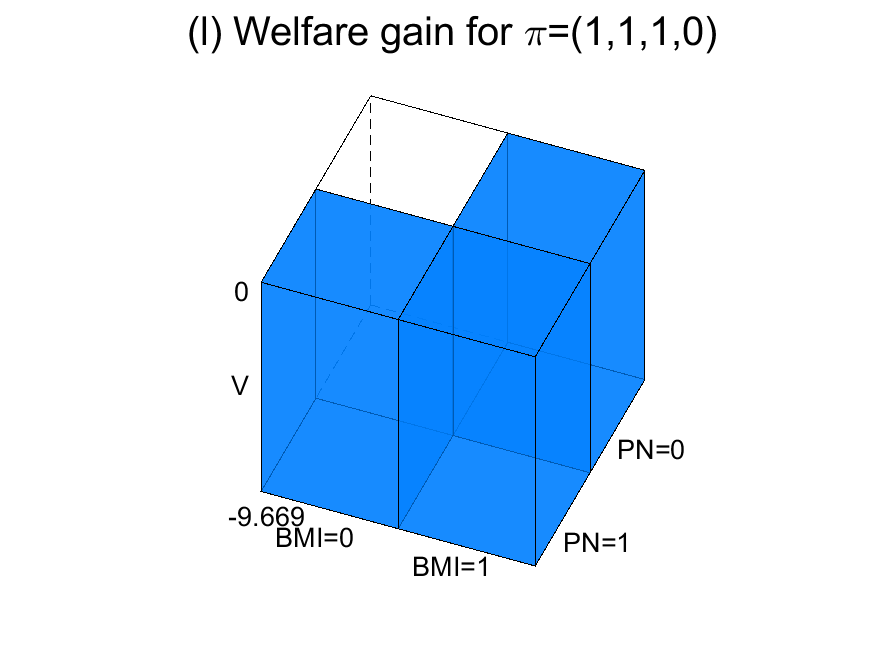}
		\includegraphics[width=2.1in]{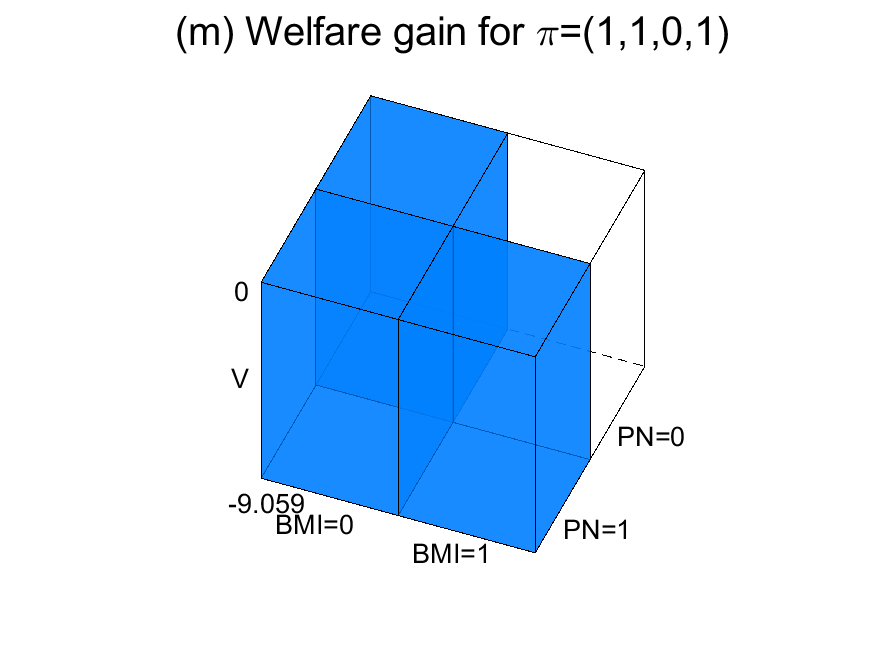}
		\includegraphics[width=2.1in]{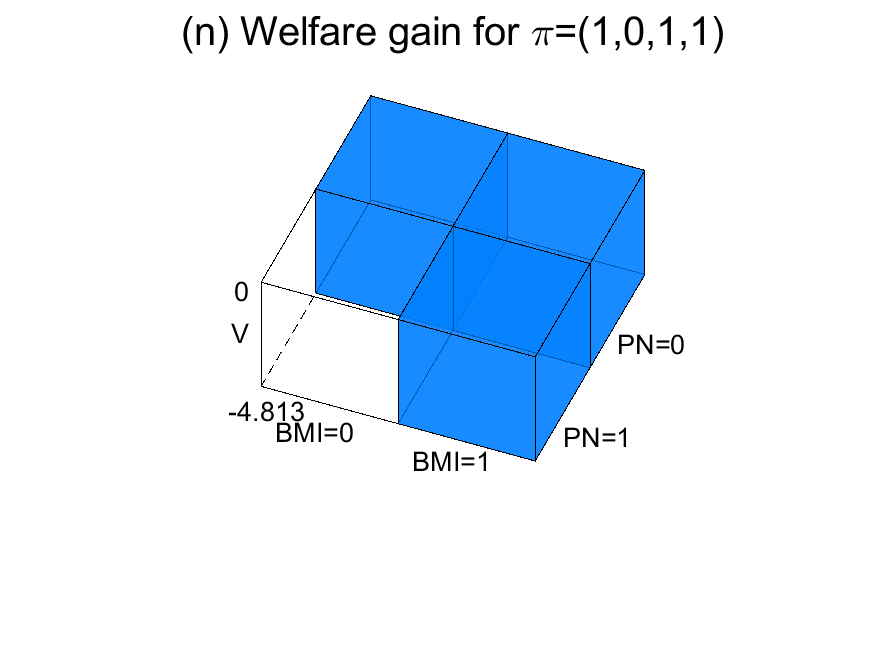}
		\includegraphics[width=2.1in]{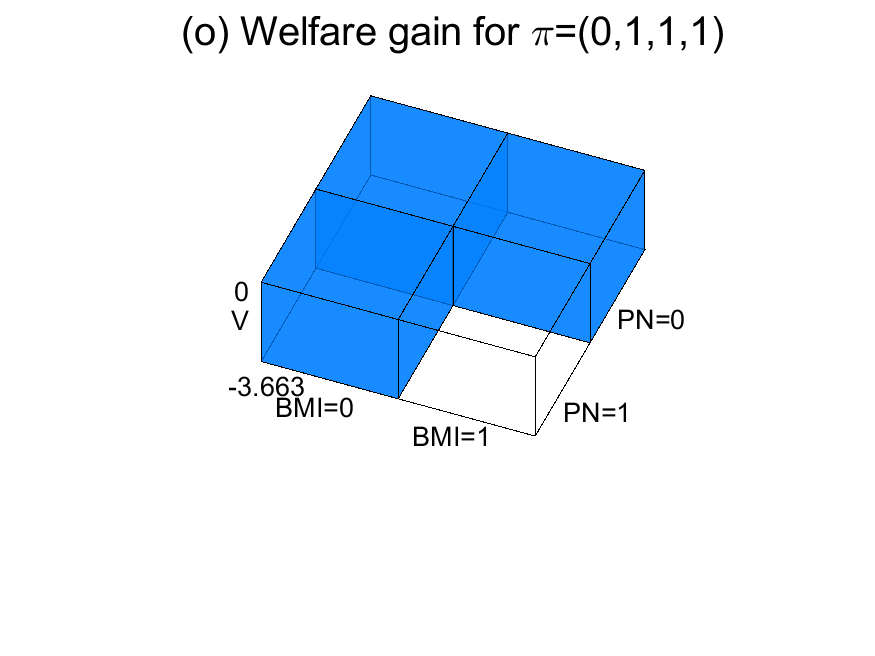}
		\caption*{\textbf{Figure 6.1}: The estimated welfare gain $\widehat{V}_n(0.01,0.197,\pi,\mathcal{G})$ for different $\pi$ and scale intervention $\mathcal{G}_\delta(D)=D/(1+\delta)$. The highlighted sections indicate which subpopulation receives the intervention. The height of each cuboid represents the corresponding welfare gain for low birth weight infants. Positive welfare gains are depicted in green, while negative gains are shown in blue.}
	\end{figure}

\section{Conclusions}

This paper offers OCPPEs as novel
and $\sqrt{n}$-estimable causal quantities to evaluate the effect of a general counterfactual change in a target
covariate that is heterogeneous across the unconditional distribution of $Y$. We propose a debiased machine learning (DML) estimator for an OCPPE, compatible
with high-dimensional settings. We show the estimator is
$\sqrt{n}$-consistent and asymptotically normal uniformly in $\mathcal{G}_{\delta}(\cdot)$
in a compact function space. We prove uniform validity of the multiplier bootstrap  for inference of
an OCPPE process.  Our work is the
first to develop uniform limiting theories for causal quantities measuring distributional impacts
of counterfactual changes in covariates.

The derived doubly robust score for an OCPPE estimand paves the way for learning optimal intervention assignment rules
within the framework of Empirical Welfare Maximization.  We utilize the OCPPE to analyze how anti-smoking policies impact low percentiles of infants'
birth weight. We find these policies increase infants'
birth weight over almost the entire range of quantiles considered except for extremal low quantiles. Low birth weight infants hardly benefit by an anti-smoking policy that treats everyone uniformly without accounting for personal characteristics.
We then make a preliminary attempt at conditioning intervention assignment on a small number of covariates, in order to achieve higher welfare gain for the target subpopulation (infants with birth weight lower than 2500 grams). We find assigning intervention to mothers who paid no visit at all can raise the baseline welfare gain (by assigning intervention to every mother in the sample) by 80\%.

	\newpage
	
	\section*{Appendix A: Proofs}

	\subsection*{A.1. Notations and Assumptions}

	\noindent Denote nuisance parameters in Proposition 2.2 as $\eta_1(\cdot)$, $\eta_2(D,X;\cdot)$, $\eta_3(D,X;\cdot)$ and $\eta_4(\cdot)$, respectively, namely,
	\[
	\begin{aligned}
		\eta(W;)&=\left(Q_{Y}(\cdot),F_Y(\cdot|D,X),\frac{\partial_D\Big(\vartheta(D;\cdot)f(D,X)\Big)}{f(D,X)},f_Y(\cdot),E\Big[\vartheta(D;\cdot)\partial_DF_Y\left(Q_Y(\cdot)|D,X\right)\Big]\right),\\
		&\equiv\left(\eta_1(\cdot),\eta_2(D,X;\cdot),\eta_3(D,X;\cdot),\eta_4(\cdot),E\left[\vartheta(D;\cdot)\partial_D\eta_2\Big(D,X;\eta_1(\cdot)\Big)\right]\right).
	\end{aligned}
	\]
	Notice that $\eta_{5}$ is expressed as the combination of $\eta_{1}$ and $\eta_{2}$. The following regularity conditions are needed to derive Propositions 2.1.
	
	\noindent{\textbf{Assumption A.1.}} The conditional CDF $F_Y(y|d,x)$ is absolutely continuous with respect to the Lebesgue measure for $d$ in a neighborhood of $d$ given $x$.  The density $f_Y(y|d,x)$ is continuous at $(y,d)=\left(Q_Y(\tau|d,x),d\right)$ and bounded in $y\in\mathbb{R}$.
	
	\noindent{\textbf{Assumption A.2.}}	$Q_Y(\tau|d,x)$ is partially differentiable with respect to $d$. There exists a measurable function $m^{(1)}$ that satisfies
	\begin{eqnarray*}
		P\bigg(\left|m(d+t,x,U)-m(d,x,U)-tm^{(1)}(d,x,U)\right|\geq t\epsilon\bigg|D=d,X=x\bigg)=o(t)
	\end{eqnarray*}
	for $t\rightarrow 0^+$ and any fixed $\epsilon>0$, where $m^{(1)}(d,x,u)=\partial_dm(d,x,u)$.
	
	\noindent{\textbf{Assumption A.3.}}
	The conditional distribution of $\left(Y,\partial_dm\left(d,x,U\right)\right)$ given $D=d$ and $X=x$ is absolutely continuous with respect to the Lebesgue measure. For the conditional density $f_{Y,\partial_dm\left(d,x,U\right)|D,X}$ of $(Y,\partial_dm\left(d,x,U\right))$ given $D$ and $X$, we require that
	\[
	f_{Y,\partial_dm\left(d,x,U\right)|D,X}(y,y'|d,x)\leq Cg(y'),
	\]
	where $C$ is a positive constant and $g$ is a positive density on $\mathbb{R}$ with finite mean (i.e., $\int |y'|g(y')dy'<\infty$).
	
%
	
	\subsection*{A.2. Proofs for Section 2}
	
	\noindent\textbf{Proof of Proposition 2.1.} Notice that
\[\begin{aligned}
		\theta(\tau_1,\tau_2,\mathcal{G})=&E\left[\lim_{\delta\rightarrow 0}\frac{m\left(\mathcal{G}_{\delta}(D),X,U\right)-m\left(D,X,U\right)}{\delta}\bigg|Y\in\left(Q_Y(\tau_1),Q_Y(\tau_2)\right)\right] \\
	=&E\left[\frac{\partial \mathcal{G}_\delta(D)}{\partial \delta}\Big|_{\delta=0}\cdot\partial_D m(D,X,U)\bigg|Y\in\left(Q_Y(\tau_1),Q_Y(\tau_2)\right)\right] \\
	=&E\left[\vartheta\left(D;\mathcal{G}\right)\partial_D m(D,X,U)\bigg|Y\in\left(Q_Y(\tau_1),Q_Y(\tau_2)\right)\right]\\
	=&\frac{1}{\tau_2-\tau_1}E\left[1\Big\{Q_Y(\tau_1)<Y<Q_Y(\tau_2)\Big\}\vartheta\left(D;\mathcal{G}\right)\partial_D m(D,X,U)\right]\\
	=&\frac{1}{\tau_2-\tau_1}\int_{Q_Y(\tau_1)}^{Q_Y(\tau_2)}E\left[\vartheta\left(D;\mathcal{G}\right)\partial_D m(D,X,U)\bigg|Y=y\right]f_Y(y)dy. \\
\end{aligned}\]
According to the proof of Proposition 2.1 in Jin et al. (2024), we have
\[
E\left[\partial_D m(D,X,U)\bigg|Y=y,D=d,X=x\right]=-\frac{\partial_dF_Y(y|d,x)}{f_Y(y|d,x)}.
\]
Thus, we can conclude that
\[\begin{aligned}
	\theta(\tau_1,\tau_2,\mathcal{G})=&\frac{1}{\tau_2-\tau_1}\int_{Q_Y(\tau_1)}^{Q_Y(\tau_2)}E\left[\vartheta\left(D;\mathcal{G}\right)\cdot\left(-\frac{\partial_{D}F_Y(y|D,X)}{f_Y(y|D,X)}\right)\bigg|Y=y\right]f_Y(y)dy \\
	=&-\frac{1}{\tau_2-\tau_1}\int\vartheta\left(D;\mathcal{G}\right)\left[\int_{Q_Y(\tau_1)}^{Q_Y(\tau_2)}\frac{\partial_{D}F_Y(Y|D,X)}{f_Y(Y|D,X)}f_Y(Y|D,X)dY\right]f(D,X)dDdX \\
	=&-\frac{1}{\tau_2-\tau_1}E\left[\vartheta\left(D;\mathcal{G}\right)\int_{Q_Y(\tau_1)}^{Q_Y(\tau_2)}\partial_DF_Y(y|D,X)dy\right],
\end{aligned}\]
which completes the proof. $\blacksquare$
\\

	\noindent\textbf{Proof of Proposition 2.2.} By definition, the score function $\psi$ can be rewritten as
	\[\begin{aligned}
		\psi\bigg(W,\theta,\eta;\tau_1,\tau_2,\mathcal{G}\bigg)=&\frac{-1}{\tau_2-\tau_1}\vartheta\left(D;\mathcal{G}\right)\int_{\eta_1(\tau_1)}^{\eta_1(\tau_2)}\partial_D\eta_2(D,X;y)dy-\theta\\
		&+\frac{1}{\tau_2-\tau_1}\eta_3(D,X;\mathcal{G})\left(\int_{\eta_1(\tau_1)}^{\eta_1(\tau_2)}\eta_2(D,X;y)dy-\int_{\eta_1(\tau_1)}^{\eta_1(\tau_2)}1\{Y<y\}dy\right)\\
		&-\frac{1}{\tau_2-\tau_1}\frac{E\left[\vartheta\left(D;\mathcal{G}\right)\partial_D\eta_2\Big(D,X;\eta_1(\tau_1)\Big)\right]}{\eta_4\Big(\eta_1(\tau_1)\Big)}\left(1\Big\{Y\leq \eta_1(\tau_1)\Big\}-\tau_1\right)\notag\\
		&+\frac{1}{\tau_2-\tau_1}\frac{E\left[\vartheta\left(D;\mathcal{G}\right)\partial_D\eta_2\Big(D,X;\eta_1(\tau_2)\Big)\right]}{\eta_4\Big(\eta_1(\tau_2)\Big)}\left(1\Big\{Y\leq\eta_1(\tau_2)\Big\}-\tau_2\right).\\
	\end{aligned}\]
	It is straightforward to show that $E\psi\bigg(W,\theta(\tau_1,\tau_2,\mathcal{G}),\eta;\tau_1,\tau_2,\mathcal{G}\bigg)=0$, and then part (i) is proved.

	For part (ii), notice that $E\left[\eta_2(D,X;y)-1\Big\{Y<y\Big\}\Big|D,X\right]=E\left[1\Big\{Y\leq \eta_1(\tau_1)\Big\}-\tau_1\right]=E\left[1\Big\{Y\leq \eta_1(\tau_2)\Big\}-\tau_2\right]=0$ and $\eta_4(\cdot)=f_Y(\cdot)$. Thus,
	\[\begin{aligned}
		&\frac{\partial E\psi\bigg(W,\theta(\tau_{1},\tau_{2},\mathcal{G}),\eta_1+r(\widetilde{\eta}_1-\eta_1),\eta_2,\eta_3,\eta_4;\tau_1,\tau_2,\mathcal{G}\bigg)}{\partial r}\bigg|_{r=0}\\
		=&-\frac{1}{\tau_2-\tau_1}\sum_{j=1}^{2}(-1)^{j}E\left[\vartheta\left(D;\mathcal{G}\right)\partial_D\eta_2\Big(D,X;\eta_1(\tau_{j})\Big)\right]\Big(\widetilde{\eta}_1(\tau_{j})-\eta_1(\tau_{j})\Big)\\
		&+\frac{1}{\tau_2-\tau_1}\sum_{j=1}^{2}(-1)^{j}\frac{E\left[\vartheta\left(D;\mathcal{G}\right)\partial_D\eta_2\Big(D,X;\eta_1(\tau_j)\Big)\right]}{\eta_4\Big(\eta_1(\tau_j)\Big)}f_Y\Big(\eta_1(\tau_j)\Big)\Big(\widetilde{\eta}_1(\tau_j)-\eta_1(\tau_j)\Big)\notag\\
		&-\frac{1}{\tau_2-\tau_1}E\left\{\eta_3(D,X;\mathcal{G})\sum_{j=1}^{2}(-1)^{j}E\left[\eta_2(D,X;y)-1\Big\{Y<y\Big\}|D,X\right]\Big|_{y=\eta_1(\tau_j)}\right\}\Big(\widetilde{\eta}_1(\tau_j)-\eta_1(\tau_j)\Big) \\
		&+\frac{1}{\tau_2-\tau_1}\sum_{j=1}^{2}(-1)^{j}\frac{\partial}{\partial{t}}\left(\frac{E\left[\vartheta\left(D;\mathcal{G}\right)\partial_D\eta_2\Big(D,X;t\Big)\right]}{\eta_4\Big(t\Big)}\right)\Big|_{t=\eta_1(\tau_{j})} \\
		&\times\Big(\widetilde{\eta}_1(\tau_{j})-\eta_1(\tau_{j})\Big)E\left[1\Big\{Y\leq\eta_1(\tau_j)\Big\}-\tau_j\right], \\
	\end{aligned}\]
	which is equal to $0$. Similarly, we can show that
	\[\begin{aligned}
		&\frac{\partial E\psi\bigg(W,\theta(\tau_{1},\tau_{2},\mathcal{G}),\eta_1,\eta_2+r(\widetilde{\eta}_2-\eta_2),\eta_3,\eta_4;\tau_1,\tau_2,\mathcal{G}\bigg)}{\partial r}\bigg|_{r=0}\\
		=&\frac{-1}{\tau_2-\tau_1}E\bigg[\vartheta\left(D;\mathcal{G}\right)\int_{\eta_1(\tau_1)}^{\eta_1(\tau_2)}\partial_D\left(\widetilde{\eta}_2(D,X,y)-\eta_2(D,X;y)\right)dy\\
		&+\eta_3(D,X;\mathcal{G})\int_{\eta_1(\tau_1)}^{\eta_1(\tau_2)}\left(\widetilde{\eta}_2(D,X,y)-\eta_2(D,X;y)\right)dy\bigg]. \\
	\end{aligned}\]
	After applying integration by part, it is not difficult to show that 
	\[\begin{aligned}
		E\bigg[\vartheta\left(D;\mathcal{G}\right)\int_{\eta_1(\tau_1)}^{\eta_1(\tau_2)}\partial_D&\left(\widetilde{\eta}_2(D,X;y)-\eta_2(D,X;y)\right)dy\\
		&+\eta_3(D,X;\mathcal{G})\int_{\eta_1(\tau_1)}^{\eta_1(\tau_2)}\left(\widetilde{\eta}_2(D,X;y)-\eta_2(D,X;y)\right)dy\bigg]=0. \\
	\end{aligned}\]
	Moreover, we can similarly show that
	\[
	\frac{\partial E\psi\bigg(W,\theta(\tau_{1},\tau_{2},\mathcal{G}),\eta_1,\eta_2,\eta_3+r(\widetilde{\eta}_3-\eta_3),\eta_4;\tau_1,\tau_2,\mathcal{G}\bigg)}{\partial r}\bigg|_{r=0}=0
	\]
	and
	\[
	\frac{\partial E\psi\bigg(W,\theta(\tau_{1},\tau_{2},\mathcal{G}),\eta_1,\eta_2,\eta_3,\eta_4+r(\widetilde{\eta}_4-\eta_4);\tau_1,\tau_2,\mathcal{G}\bigg)}{\partial r}\bigg|_{r=0}=0
	\]
	by the fact that $E\left[\eta_2(D,X;y)-1\Big\{Y<y\Big\}\Big|D,X\right]=0$ and $E\left[1\Big\{Y\leq \eta_1(\tau_1)\Big\}-\tau_1\right]=0$, \\
	\noindent$E\left[1\Big\{Y\leq \eta_1(\tau_2)\Big\}-\tau_2\right]=0$, respectively. This completes the proof of part (ii). 
	
	Finally, we turn to analyze part (iii). It is straightforward to show that
	\[
	E\Big[\psi\Big(W,\theta,{\eta};\tau_1,\tau_2,\mathcal{G}\Big)\Big]=E\Big[\psi\Big(W,\theta,{\eta}_{1},{\eta}_{2},\widetilde{\eta}_3,\widetilde{\eta}_4;\tau_1,\tau_2,\mathcal{G}\Big)\Big]
	\]
	by the fact that $E\left[\eta_2(D,X;y)-1\Big\{Y<y\Big\}\Big|D,X\right]=0$ and $E\left[1\Big\{Y\leq \eta_1(\tau_1)\Big\}-\tau_1\right]=0$, \\
	\noindent$E\left[1\Big\{Y\leq \eta_1(\tau_2)\Big\}-\tau_2\right]=0$. We then prove another equality. Notice that
		\[\begin{aligned}
		&E\psi\bigg(W,\theta,{\eta}_{1},\widetilde{\eta}_{2},\eta_3,\widetilde{\eta}_{4};\tau_1,\tau_2,\mathcal{G}\bigg)\\
		=&-\frac{1}{\tau_2-\tau_1}E\bigg[\vartheta\left(D;\mathcal{G}\right)\int_{\eta_1(\tau_1)}^{\eta_1(\tau_2)}\partial_D\widetilde{\eta}_2(D,X;y)dy\bigg]-\theta\\
		&-\frac{1}{\tau_2-\tau_1}E\left[\eta_3(D,X;\mathcal{G})\left(\int_{\eta_1(\tau_1)}^{\eta_1(\tau_2)}\widetilde{\eta}_2(D,X;y)dy-\int_{\eta_1(\tau_1)}^{\eta_1(\tau_2)}1\{Y<y\}dy\right)\right].\\
	\end{aligned}\]
	Applying integration by part, we have
	\[
	E\bigg[\vartheta\left(D;\mathcal{G}\right)\int_{\eta_1(\tau_1)}^{\eta_1(\tau_2)}\partial_D\widetilde{\eta}_2(D,X;y)dy+\eta_3(D,X;\mathcal{G})\int_{\eta_1(\tau_1)}^{\eta_1(\tau_2)}\widetilde{\eta}_2(D,X;y)dy\bigg]=0.
	\]
	Then we can conclude that
	\[\begin{aligned}
		E\psi\bigg(W,\theta,{\eta}_{1},\widetilde{\eta}_{2},{\eta}_{3},\widetilde{\eta}_4;\tau_1,\tau_2,\mathcal{G}\bigg)=&\frac{1}{\tau_2-\tau_1}E\bigg[\eta_3(D,X;\mathcal{G})\int_{\eta_1(\tau_1)}^{\eta_1(\tau_2)}1\{Y<y\}dy\bigg]-\theta \\
		=&-\frac{1}{\tau_2-\tau_1}E\bigg[\vartheta\left(D;\mathcal{G}\right)\int_{\eta_1(\tau_1)}^{\eta_1(\tau_2)}\partial_D\eta_2(D,X;y)dy\bigg]-\theta\\
		=&E\psi\bigg(W,\theta,{\eta};\tau_1,\tau_2,\mathcal{G}\bigg), \\
	\end{aligned}\]
	which completes the proof. $\blacksquare$

	\subsection*{A.3. Proofs for Section 4}

	In the proof $a\lesssim{b}$ means that $a\leq{cb}$, where the constant $c$ depends on the constants in Assumptions 4.1-4.3 only, but not on $n$. Let $\Delta_{n}$ and $\delta_{n}$ be fixed sequences of numbers satisfying $\Delta_{n}\to0^+$ and $\delta_{n}\to0^+$ at a speed at most polynomial in $n$, and $C$ be a generic positive constant. We suppress the claim ``uniformly over $u\in\mathcal{U}$" throughout the proof.
	
	\noindent\textbf{Proof of Theorem 4.1.}
	We abbreviate the process $\widehat{\theta}(u)$ and ${\theta}(u)$ into $\widehat{\theta}$ and ${\theta}$. 
	
	\noindent\textbf{Step 1. (Linearization)} In this step, we establish the claim that the pre-estimators have no first order effects, namely  
	\[
	\sqrt{n}\left(\widehat{\theta}-\theta\right)=Z_{n}+o_{P}(1) \quad in \quad \mathbb{D}=\ell^{\infty}\left(\mathcal{U}\right),
	\]
	where $Z_{n}=\mathbb{G}_{n}\psi(W,\theta,\eta;u)$. 
	
	For $t\in\{1,2\}$, define the following spaces of functions:
	\[
	\mathcal{Q}_{1}^{t}=\left\{
	\begin{array}{l}
		u\mapsto\widetilde{Q}_{Y}(\tau_{t}): \left\Vert\widetilde{Q}_{Y}(\tau_{t})\right\Vert_{0}=1 \\
		\left\Vert \widetilde{Q}_{Y}(\tau_{t})-Q_{Y}(\tau_{t})\right\Vert_{P,2}=o_{p}\left(n^{-1/4}\right) \\
		\left\Vert \widetilde{Q}_{Y}(\tau_{t})-Q_{Y}(\tau_{t})\right\Vert_{P,\infty}=o_{p}(1) \\
	\end{array}
	\right\}
	\]
	and
	\[
	\mathcal{Q}_{2}^{t}=\left\{
	\begin{array}{l}
		u\mapsto\widetilde{f}_{Y}\left(\widetilde{Q}_{Y}(\tau_{t})\right): \widetilde{Q}_{Y}(\tau_{t})\in\mathcal{Q}_{1}^{t}  \\
		\left\Vert \widetilde{f}_{Y}\left(\widetilde{Q}_{Y}(\tau_{t})\right)-f_{Y}\Big(Q_{Y}(\tau_{t})\Big)\right\Vert_{P,2}=o_{p}\left(n^{-1/4}\right) \\
		\left\Vert \widetilde{f}_{Y}\left(\widetilde{Q}_{Y}(\tau_{t})\right)-f_{Y}\Big(Q_{Y}(\tau_{t})\Big)\right\Vert_{P,\infty}=o_{p}(1) \\
	\end{array}
	\right\}.
	\]
	With probability no less than $1-\Delta_{n}$, for $t\in\{1,2\}$, it is obvious that 
	\[
	\widehat{Q}_{Y}(\tau_{t})\in\mathcal{Q}_{1}^{t}, \quad \widehat{f}_{Y}\left(\widehat{Q}_{Y}(\tau_{t})\right)\in\mathcal{Q}_{2}^{t}.
	\]
	by the standard results of quantile regression, kernel regression and Assumption 4.3(i).
	
	Subsequently, for $t\in\{1,2\}$, define the following spaces of functions:
	\[
	\mathcal{F}_{1}=\left\{
	\begin{array}{l}
		(d,x,u)\mapsto\displaystyle\int_{\widetilde{Q}_{Y}(\tau_{1})}^{\widetilde{Q}_{Y}(\tau_{2})}\Lambda\bigg(b(d,x)^{\prime}\widetilde{\beta}(y)\bigg)dy: \left\Vert\widetilde{\beta}(y)\right\Vert_{0}\leq{Cs_{\beta}}, \widetilde{Q}_{Y}(\tau_{1})\in\mathcal{Q}_{1}^{1}, \widetilde{Q}_{Y}(\tau_{2})\in\mathcal{Q}_{1}^{2} \\
		\left\Vert \displaystyle\int_{\widetilde{Q}_{Y}(\tau_{1})}^{\widetilde{Q}_{Y}(\tau_{2})}\Lambda\bigg(b(D,X)^{\prime}\widetilde{\beta}(y)\bigg)dy-\displaystyle\int_{Q_{Y}(\tau_{1})}^{Q_{Y}(\tau_{2})}F_{Y}(y|D,X)dy\right\Vert_{P,2}=o_{p}\left(n^{-1/4}\right) \\
		\left\Vert \displaystyle\int_{\widetilde{Q}_{Y}(\tau_{1})}^{\widetilde{Q}_{Y}(\tau_{2})}\Lambda\bigg(b(D,X)^{\prime}\widetilde{\beta}(y)\bigg)dy-\displaystyle\int_{Q_{Y}(\tau_{1})}^{Q_{Y}(\tau_{2})}F_{Y}(y|D,X)dy\right\Vert_{P,\infty}=o_{p}(1) \\
	\end{array}
	\right\},
	\]
	\[
	\mathcal{F}_{2}=\left\{
	\begin{array}{l}
		(d,x,u)\mapsto \partial_{d}\widetilde{F}(d,x,y): \widetilde{F}\in\mathcal{F}_{1} \\
		\left\Vert \partial_{D}\widetilde{F}(D,X,y)-\partial_{D}\displaystyle\int_{Q_{Y}(\tau_{1})}^{Q_{Y}(\tau_{2})}F_{Y}(y|D,X)dy\right\Vert_{P,2}=o_{p}\left(n^{-1/4}\right) \\
		\left\Vert \partial_{D}\widetilde{F}(D,X,y)-\partial_{D}\displaystyle\int_{Q_{Y}(\tau_{1})}^{Q_{Y}(\tau_{2})}F_{Y}(y|D,X)dy\right\Vert_{P,\infty}=o_{p}(1) \\
	\end{array}
	\right\},
	\]
	\[
	\mathcal{F}_{3}^{t}=\left\{
	\begin{array}{l}
		u\mapsto \mathbb{E}_{n}\left[\vartheta(D_{i};\sigma)\cdot\partial_{D_{i}}\Lambda\bigg(b(D_{i},X_{i})^{\prime}\widetilde{\beta}\left(\widetilde{Q}_{Y}(\tau_{t})\right)\bigg)\right]=\mathbb{E}_{n}\widetilde{\Gamma}(D_{i},X_{i};\tau_{t},\sigma): \\
		\left\Vert\widetilde{\beta}(y)\right\Vert_{0}\leq{Cs_{\beta}}, \widetilde{Q}(\tau_{t})\in\mathcal{Q}_{1}^{t} \\
		\left\Vert \mathbb{E}_{n}\widetilde{\Gamma}(D_{i},X_{i};\tau_{t},\sigma)-E\left[\vartheta(D;\sigma)\cdot\partial_{D}F_{Y}\Big(Q_{Y}(\tau_{t})|D,X\Big)\right]\right\Vert_{P,2}=o_{p}\left(n^{-1/4}\right) \\
		\left\Vert \mathbb{E}_{n}\widetilde{\Gamma}(D_{i},X_{i};\tau_{t},\sigma)-E\left[\vartheta(D;\sigma)\cdot\partial_{D}F_{Y}\Big(Q_{Y}(\tau_{t})|D,X\Big)\right]\right\Vert_{P,\infty}=o_{p}(1) \\
	\end{array}
	\right\},
	\]
	\[
	\mathcal{L}=\left\{
	\begin{array}{l}
		(d,x,u)\mapsto h(d,x)^{\prime}\widetilde{\gamma}(\sigma): \Big\Vert\widetilde{\gamma}(\sigma)\Big\Vert_{0}\leq Cs_{\gamma} \\
		\Big\Vert h(D,X)^{\prime}\widetilde{\gamma}(\sigma)-L(D,X;\sigma)\Big\Vert_{P,2}=o_{p}\left(n^{-1/4}\right) \\
		\Big\Vert h(D,X)^{\prime}\widetilde{\gamma}(\sigma)-L(D,X;\sigma)\Big\Vert_{P,\infty}=o_{p}(1) \\
	\end{array}
	\right\}.
	\]
	Applying similar arguments as the proof of Theorem 4.1 in Belloni et al. (2017), Lemma D.2 in Sasaki et al. (2022) and Theorem 4.1 in Jin et al. (2024), we can derive that under Assumptions 4.2-4.3, with probability no less than $1-\Delta_{n}$ and for $t\in\{1,2\}$,
	\[
	\widehat{IF}(d,x;u)\in\mathcal{F}_{1}, \ \widehat{IDF}(d,x;u)\in\mathcal{F}_{2}, \ \mathbb{E}_{n}\left[\vartheta(D_{i};\sigma)\widehat{DF}\left(D_{i},X_{i};\widehat{Q}_{Y}(\tau_{t})\right)\right]\in\mathcal{F}_{3}^{t} \ \ \text{and} \ \  \widehat{L}(d,x;\sigma)\in\mathcal{L}.\eqno{(T.4.1.1)}
	\]

	We have that
	\[\begin{aligned}
		\sqrt{n}\left(\widehat{\theta}(u)-\theta(u)\right)
		=&\underbrace{\mathbb{G}_{n}\psi(W,\theta,\eta;u)}_{\dagger_{4.1.1}}+\underbrace{\mathbb{G}_{n}\bigg[\psi\left(W,\theta,\widetilde{\eta};u\right)-\psi(W,\theta,\eta;u)\bigg]}_{\dagger_{4.1.2}} \\
		&+\underbrace{\sqrt{n}E\bigg[\psi\left(W,\theta,\widetilde{\eta};u\right)-\psi\left(W,\theta,{\eta};u\right)\bigg]}_{\dagger_{4.1.3}},
	\end{aligned}\]
	with $\widetilde{\eta}$ evaluated at $\widetilde{\eta}=\widehat{\eta}$.
	
	Firstly, we consider $\dagger_{4.1.3}$. After applying a series of Taylor expansions, orthogonality property, convergence rates described in $(T.4.1.1)$ and similar arguments as the proof of Theorem 4.1 in Jin et al. (2024), we can conclude that with probability no less than $1-\Delta_{n}$,
	\[
	P\bigg(\left|\dagger_{4.1.3}\right|\lesssim\delta_{n}\bigg)\geq1-\Delta_{n}.
	\]
	
	Then we consider $\dagger_{4.1.2}$. Define $\bar{\psi}\left(W,\widetilde{\eta};u\right)={\psi}\left(W,\theta,\widetilde{\eta};u\right)+\theta$ and $\mathcal{R}=\left(\mathcal{Q}_{1}^{1}\cup\mathcal{Q}_{1}^{2}\right)\times\left(\mathcal{F}_{1}\cup\mathcal{F}_{2}\right)\times\mathcal{L}\times\left(\mathcal{Q}_{2}^{1}\cup\mathcal{Q}_{2}^{2}\right)\times\left(\mathcal{F}_{3}^{1}\cup\mathcal{F}_{3}^{2}\right)$. Thus, with probability no less than $1-\Delta_{n}$,
	\[
	\left|\dagger_{4.1.2}\right|\leq\sup_{\widetilde{\eta}\in\mathcal{R}}\left|\mathbb{G}_{n}\bigg[\bar{\psi}\left(W,\widetilde{\eta};u\right)-\bar{\psi}(W,\eta;u)\bigg]\right|.
	\]
	To bound the term at the right-hand side, we further decompose the moment function $\bar{\psi}\left(W,\widetilde{\eta};u\right)$ into
	\[
	\bar{\psi}\left(W,\widetilde{\eta};u\right)=\bar{\psi}_{1}\left(W,\widetilde{\eta};u\right)+\sum_{t=1}^{2}\bar{\psi}_{2}^{t}\left(W,\widetilde{\eta};u\right),
	\]
	where
	\[\begin{aligned}
		\bar{\psi}_{1}\left(W,\widetilde{\eta};u\right)=&\frac{-1}{\tau_{2}-\tau_{1}}\vartheta(D;\sigma)\cdot\partial_D\int_{\widetilde{Q}_{Y}(\tau_{1})}^{\widetilde{Q}_{Y}(\tau_{2})}\widetilde{F}_Y(y|D,X)dy\notag\\
		&-\frac{1}{\tau_{2}-\tau_{1}}\widetilde{L}(D,X;\sigma)\int_{\widetilde{Q}_{Y}(\tau_{1})}^{\widetilde{Q}_{Y}(\tau_{2})}\left(\widetilde{F}_Y\left(y\big|D,X\right)-1\big\{Y<y\big\}\right)dy \\
	\end{aligned}\]
	and
	\[
	\bar{\psi}_{2}^{t}\left(W,\widetilde{\eta};u\right)=\frac{(-1)^{t}}{\tau_{2}-\tau_{1}}\cdot\frac{\mathbb{E}_{n}\left[\vartheta(D_{i};\sigma)\cdot\partial_D \widetilde{F}_Y\Big(\widetilde{Q}_{Y}(\tau_{t})|D_{i},X_{i}\Big)\right]}{\widetilde{f}_{Y}\Big(\widetilde{Q}_{Y}(\tau_{t})\Big)}\Big(1\left\{Y<\widetilde{Q}_{Y}(\tau_{t})\right\}-\tau_{t}\Big). 
	\]
Thus, with probability no less than $1-\Delta_{n}$, we have
	\[\begin{aligned}
		\left|\dagger_{4.1.2}\right|\leq&\sup_{\widetilde{\eta}\in\mathcal{R}}\left|\mathbb{G}_{n}\bigg[\bar{\psi}_{1}\left(W,\widetilde{\eta};u\right)-\bar{\psi}_{1}(W,\eta;u)\bigg]\right|+\sup_{\widetilde{\eta}\in\mathcal{R},t\in\{1,2\}}\left|\mathbb{G}_{n}\bigg[\bar{\psi}_{2}^{t}\left(W,\widetilde{\eta};u\right)-\bar{\psi}_{2}^{t}(W,\eta;u)\bigg]\right|. 
	\end{aligned}\]
	
	To bound the above empirical processes, we should calculate the entropy of each term. We first analyze the term with true nuisance functions $\eta$. For $t\in\{1,2\}$, the classes of functions
	\[
	\mathcal{V}_{1}=\left\{\displaystyle\int_{Q_{Y}(\tau_{1})}^{Q_{Y}(\tau_{2})}1\{Y\leq{y}\}dy:u\in\mathcal{U}\right\}
	\]
	and
	\[
	\mathcal{V}_{2}^{t}=\bigg\{1\{{Y}<{Q_{Y}(\tau_{t})}\}:u\in\mathcal{U}\bigg\}
	\]
	viewed as maps from the sample space $\mathcal{W}$ to the real line, are bounded by constant envelops and have finite VC dimensions. According to Theorem 2.6.7 in van der Vaart and Wellner (1996), we can deduce that
	\[\begin{aligned}
		&\sup_{Q}\log N\left(\epsilon,\mathcal{V}_{1},\Vert\cdot\Vert_{Q,2}\right)\lesssim\log(e/\epsilon),\\
		&\sup_{Q}\log N\left(\epsilon,\mathcal{V}_{2}^{t},\Vert\cdot\Vert_{Q,2}\right)\lesssim\log(e/\epsilon),\\
	\end{aligned}\]
	with the supremum taken over all finitely discrete probability measures $Q$ on $(\mathcal{W},\mathcal{A}_{\mathcal{W}})$. According to Lemma L.2 in Belloni et al. (2017), for $t\in\{1,2\}$, the following classes of functions
	\[
	\mathcal{V}_{3}=\left\{\displaystyle\int_{Q_{Y}(\tau_{1})}^{Q_{Y}(\tau_{2})}F_{Y}(y|D,X)dy:u\in\mathcal{U}\right\}
	\]
	and
	\[
	\mathcal{V}_{4}^{t}=\bigg\{\tau_{t}\equiv E1\{{Y}<{Q_{Y}(\tau_{t})}\}:u\in\mathcal{U}\bigg\}
	\]
	are bounded by constant envelops and obey
	\[\begin{aligned}
		&\sup_{Q}\log N\left(\epsilon,\mathcal{V}_{3},\Vert\cdot\Vert_{Q,2}\right)\lesssim\log(e/\epsilon), \\
		&\sup_{Q}\log N\left(\epsilon,\mathcal{V}_{4}^{t},\Vert\cdot\Vert_{Q,2}\right)\lesssim\log(e/\epsilon). \\
	\end{aligned}\]
	Let $\vartheta'(d;\sigma)=\partial_{d}\vartheta(d;\sigma)$. According to Lemma 2.6.15 in van der Vaart and Wellner (1996), the class of functions
	\[
	\mathcal{V}_{5}=\Big\{\vartheta(D;\sigma):u\in\mathcal{U}\Big\}
	\]
	and
	\[
	\mathcal{V}_{6}=\Big\{\vartheta'(D;\sigma):u\in\mathcal{U}\Big\}
	\]
	viewed as maps from the sample space $\mathcal{W}$ to the real line, are bounded by constant envelops and have finite VC dimensions. Again by Theorem 2.6.7 in van der Vaart and Wellner (1996), we can deduce that
	\[\begin{aligned}
		&\sup_{Q}\log N\left(\epsilon,\mathcal{V}_{5},\Vert\cdot\Vert_{Q,2}\right)\lesssim\log(e/\epsilon),\\
		&\sup_{Q}\log N\left(\epsilon,\mathcal{V}_{6},\Vert\cdot\Vert_{Q,2}\right)\lesssim\log(e/\epsilon).\\
	\end{aligned}\]
	 By simple calculation, we have
	\[
	\partial_{D}\displaystyle\int_{Q_{Y}(\tau_{1})}^{Q_{Y}(\tau_{2})}F_{Y}(y|D,X)dy=E\left[\int_{Q_{Y}(\tau_{1})}^{Q_{Y}(\tau_{2})}1\{Y\leq{y}\}dy\cdot\frac{\partial_Df_Y(Y|D,X)}{f_Y(Y|D,X)}\bigg|D,X\right].
	\]
	According to Lemma 2.6.18 in van der Vaart and Wellner (1996), for some positive constant $C$, the class of functions
	\[
	\mathcal{V}_{7}=\left\{\vartheta(D;\sigma)\cdot\int_{Q_{Y}(\tau_{1})}^{Q_{Y}(\tau_{2})}1\{Y\leq{y}\}dy\cdot\frac{\partial_Df_Y(Y|D,X)}{f_Y(Y|D,X)}:u\in\mathcal{U}\right\}\subset\mathcal{V}_{5}\cdot\mathcal{V}_{1}\cdot\left\{\frac{\partial_Df_Y(Y|D,X)}{f_Y(Y|D,X)}\right\}
	\]
	is bounded by measurable envelop  $T_{7}\geq{C}\left|\frac{\partial_Df_Y(Y|D,X)}{f_Y(Y|D,X)}\right|$ with $\Vert{T}_{7}\Vert_{P,2}<{\infty}$, and obey
	\[
	\sup_{Q}\log N\left(\epsilon\Vert{T_{7}}\Vert_{Q,2},\mathcal{V}_{7},\Vert\cdot\Vert_{Q,2}\right)\lesssim\log(e/\epsilon). 
	\]
	Again, as per Lemma L.2 of Belloni et al. (2017), the class of functions
	\[
	\mathcal{V}_{8}=\left\{\vartheta(D;\sigma)\cdot\partial_{D}\displaystyle\int_{Q_{Y}(\tau_{1})}^{Q_{Y}(\tau_{2})}F_{Y}(y|D,X)dy:u\in\mathcal{U}\right\}
	\]
	is bounded by a measurable envelop $T_{8}\geq{E}\left[C\left|\frac{\partial_Df_Y(Y|D,X)}{f_Y(Y|D,X)}\right|\big|D,X\right]$ with $\Vert{T}_{8}\Vert_{P,2}<{\infty}$, and obey
	\[
	\sup_{Q}\log N\left(\epsilon\Vert{T}_{8}\Vert_{Q,2},\mathcal{V}_{8},\Vert\cdot\Vert_{Q,2}\right)\lesssim\log(e/\epsilon).
	\]
	Recall that 
	\[
	L(D,X;\sigma)=\frac{\partial_{D}\Big(\vartheta(D;\sigma)f(D,X)\Big)}{f(D,X)}=\vartheta'(D;\sigma)+\vartheta(D;\sigma)\frac{\partial_{D} f(D,X)}{f(D,X)}.
	\]
	Again, by Lemma L.2 of Belloni et al. (2017), the class of functions
	\[
	\mathcal{V}_{9}=\left\{L(D,X;\sigma)\int_{Q_Y(\tau_1)}^{Q_Y(\tau_2)}\Big(F_Y(y|D,X)-1\{Y\leq{y}\}\Big)dy:u\in\mathcal{U}\right\}
	\]
	satisfying
	\[
	\mathcal{V}_{9}\subset\left(\mathcal{V}_{6}+\mathcal{V}_{5}\cdot\left\{\frac{\partial_{D} f(D,X)}{f(D,X)}\right\}\right)\cdot\Big(\mathcal{V}_{3}-\mathcal{V}_{1}\Big)
	\]
	is bounded by a measurable envelop $T_{9}\geq{C}\left|\frac{\partial_{D} f(D,X)}{f(D,X)}\right|$ with $\Vert{T}_{9}\Vert_{P,2}<{\infty}$, and obey
	\[
	\sup_{Q}\log N\left(\epsilon\Vert{T}_{9}\Vert_{Q,2},\mathcal{V}_{9},\Vert\cdot\Vert_{Q,2}\right)\lesssim\log(e/\epsilon).
	\]
	For $t\in\{0,1\}$, we have
	\[
	\partial_{D}F_{Y}\Big(Q_{Y}(\tau_{t})|D,X\Big)=E\left[1\Big\{Y\leq{Q_{Y}(\tau_{t})}\Big\}\frac{\partial_{D}f_{Y}(Y|D,X)}{f_{Y}(Y|D,X)}\Big|D,X\right].
	\]
	Applying similar arguments as the proceeding one, we can conclude that the class of functions
	\[
	\mathcal{V}_{10}^{t}=\Big\{\partial_{D}F_{Y}\Big(Q_{Y}(\tau_{t})|D,X\Big):u\in\mathcal{U}\Big\}
	\]
	is bounded by a measurable envelop $T_{10}^{t}\geq{E}\left[\left|\frac{\partial_Df_Y(Y|D,X)}{f_Y(Y|D,X)}\right|\big|D,X\right]$ with $\Vert{T}_{10}^{t}\Vert_{P,2}<{\infty}$ and obeys
	\[
	\sup_{Q}\log N\left(\epsilon\Vert{T_{10}^{t}}\Vert_{Q,2},\mathcal{V}_{10}^{t},\Vert\cdot\Vert_{Q,2}\right)\lesssim\log(e/\epsilon).
	\]
	Thus, the class of functions
	\[
	\mathcal{V}_{11}^{t}=\left\{\vartheta(D;\sigma)\cdot\partial_{D}F_{Y}\Big(Q_{Y}(\tau_{t})|D,X\Big):u\in\mathcal{U}\right\}\subset\mathcal{V}_{5}\cdot\mathcal{V}_{10}^{t}
	\]
	is bounded by a measurable envelop $T_{11}^{t}\geq{C}{T}_{10}^{t}$ with $\Vert{T}_{11}^{t}\Vert_{P,2}<{\infty}$ and obeys
	\[
	\sup_{Q}\log N\left(\epsilon\Vert{T_{11}^{t}}\Vert_{Q,2},\mathcal{V}_{11}^{t},\Vert\cdot\Vert_{Q,2}\right)\lesssim\log(e/\epsilon).
	\]
	As per Lemma L.2 of Belloni et al. (2017), for $t\in\{1,2\}$, the class of functions
	\[
	\mathcal{V}_{12}^{t}=\left\{E\left[\vartheta(D;\sigma)\cdot\partial_{D}F_{Y}\Big(Q_{Y}(\tau_{t})|D,X\Big)\right]:u\in\mathcal{U}\right\}
	\]
	is bounded by a constant envelop and obeys
	\[
	\sup_{Q}\log N\left(\epsilon,\mathcal{V}_{12}^{t},\Vert\cdot\Vert_{Q,2}\right)\lesssim\log(e/\epsilon).
	\]
	According to Lemma 2.6.15 in van der Vaart and Wellner (1996), for $t\in\{0,1\}$, the class of functions
	\[
	\mathcal{V}_{13}^{t}=\left\{f_{Y}\Big(Q_{Y}(\tau_{t})\Big):u\in\mathcal{U}\right\}
	\]
	is bounded by constant envelops and has finite VC dimensions. Again, as per Theorem 2.6.7 in van der Vaart and Wellner (1996), we can deduce that
	\[
	\sup_{Q}\log N\left(\epsilon,\mathcal{V}_{13}^{t},\Vert\cdot\Vert_{Q,2}\right)\lesssim\log(e/\epsilon). 
	\]
	
	Second, we turn to analyze the term with other nuisance functions $\widetilde{\eta}$. For $t\in\{1,2\}$, the uniform covering entropy of the function set $\mathcal{Q}_{1}^{t}$ is trivially bounded by $\log(e/\epsilon)$. The class of functions $\mathcal{F}_{1}$ has a constant envelop and is a subset of 
	\[
	\widetilde{\mathcal{F}}_{1}=\left\{\begin{array}{c}
		(d,x,u)\mapsto\displaystyle\int1\left\{\widetilde{Q}_{Y}(\tau_{1})<y<\widetilde{Q}_{Y}(\tau_{2})\right\}\Lambda\bigg(b(d,x)^{\prime}\widetilde{\beta}(y)\bigg)dy: \\
		\left\Vert\widetilde{\beta}(y)\right\Vert_{0}\leq{Cs_{\beta}}, \widetilde{Q}_{Y}(\tau_{1})\in\mathcal{Q}_{1}^{1}, \widetilde{Q}_{Y}(\tau_{2})\in\mathcal{Q}_{1}^{2}
	\end{array} \right\}.
	\]
	Define the following classes of functions
	\[
	\widetilde{\mathcal{F}}_{1a}=\left\{\begin{array}{c}
		u\mapsto1\left\{\widetilde{Q}_{Y}(\tau_{1})<y<\widetilde{Q}_{Y}(\tau_{2})\right\}: \widetilde{Q}_{Y}(\tau_{1})\in\mathcal{Q}_{1}^{1}, \widetilde{Q}_{Y}(\tau_{2})\in\mathcal{Q}_{1}^{2}
	\end{array} \right\}
	\]
	and
	\[
	\widetilde{\mathcal{F}}_{1b}=\left\{\begin{array}{c}
		(d,x)\mapsto\Lambda\bigg(b(d,x)^{\prime}\widetilde{\beta}(y)\bigg): 
		\left\Vert\widetilde{\beta}(y)\right\Vert_{0}\leq{Cs_{\beta}}
	\end{array} \right\}.
	\]
	The uniform covering entropy of the function set $\widetilde{\mathcal{F}}_{1a}$ is trivially bounded by $\log(e/\epsilon)$. Notice that the fixed monotone transformation $\Lambda$ preserves the VC-subgraph property (e.g., Lemma 2.6.18 in van der Vaart and Wellner (1996)). The function set $\widetilde{\mathcal{F}}_{1b}$  is the unions of at most $\tbinom{p_{b}}{Cs_{\beta}}$ VC-subgraph classes of functions with VC indices bounded by $Cs_{\beta}$. Thus, 
	\[
	\sup_{Q}\log N\left(\epsilon,\widetilde{\mathcal{F}}_{1b},\Vert\cdot\Vert_{Q,2}\right)\lesssim{s_{\beta}}\log{p_{b}}+s_{\beta}\log(e/\epsilon).
	\]
	Combining Lemma L.1 of Belloni et al. (2017) and Lemma A.2 in Ghosal et al. (2000), we can deduce that
	\[
	\sup_{Q}\log N\left(\epsilon,\mathcal{F}_{1},\Vert\cdot\Vert_{Q,2}\right)\lesssim\sup_{Q}\log N\left(\epsilon,\widetilde{\mathcal{F}}_{1},\Vert\cdot\Vert_{Q,2}\right)\lesssim{s_{\beta}}\log{p_{b}}+s_{\beta}\log(e/\epsilon).
	\]
	Similarly, the classes of functions $\mathcal{F}_{2}$ and $\mathcal{L}$ are bounded by measurable envelops
	\[\begin{aligned}
		&\widetilde{T}_{2}\geq{C}\sup_{u\in\mathcal{U}}\left|\partial_{D}\displaystyle\int_{Q_{Y}(\tau_{1})}^{Q_{Y}(\tau_{2})}F_{Y}(y|D,X)dy\right|, \\
		&\widetilde{T}_{\mathcal{L}}\geq{C}\sup_{u\in\mathcal{U}}\left|L(D,X;\sigma)\right|, \\
	\end{aligned}\]
	with $\left\Vert\widetilde{T}_{2}\right\Vert_{P,2}<{\infty}$ and $\left\Vert\widetilde{T}_{\mathcal{L}}\right\Vert_{P,2}<{\infty}$. Applying similar arguments as the proceeding one, we can conclude that
	\[\begin{aligned}
		&\sup_{Q}\log N\left(\epsilon\left\Vert\widetilde{T}_{2}\right\Vert_{Q,2},\mathcal{F}_{2},\Vert\cdot\Vert_{Q,2}\right)\lesssim{s_{\beta}}\log{p_{b}}+s_{\beta}\log(e/\epsilon), \\
		&\sup_{Q}\log N\left(\epsilon\left\Vert\widetilde{T}_{\mathcal{L}}\right\Vert_{Q,2},\mathcal{L},\Vert\cdot\Vert_{Q,2}\right)\lesssim{s_{\gamma}}\log{p_{h}}+s_{\gamma}\log(e/\epsilon). \\
	\end{aligned}\]
	Applying similar arguments as the discussion of $\mathcal{F}_{1}$, for $t\in\{1,2\}$, the uniform covering entropy of the function set
	\[
	\mathcal{F}_{4}=\left\{\begin{array}{c}
		(y,u)\mapsto\displaystyle\int1\left\{\widetilde{Q}_{Y}(\tau_{1})<t<\widetilde{Q}_{Y}(\tau_{2})\right\}1\Big\{y<t\Big\}dt: 
		\widetilde{Q}_{Y}(\tau_{1})\in\mathcal{Q}_{1}^{1}, \widetilde{Q}_{Y}(\tau_{2})\in\mathcal{Q}_{1}^{2}
	\end{array} \right\}
	\]
	and
	\[
	\mathcal{F}_{5}^{t}=\left\{\begin{array}{c}
		(y,u)\mapsto1\left\{y<\widetilde{Q}_{Y}(\tau_{t})\right\}: 
		\widetilde{Q}_{Y}(\tau_{t})\in\mathcal{Q}_{1}^{t}
	\end{array} \right\}
	\]
	are bounded by $\log(e/\epsilon)$.
	
	We now analyze the first term $\mathbb{G}_{n}\bigg[\bar{\psi}_{1}\left(W,\widetilde{\eta};u\right)-\bar{\psi}_{1}(W,\eta;u)\bigg]$. Notice that the set of functions
	\[
	\mathcal{J}_{1}=\Big\{\begin{array}{c}
		\bar{\psi}_{1}\left(W,\widetilde{\eta};u\right)-\bar{\psi}_{1}(W,\eta;u): 
		u\in\mathcal{U}, \widetilde{\eta}\in\mathcal{R}
	\end{array} \Big\}
	\]
	is a Lipschitz transform of function sets $\mathcal{V}_{4}^{t}$, $\mathcal{V}_{8}$, $\mathcal{V}_{9}$,  $\mathcal{F}_{1}$, $\mathcal{F}_{2}$,  $\mathcal{F}_{4}$, $\mathcal{L}$, for $t\in\{1,2\}$, with a measurable envelop $T_{\mathcal{J}_{1}}$ satisfying $\left\Vert T_{\mathcal{J}_{1}}\right\Vert_{P,2}<\infty$\footnote{We can construct such envelop by applying the similar way before Theorem 2.10.20 of van der Vaart and Wellner (1996).}. Thus, 
	\[
	\sup_{Q}\log N\left(\epsilon\left\Vert{T}_{\mathcal{J}_{1}}\right\Vert_{Q,2},\mathcal{J}_{1},\Vert\cdot\Vert_{Q,2}\right)\lesssim{(s_{\beta}\vee s_{\gamma})}\log\left(p_{b}\vee{p}_{h}\right)+(s_{\beta}\vee s_{\gamma})\log(e/\epsilon).
	\]
	Notice that
	\[
	\sup_{f\in\mathcal{J}_{1}}\Vert{f}\Vert_{P,2}\lesssim\sup_{\widetilde{\eta}\in\mathcal{R}}\Vert{\widetilde{\eta}-\eta}\Vert_{P,2}\lesssim\delta_{n}n^{-1/4}.
	\]
	Applying Lemma C.1 in Belloni et al. (2017) with $\nu_{n}=\left\Vert T_{\mathcal{J}_{1}}\right\Vert_{P,2}\delta_{n}n^{-1/4}$ and the envelop $T_{\mathcal{J}_{1}}$, with probability no less than $1-\Delta_{n}$,
	\[\begin{aligned}
		&\sup_{\widetilde{\eta}\in\mathcal{R}}\left|\mathbb{G}_{n}\bigg[\bar{\psi}_{1}\left(W,\widetilde{\eta};u\right)-\bar{\psi}_{1}(W,\eta;u)\bigg]\right|
		=\sup_{f\in\mathcal{J}_{1}}\left|\mathbb{G}_{n}f\right| \\
		\lesssim&\sqrt{(s_{\beta}\vee s_{\gamma})\nu_{n}^{2}\log(p_{b}\vee{p}_{h}\vee{n}\vee\nu_{n}^{-1})}+\sqrt{(s_{\beta}\vee s_{\gamma})^{2}\log^{2}(p_{b}\vee{p}_{h}\vee{n}\vee\nu_{n}^{-1})/n} \\
		\lesssim&\sqrt{(s_{\beta}\vee s_{\gamma})\delta_{n}^{2}n^{-1/2}\log(p_{b}\vee{p}_{h}\vee{n})}+\sqrt{(s_{\beta}\vee s_{\gamma})^{2}\log^{2}(p_{b}\vee{p}_{h}\vee{n})/n} \\
		=&o_{p}(1), \\
	\end{aligned}\]
	according to the fact that $\log(1/\delta_{n})\lesssim\log(n)$ by the assumption on $\delta_{n}$ and $(s_{\beta}\vee s_{\gamma})^{2}\log^{2}((p_{b}\vee{p}_{h}\vee{n})/n=o(1)$ by Assumption 4.3(i). 
	
	Next we consider the second term $\mathbb{G}_{n}\bigg[\bar{\psi}_{2}^{t}\left(W,\widetilde{\eta};u\right)-\bar{\psi}_{2}^{t}(W,\eta;u)\bigg]$. Define 
	\[
	\zeta^{t}\left(\widetilde{\eta};u\right)=\frac{(-1)^{t}}{\tau_{2}-\tau_{1}}\cdot\frac{\mathbb{E}_{n}\left[\vartheta(D_{i};\sigma)\cdot\partial_D \widetilde{F}_Y\Big(\widetilde{Q}_{Y}(\tau_{t})|D_{i},X_{i}\Big)\right]}{\widetilde{f}_{Y}\Big(\widetilde{Q}_{Y}(\tau_{t})\Big)}
	\]
	and
	\[
	\zeta^{t}\left({\eta};u\right)=\frac{(-1)^{t}}{\tau_{2}-\tau_{1}}\cdot\frac{E\left[\vartheta(D;\sigma)\cdot\partial_D{F}_Y\Big({Q}_{Y}(\tau_{t})|D,X\Big)\right]}{f_{Y}\Big(Q_{Y}(\tau_{t})\Big)}.
	\]
	Then uniformly over $\widetilde{\eta}\in\mathcal{R}$ and $t\in\{1,2\}$, we can deduce that
	\[\begin{aligned}
		&\mathbb{G}_{n}\bigg[\bar{\psi}_{2}^{t}\left(W,\widetilde{\eta};u\right)-\bar{\psi}_{2}^{t}(W,\eta;u)\bigg]\\
		=&\mathbb{G}_{n}\bigg[\zeta^{t}\left(\widetilde{\eta};u\right)\Big(1\left\{Y<\widetilde{Q}_{Y}(\tau_{t})\right\}-\tau_{t}\Big)-\zeta^{t}\left({\eta};u\right)\Big(1\left\{Y<Q_{Y}(\tau_{t})\right\}-\tau_{t}\Big)\bigg] \\
		=&\Big(\zeta^{t}\left(\widetilde{\eta};u\right)-\zeta^{t}\left({\eta};u\right)\Big)\mathbb{G}_{n}1\left\{Y<Q_{Y}(\tau_{t})\right\}+\zeta^{t}\left({\eta};u\right)\mathbb{G}_{n}\bigg[1\left\{Y<\widetilde{Q}_{Y}(\tau_{t})\right\}-1\left\{Y<Q_{Y}(\tau_{t})\right\}\bigg] \\
		&+o_{p}(1). \\
	\end{aligned}\]
	Applying Lemma C.1 in Belloni et al. (2017) again, it is straightforward to show that for $t\in\{1,2\}$,
	\[
	\mathbb{G}_{n}1\left\{Y<Q_{Y}(\tau_{t})\right\}=O_{p}(1),
	\]
	and
	\[
	\sup_{\widetilde{\eta}\in\mathcal{R},t\in\{1,2\}}\left|\mathbb{G}_{n}\bigg[1\left\{Y<\widetilde{Q}_{Y}(\tau_{t})\right\}-1\left\{Y<Q_{Y}(\tau_{t})\right\}\bigg]\right|=o_{p}(1).
	\]
	Thus,
	\[\begin{aligned}
		&\sup_{\widetilde{\eta}\in\mathcal{R},t\in\{1,2\}}\left|\mathbb{G}_{n}\bigg[\bar{\psi}_{2}^{t}\left(W,\widetilde{\eta};u\right)-\bar{\psi}_{2}^{t}(W,\eta;u)\bigg]\right|\\
		\leq&\sup_{\widetilde{\eta}\in\mathcal{R},t\in\{1,2\}}\left|\Big(\zeta^{t}\left(\widetilde{\eta};u\right)-\zeta^{t}\left({\eta};u\right)\Big)\mathbb{G}_{n}1\left\{Y<Q_{Y}(\tau_{t})\right\}\right|\\
		&+\sup_{\widetilde{\eta}\in\mathcal{R},t\in\{1,2\}}\left|\zeta^{t}\left({\eta};u\right)\mathbb{G}_{n}\bigg[1\left\{Y<\widetilde{Q}(\tau_{t})\right\}-1\left\{Y<Q_{Y}(\tau_{t})\right\}\bigg]\right|+o_{p}(1) \\
		\leq&\sup_{\widetilde{\eta}\in\mathcal{R}}\left\Vert\widetilde{\eta}-\eta\right\Vert_{P,\infty}\cdot{O}_{p}(1)+O(1)\cdot{o}_{p}(1)+o_{p}(1)={o}_{p}(1). \\
	\end{aligned}\]
	Combining the results above, we can conclude that
	\[
	\left|\dagger_{4.1.2}\right|\lesssim {o}_{p}(1)+o_{p}(1)={o}_{p}(1).
	\]
	Thus, with probability no less than $1-\Delta_{n}$,
	\[
	P\bigg(\left|\dagger_{4.1.2}\right|\lesssim\delta_{n}\bigg)\geq1-\Delta_{n}.
	\]

	\noindent\textbf{Step 2. (Uniform Donskerness)} Here we claim that Assumptions 4.1-4.3 imply that the set of functions $\{\psi\left(W,\theta,{\eta};u\right)\}_{u\in\mathcal{U}}$ is $P$-Donsker, namely
	\[
	Z_{n}(u)\leadsto Z(u) \quad in \quad \mathbb{D}=\ell^{\infty}\left(\mathcal{U}\right),
	\]
	where $Z(u)=\mathbb{G}\psi\left(W,\theta,{\eta};u\right)$. 
	
	We apply Theorem B.1 in Belloni et al. (2017) to verify this claim. Recall that by definition $\bar{\psi}\left(W,{\eta};u\right)=\psi\left(W,\theta,{\eta};u\right)+\theta$, we have $\mathbb{G}_{n}\psi\left(W,\theta,{\eta};u\right)=\mathbb{G}_{n}\bar{\psi}\left(W,{\eta};u\right)$. Define the class of functions
	\[
	\mathcal{Z}=\Big\{\bar{\psi}\left(W,{\eta};u\right):u\in\mathcal{U}\Big\}.
	\]
	Notice that $\mathcal{Z}$ is formed as a uniform Lipschitz transform of the function sets $\mathcal{V}_{2}^{t}$, $\mathcal{V}_{4}^{t}$, $\mathcal{V}_{8}$, $\mathcal{V}_{9}$, $\mathcal{V}_{12}^{t}$ and $\mathcal{V}_{13}^{t}$ for $t\in\{1,2\}$. These function sets are uniformly bounded classes that have uniform covering entropy bounded by $\log(e/\epsilon)$ up to a multiplicative constant. Let $T_{\mathcal{Z}}$ be a measurable envelop of $\mathcal{Z}$ with $\left\Vert T_{\mathcal{Z}}\right\Vert_{P,2}<{\infty}$.
	According to Theorem 2.10.20 of van der Vaart and Wellner (1996), the class of functions $\mathcal{Z}$ obeys
	\[
	\sup_{Q}\log N\left(\epsilon\left\Vert{T_{\mathcal{Z}}}\right\Vert_{Q,2},\mathcal{Z},\Vert\cdot\Vert_{Q,2}\right)\lesssim\log(e/\epsilon).
	\]
	Since
	\[
	\lim_{\delta\to{0}}\int_{0}^{\delta}\sqrt{\log(e/\epsilon)} d\epsilon\to{0},
	\]
	the entropy condition (B.2) in Theorem B.1 of Belloni et al. (2017) holds.
	
	The first condition in (B.1) is trivially satisfied. We demonstrate the second condition in (B.1). Consider a sequence of positive constants $\epsilon$ approaching zero, and it suffice to verify that
	\[
	\lim_{\epsilon\to{0}^{+}}\sup_{d_{\mathcal{U}}\left(u,\widetilde{u}\right)\leq{\epsilon}}\left\Vert\psi\left(W,\theta,{\eta};u\right)-\psi\left(W,\theta,{\eta};\widetilde{u}\right)\right\Vert_{P,2}=0.
	\]
	Notice that
	\[\begin{aligned}
		&\left\Vert\psi\left(W,\theta,{\eta};u\right)-\psi\left(W,\theta,{\eta};\widetilde{u}\right)\right\Vert_{P,2}\\
		\lesssim&\underbrace{\left|{\frac{1}{\tau_{2}-\tau_{1}}}-\frac{1}{\widetilde{\tau}_{2}-\widetilde{\tau}_{1}}\right|+\sum_{t=1}^{2}\left|\tau_{t}-\widetilde{\tau}_{t}\right|}_{\dagger_{4.1.4}}
		+\sum_{t=1}^{2}\underbrace{\left\Vert1\bigg\{{Y}\leq{Q_{Y}(\tau_{t})}\bigg\}-1\bigg\{{Y}\leq{Q_{Y}\left(\widetilde{\tau}_{t}\right)}\bigg\}\right\Vert_{P,2}}_{\dagger_{4.1.5}} \\
		+&\underbrace{\left\Vert{\displaystyle\int_{Q_{Y}(\tau_{1})}^{Q_{Y}(\tau_{2})}1\{Y<y\}dy-\displaystyle\int_{Q_{Y}\left(\widetilde{\tau}_{1}\right)}^{Q_{Y}\left(\widetilde{\tau}_{2}\right)}1\{Y<y\}dy}\right\Vert_{P,2}}_{\dagger_{4.1.6}}\\
		+&\underbrace{\left\Vert{\displaystyle\int_{Q_{Y}(\tau_{1})}^{Q_{Y}(\tau_{2})}F_{Y}(y|D,X)dy-\displaystyle\int_{Q_{Y}\left(\widetilde{\tau}_{1}\right)}^{Q_{Y}\left(\widetilde{\tau}_{2}\right)}F_{Y}(y|D,X)dy}\right\Vert_{P,2}}_{\dagger_{4.1.7}} \\
		+&\underbrace{\left\Vert{\displaystyle\int_{Q_{Y}(\tau_{1})}^{Q_{Y}(\tau_{2})}\partial_{D}F_{Y}(y|D,X)dy-\displaystyle\int_{Q_{Y}\left(\widetilde{\tau}_{1}\right)}^{Q_{Y}\left(\widetilde{\tau}_{2}\right)}\partial_{D}F_{Y}(y|D,X)dy}\right\Vert_{P,2}}_{\dagger_{4.1.8}} \\
		+&\underbrace{\left\Vert{\vartheta(D;{\sigma})-\vartheta(D;\widetilde{\sigma})}\right\Vert_{P,2}+\left\Vert{\partial_{D}\vartheta(D;{\sigma})-\partial_{D}\vartheta(D;\widetilde{\sigma})}\right\Vert_{P,2}}_{\dagger_{4.1.9}}+\sum_{t=1}^{2}\underbrace{\left|f_{Y}\Big(Q_{Y}(\tau_{t})\Big)-f_{Y}\Big(Q_{Y}\left(\widetilde{\tau}_{t}\right)\Big)\right|}_{\dagger_{4.1.10}} \\
		+&\sum_{t=1}^{2}\underbrace{\left|E\left[\vartheta(D;\sigma)\partial_{D}F_{Y}\Big(Q_{Y}(\tau_{t})|D,X\Big)\right]-E\left[\vartheta(D;\widetilde{\sigma})\partial_{D}F_{Y}\Big(Q_{Y}\left(\widetilde{\tau}_{t}\right)|D,X\Big)\right]\right|}_{\dagger_{4.1.11}}. \\
	\end{aligned}\]
	$\dagger_{4.1.4}$ trivially converges to $0$ as $d_{\mathcal{U}}\left(u,\widetilde{u}\right)\to{0}$. Under Assumption 4.2 and 4.3(iii), $\dagger_{4.1.5}$, $\dagger_{4.1.6}$, $\dagger_{4.1.9}$ and $\dagger_{4.1.10}$ converge to $0$ as $d_{\mathcal{U}}\left(u,\widetilde{u}\right)\to{0}$. Note that
	\[
	\displaystyle\int_{Q_{Y}(\tau_{1})}^{Q_{Y}(\tau_{2})}F_{Y}(y|D,X)dy=E\left[\displaystyle\int_{Q_{Y}(\tau_{1})}^{Q_{Y}(\tau_{2})}1\{Y\leq{y}\}dy\bigg|D,X\right]
	\]
	By the contradiction property of the conditional expectation,
	\[
	\dagger_{4.1.7}\leq\dagger_{4.1.6}\to{0}
	\]
	as $d_{\mathcal{U}}\left(u,\widetilde{u}\right)\to{0}$. For $\dagger_{4.1.8}$,
	\[\begin{aligned}
		\dagger_{4.1.8}=&\left\Vert{\displaystyle\int_{Q_{Y}(\tau_{1})}^{Q_{Y}(\tau_{2})}\partial_{D}F_{Y}(y|D,X)dy-\displaystyle\int_{Q_{Y}\left(\widetilde{\tau}_{1}\right)}^{Q_{Y}\left(\widetilde{\tau}_{2}\right)}\partial_{D}F_{Y}(y|D,X)dy}\right\Vert_{P,2} \\
		\leq&\left\Vert\displaystyle\int_{Q_{Y}\left(\widetilde{\tau}_{2}\right)}^{Q_{Y}(\tau_{2})}\partial_{D}F_{Y}(y|D,X)dy\right\Vert_{P,2}+\left\Vert\displaystyle\int_{Q_{Y}\left(\widetilde{\tau}_{1}\right)}^{Q_{Y}(\tau_{1})}\partial_{D}F_{Y}(y|D,X)dy\right\Vert_{P,2} \\
		\leq&\Big\{\left|Q_{Y}(\tau_{2})-Q_{Y}\left(\widetilde{\tau}_{2}\right)\right|+\left|Q_{Y}(\tau_{1})-Q_{Y}\left(\widetilde{\tau}_{1}\right)\right|\Big\}\sup_{y\in\mathcal{H}}\left\Vert\partial_{D}F_{Y}(y|D,X)\right\Vert_{P,2}
		\to{0} \\
	\end{aligned}\]
	as $d_{\mathcal{U}}\left(u,\widetilde{u}\right)\to{0}$ by Assumption 4.3(iii). Finally,
	\[\begin{aligned}
		\dagger_{4.1.11}\lesssim&{E}\left|\vartheta(D;{\sigma})-\vartheta(D;\widetilde{\sigma})\right|+E\left|\Big(1\{Y\leq{Q_{Y}(\tau_{t})}\}-1\{Y\leq{Q_{Y}\left(\widetilde{\tau}_{t}\right)}\}\Big)\cdot\frac{\partial_Df_Y(Y|D,X)}{f_Y(Y|D,X)}\right|\to{0} \\
	\end{aligned}\]
	 as $d_{\mathcal{U}}\left(u,\widetilde{u}\right)\to{0}$. Thus, the proof is completed.
	$\blacksquare$
	\\
	
\noindent\textbf{Proof of Theorem 4.2.} 		The joint density of the observed variables $W=(Y,D,X)$ can be written as
\[
f(y,d,x)=f_Y(y|d,x)f_X(x|d)f_{D}(d).
\]
Consider a regular parametric submodel indexed by $\epsilon$ with $\epsilon_0$ corresponding to the true model: $f(y,d,x;\epsilon_0)\equiv f(y,d,x)$. The density of $f(y,d,x;\epsilon)$ can be written as
\[
	f(y,d,x;\epsilon)=f_Y(y|d,x;\epsilon)f_X(x|d;\epsilon)f_{D}(d;\epsilon).
\]
We will assume that all terms of the previous equation admit an interchange of the order of integration and differentiation, which will hold under sufficient condition given by Theorem 1.3.2 of Amemiya (1985) such that
\[
\int\frac{\partial f(y,d,x;\epsilon)}{\partial\epsilon}dxdddy=\frac{\partial}{\partial\epsilon}\underbrace{\int f(y,d,x;\epsilon)dxdddy}_1=0. \eqno{(A.1)}
\]
The corresponding score of $f(y,d,x;\epsilon)$ is
\[
s(y,d,x;\epsilon)=\frac{\partial \log f(y,d,x;\epsilon)}{\partial\epsilon}=\check{f}_Y(y|d,x;\epsilon)+\check{f}_X(x|d;\epsilon)+\check{f}_{D}(d;\epsilon),
\]
where $\check{f}$ defines a derivative of the log, that is, 
\[
\check{f}_Y(y|d,x;\epsilon)\equiv\frac{\partial \log f_Y(y|d,x;\epsilon)}{\partial\epsilon}, \quad  \check{f}_X(x|d;\epsilon)\equiv\frac{\partial \log f_X(x|d;\epsilon)}{\partial\epsilon} \quad\text{and}\quad \check{f}_{D}(d;\epsilon)\equiv\frac{\partial \log f_{D}(d;\epsilon)}{\partial\epsilon}.
\]
Notice that the expectation of the score is zero if $\epsilon$ is evaluated at the true value $\epsilon_{0}$.

According to Proposition 2.1, we have
\[
\theta\left(u\right)=-\frac{\displaystyle\int\vartheta\left(d;\sigma\right)
\left(\int1\{Q_Y(\tau_1)<y<Q_Y(\tau_2)\}1\{t\leq y\}\partial_df_Y(t|d,x)dtdy\right)f_{X}(x|d)f_{D}(d)dxdd}{\tau_2-\tau_1}.
\]
Therefore, the parameter $\theta(u;\epsilon)$ induced by the submodel $f(y,d,x;\epsilon)$ satisfies
\[
\theta\left(u\right)=-\frac{\displaystyle\int\vartheta\left(d;\sigma\right)
	\left(\int_{Q_{Y}(\tau_{1};\epsilon)}^{Q_{Y}(\tau_{2};\epsilon)}\partial_{d}F_Y(y|d,x;\epsilon)dy\right)f_{X}(x|d;\epsilon)f_{D}(d;\epsilon)dxdd}{\tau_2-\tau_1}.
\]
The tangent space of the model is the set of functions that are mean zero and satisfy the additive structure of the score:
\begin{eqnarray}
	\Im=\Big\{s_y(y|d,x)+s_x(x|d)+s_d(d)\Big\}\notag
\end{eqnarray}
for any functions $s_y$, $s_x$ and $s_d$ satisfying the mean zero property
\[
E[s_y(Y|D,X)|D,X]=E[s_x(X|D)|D]=Es_d(D)=0.
\]
Then the semiparametric variance bound of $\theta\left(u\right)$ is the variance of the projection on $\Im$ of a function $\Gamma(W;u)$(with $E\Gamma(\cdot;u)=0$ and $E\left\|\Gamma^2(\cdot;u)\right\|<\infty$ for any $u\in\mathcal{U}$) that satisfies for all regular parametric submodels
\[
\frac{\partial\theta(u;\epsilon)}{\partial\epsilon}\bigg|_{\epsilon=\epsilon_0}=E\Big[\Gamma(W;u)\cdot s(W;\epsilon_0)\Big].
\]
If $\Gamma(W;u)$ itself already lies in the tangent space, the variance bound is given by $E\Gamma^2(W;u)$ for any $u\in\mathcal{U}$.

We first calculate $\dfrac{\partial\theta(u;\epsilon)}{\partial\epsilon}\big|_{\epsilon=\epsilon_0}$.
\[\begin{aligned}
	\frac{\partial\theta(u;\epsilon)}{\partial\epsilon}\bigg|_{\epsilon=\epsilon_0}
	=&\frac{-1}{\tau_2-\tau_1}\int\vartheta\left(d;\sigma\right)\frac{\partial}{\partial\epsilon}\bigg(\int_{Q_Y(\tau_1;\epsilon)}^{Q_Y(\tau_2)}\partial_dF_Y(y|d,x)dy\bigg)\bigg|_{\epsilon=\epsilon_0}f_{X}(x|d)f_D(d)dxdd\\
	&-\frac{1}{\tau_2-\tau_1}\int\vartheta\left(d;\sigma\right)\frac{\partial}{\partial\epsilon}\bigg(\int_{Q_Y(\tau_1)}^{Q_Y(\tau_2;\epsilon)}\partial_dF_Y(y|d,x)dy\bigg)\bigg|_{\epsilon=\epsilon_0}f_{X}(x|d)f_D(d)dxdd\\
	&-\frac{1}{\tau_2-\tau_1}\int\vartheta\left(d;\sigma\right)\left[\int_{Q_{Y}(\tau_{1})}^{Q_{Y}(\tau_{2})}\partial_{d}\left(\frac{\partial F_Y(y|d,x;\epsilon)}{\partial{\epsilon}}\bigg|_{\epsilon=\epsilon_0}\right)dy\right]f_{X}(x|d)f_D(d)dxdd \\
	&-\frac{1}{\tau_2-\tau_1}\int\vartheta\left(d;\sigma\right)\left(\int_{Q_Y(\tau_1)}^{Q_Y(\tau_2)}\partial_dF_Y(y|d,x)dy\right)\frac{\partial f_X(x|d;\epsilon)}{\partial\epsilon}\bigg|_{\epsilon=\epsilon_0}f_D(d)dxdd\\
	&-\frac{1}{\tau_2-\tau_1}\int\vartheta\left(d;\sigma\right)\left(\int_{Q_Y(\tau_1)}^{Q_Y(\tau_2)}\partial_dF_Y(y|d,x)dy\right)f_X(x|d)\frac{\partial f_D(d;\epsilon)}{\partial\epsilon}\bigg|_{\epsilon=\epsilon_0}dxdd. 
\end{aligned}\eqno{(A.2)}\]
By definition,
\[
\int_{-\infty}^{Q_Y(\tau_1;\epsilon)}f_Y(y;\epsilon)dy=\tau_1.
\]
After taking derivative with respect to $\epsilon$ and evaluating at $\epsilon=\epsilon_0$, we have
\[
\frac{\partial Q_Y(\tau_1;\epsilon)}{\partial\epsilon}\bigg|_{\epsilon=\epsilon_0}f_Y\left(Q_Y(\tau_1)\right)+\int_{-\infty}^{Q_Y(\tau_1)}\frac{\partial f_Y(y;\epsilon)}{\partial\epsilon}\bigg|_{\epsilon=\epsilon_0}dy=0.
\]
Thus
\[
\frac{\partial Q_Y(\tau_1;\epsilon)}{\partial\epsilon}\bigg|_{\epsilon=\epsilon_0}=-\frac{\displaystyle\int_{-\infty}^{Q_Y(\tau_1)}\dfrac{\partial f_Y(y;\epsilon)}{\partial\epsilon}\big|_{\epsilon=\epsilon_0}dy}{f_Y\left(Q_Y(\tau_1)\right)}.
\]
It follows that
\[
\begin{aligned}
	&-\frac{1}{\tau_2-\tau_1}\displaystyle\int\vartheta\left(d;\sigma\right)\frac{\partial}{\partial\epsilon}\bigg(\int_{Q_Y(\tau_1;\epsilon)}^{Q_Y(\tau_2)}\partial_dF_Y(y|d,x)dy\bigg)\bigg|_{\epsilon=\epsilon_0}f_{X}(x|d)f_D(d)dxdd\\
	&=\frac{1}{\tau_2-\tau_1}\displaystyle\int\vartheta\left(d;\sigma\right)\frac{\partial Q_Y(\tau_1;\epsilon)}{\partial\epsilon}\bigg|_{\epsilon=\epsilon_0}\partial_dF_Y\left(Q_Y(\tau_1)|d,x\right)f_{X}(x|d)f_D(d)dxdd\\
	&=\frac{-1}{\tau_2-\tau_1}\displaystyle\int\vartheta\left(d;\sigma\right)\frac{\displaystyle\int_{-\infty}^{Q_Y(\tau_1)}\dfrac{\partial f_Y(y;\epsilon)}{\partial\epsilon}\big|_{\epsilon=\epsilon_0}dy}{f_Y\left(Q_Y(\tau_1)\right)}\partial_dF_Y\left(Q_Y(\tau_1)|d,x\right)f_{X}(x|d)f_D(d)dxdd\\
	&=\frac{-1}{\tau_2-\tau_1}\frac{\displaystyle\int_{-\infty}^{Q_Y(\tau_1)}\dfrac{\partial f_Y(y;\epsilon)}{\partial\epsilon}\big|_{\epsilon=\epsilon_0}dy}{f_Y\left(Q_Y(\tau_1)\right)}E\Big[\vartheta\left(D;\sigma\right)\partial_DF_Y\left(Q_Y(\tau_1)|D,X\right)\Big],
\end{aligned}\eqno{(A.3)}
\]
Similarly,
\[
\begin{aligned}
	&-\frac{1}{\tau_2-\tau_1}\displaystyle\int\vartheta\left(d;\sigma\right)\frac{\partial}{\partial\epsilon}\bigg(\int_{Q_Y(\tau_1)}^{Q_Y(\tau_2;\epsilon)}\partial_dF_Y(y|d,x)dy\bigg)\bigg|_{\epsilon=\epsilon_0}f_{X}(x|d)f_D(d)dxdd\\
	=&\frac{1}{\tau_2-\tau_1}\frac{\displaystyle\int_{-\infty}^{Q_Y(\tau_2)}\dfrac{\partial f_Y(y;\epsilon)}{\partial\epsilon}\big|_{\epsilon=\epsilon_0}dy}{f_Y\left(Q_Y(\tau_2)\right)}E\Big[\vartheta\left(D;\sigma\right)\partial_DF_Y\left(Q_Y(\tau_2)|D,X\right)\Big].
\end{aligned}\eqno{(A.4)}
\]
Plugging equations (A.3) and (A.4) into equation (A.2) gives
\[\begin{aligned}
	\frac{\partial\theta(u;\epsilon)}{\partial\epsilon}\bigg|_{\epsilon=\epsilon_0}
	=&\frac{-1}{\tau_2-\tau_1}\frac{\displaystyle\int_{-\infty}^{Q_Y(\tau_1)}\dfrac{\partial f_Y(y;\epsilon)}{\partial\epsilon}\big|_{\epsilon=\epsilon_0}dy}{f_Y\left(Q_Y(\tau_1)\right)}E\Big[\vartheta\left(D;\sigma\right)\partial_DF_Y\left(Q_Y(\tau_1)|D,X\right)\Big]\\
	&+\frac{1}{\tau_2-\tau_1}\frac{\displaystyle\int_{-\infty}^{Q_Y(\tau_2)}\dfrac{\partial f_Y(y;\epsilon)}{\partial\epsilon}\big|_{\epsilon=\epsilon_0}dy}{f_Y\left(Q_Y(\tau_2)\right)}E\Big[\vartheta\left(D;\sigma\right)\partial_DF_Y\left(Q_Y(\tau_2)|D,X\right)\Big]\\
	&-\frac{1}{\tau_2-\tau_1}\int\vartheta\left(d;\sigma\right)\left[\int_{Q_{Y}(\tau_{1})}^{Q_{Y}(\tau_{2})}\partial_{d}\left(\frac{\partial F_Y(y|d,x;\epsilon)}{\partial{\epsilon}}\bigg|_{\epsilon=\epsilon_0}\right)dy\right]f_{X}(x|d)f_D(d)dxdd \\
	&-\frac{1}{\tau_2-\tau_1}\int\vartheta\left(d;\sigma\right)\left(\int_{Q_Y(\tau_1)}^{Q_Y(\tau_2)}\partial_dF_Y(y|d,x)dy\right)\frac{\partial f_X(x|d;\epsilon)}{\partial\epsilon}\bigg|_{\epsilon=\epsilon_0}f_D(d)dxdd\\
	&-\frac{1}{\tau_2-\tau_1}\int\vartheta\left(d;\sigma\right)\left(\int_{Q_Y(\tau_1)}^{Q_Y(\tau_2)}\partial_dF_Y(y|d,x)dy\right)f_X(x|d)\frac{\partial f_D(d;\epsilon)}{\partial\epsilon}\bigg|_{\epsilon=\epsilon_0}dxdd. 
\end{aligned}\]

To justify our theorem, it suffice to show that (i)
\[
\dfrac{\partial\theta(u;\epsilon)}{\partial\epsilon}\bigg|_{\epsilon=\epsilon_0}=E\Big[\psi\left(W,\theta\left(u\right),\eta;u\right)\cdot s(W;\epsilon_0)\Big].
\]
and (ii) $\psi\left(W,\theta\left(u\right),\eta;u\right)$ lies in the tangent space $\Im$ for any $u\in\mathcal{U}$.

The second argument is easily verified and omitted here. Recall that the Neyman-orthogonal score is
\[\begin{aligned}
	\psi\left(W,\theta,\eta;u\right)
	=&\frac{-1}{\tau_2-\tau_1}\vartheta\left(D;\sigma\right)\int_{Q_Y(\tau_1)}^{Q_Y(\tau_2)}\partial_DF_Y(y|D,X)dy-\theta\\
	&-\frac{1}{\tau_2-\tau_1}\frac{\partial_D\left(\vartheta\left(D;\sigma\right)f(D,X)\right)}{f(D,X)}\int_{Q_Y(\tau_1)}^{Q_Y(\tau_2)}\Big(F_Y(y|D,X)-1\{Y\leq y\}\Big)dy\\
	&-\frac{1}{\tau_2-\tau_1}\frac{1\{Y\leq Q_Y(\tau_1)\}-\tau_1}{f_Y\left(Q_Y(\tau_1)\right)}E\Big[\vartheta\left(D;\sigma\right)\partial_DF_Y\left(Q_Y(\tau_1)|D,X\right)\Big]\\
	&+\frac{1}{\tau_2-\tau_1}\frac{1\{Y\leq Q_Y(\tau_2)\}-\tau_2}{f_Y\left(Q_Y(\tau_2)\right)}E\Big[\vartheta\left(D;\sigma\right)\partial_DF_Y\left(Q_Y(\tau_2)|D,X\right)\Big]. \\
\end{aligned}\]
For (i), substituting the representation of $\psi\left(W,\theta\left(u\right),\eta;u\right)$ into $E\Big[\psi\left(W,\theta\left(u\right),\eta;u\right)\cdot s(W;\epsilon_0)\Big]$ yields
\begin{eqnarray*}
	E\Big[\psi\left(W,\theta\left(u\right),\eta;u\right)\cdot s(W;\epsilon_0)\Big]
	&=&\frac{-1}{\tau_2-\tau_1}\bigg(\dagger_{4.2.1}+\dagger_{4.2.2}\bigg)\\
	&&-\frac{1}{\tau_2-\tau_1}\frac{E\Big[\vartheta\left(D;\sigma\right)\partial_DF_Y\left(Q_Y(\tau_1)|D,X\right)\Big]}{f_Y\left(Q_Y(\tau_1)\right)}\cdot\dagger_{4.2.3}\\
	&&+\frac{1}{\tau_2-\tau_1}\frac{E\Big[\vartheta\left(D;\sigma\right)\partial_DF_Y\left(Q_Y(\tau_2)|D,X\right)\Big]}{f_Y\left(Q_Y(\tau_2)\right)}\cdot\dagger_{4.2.4},
\end{eqnarray*}
where
\[
\begin{aligned}
	\dagger_{4.2.1}=E\bigg[\Big(\vartheta\left(D;\sigma\right)\int_{Q_Y(\tau_1)}^{Q_Y(\tau_2)}\partial_DF_Y(y|D,X&)dy+(\tau_2-\tau_1)\theta\left(u\right)\Big)\\
	&\times\bigg(\check{f}_Y(Y|D,X;\epsilon_{0})+\check{f}_X(X|D;\epsilon_{0})+\check{f}_D(D;\epsilon_{0})\bigg)\bigg],
\end{aligned}
\]
\[
\begin{aligned}
	\dagger_{4.2.2}=E\bigg[\frac{\partial_D\left(\vartheta\left(D;\sigma\right)f(D,X)\right)}{f(D,X)}\int_{Q_Y(\tau_1)}^{Q_Y(\tau_2)}&\bigg(F_Y\left(y\big|D,X\right)-1\big\{Y<y\big\}\bigg)dy\\
	&\times\bigg(\check{f}_Y(Y|D,X;\epsilon_{0})+\check{f}_X(X|D;\epsilon_{0})+\check{f}_D(D;\epsilon_{0})\bigg)\bigg],
\end{aligned}
\]
\[
\dagger_{4.2.3}=E\left[\Big(1\{Y\leq Q_Y(\tau_1)\}-\tau_1\Big)\bigg(\check{f}_Y(Y|D,X;\epsilon_{0})+\check{f}_X(X|D;\epsilon_{0})+\check{f}_D(D;\epsilon_{0})\bigg)\right],
\]
and
\[
\dagger_{4.2.4}=E\left[\Big(1\{Y\leq Q_Y(\tau_2)\}-\tau_2\Big)\bigg(\check{f}_Y(Y|D,X;\epsilon_{0})+\check{f}_X(X|D;\epsilon_{0})+\check{f}_D(D;\epsilon_{0})\bigg)\right].
\]
By $(A.1)$, we have
\[
E\left[\check{f}_Y(Y|D,X;\epsilon_{0})|D,X\right]=E\left[\check{f}_X(X|D;\epsilon_{0})|X\right]=E\left[\check{f}_D(D;\epsilon_{0})\right]=0.
\]
For $\dagger_{4.2.1}$, we get
\[\begin{aligned}
	\dagger_{4.2.1}=&E\left[\vartheta\left(D;\sigma\right)\int_{Q_Y(\tau_1)}^{Q_Y(\tau_2)}\partial_DF_Y(y|D,X)dy\cdot\bigg(\check{f}_Y(Y|D,X;\epsilon_{0})+\check{f}_X(X|D;\epsilon_{0})+\check{f}_D(D;\epsilon_{0})\bigg)\right] \\
	=&E\left[\vartheta\left(D;\sigma\right)\int_{Q_Y(\tau_1)}^{Q_Y(\tau_2)}\partial_DF_Y(y|D,X)dy\cdot\bigg(\check{f}_X(X|D;\epsilon_{0})+\check{f}_D(D;\epsilon_{0})\bigg)\right] \\
	=&\int\vartheta\left(d;\sigma\right)\int_{Q_Y(\tau_1)}^{Q_Y(\tau_2)}\partial_dF_Y(y|d,x)dy\left(\frac{\partial f_X(x|d;\epsilon)}{\partial\epsilon}\bigg|_{\epsilon=\epsilon_0}f_{D}(d)+f_{X}(x|d)\frac{\partial f_{D}(d;\epsilon)}{\partial\epsilon}\bigg|_{\epsilon=\epsilon_0}\right)dxdd.
\end{aligned}\]
For $\dagger_{4.2.2}$, we get
\[\begin{aligned}
	\dagger_{4.2.2}
	=&E\bigg[\frac{\partial_D\left(\vartheta\left(D;\sigma\right)f(D,X)\right)}{f(D,X)}\int_{Q_Y(\tau_1)}^{Q_Y(\tau_2)}\bigg(F_Y\left(y\big|D,X\right)-1\big\{Y<y\big\}\bigg)dy\cdot\check{f}_Y(Y|D,X;\epsilon_{0})\bigg]\\
	=&-E\bigg[\frac{\partial_D\left(\vartheta\left(D;\sigma\right)f(D,X)\right)}{f(D,X)}\int_{Q_Y(\tau_1)}^{Q_Y(\tau_2)}1\big\{Y<y\big\}dy\cdot\check{f}_Y(Y|D,X;\epsilon_{0})\bigg]\\
	=&-\int\frac{\partial_d\left(\vartheta \left(d;\sigma\right)f(d,x)\right)}{f(d,x)}\left[\int_{Q_Y(\tau_1)}^{Q_Y(\tau_2)}1\{t<y\}dy\cdot\frac{\partial}{\partial\epsilon}f_{Y}(t|d,x;\epsilon)\bigg|_{\epsilon=\epsilon_{0}}\right]f(d,x)dtdxdd\\
	=&-\int\partial_d\left(\vartheta \left(d;\sigma\right)f(d,x)\right)\left[\int_{Q_{Y}(\tau_{1})}^{Q_{Y}(\tau_{2})}\frac{\partial F_Y(y|d,x;\epsilon)}{\partial{\epsilon}}\bigg|_{\epsilon=\epsilon_0}dy\right]dxdd\\
	=&\int\vartheta\left(d;\sigma\right)f(d,x)\left[\int_{Q_{Y}(\tau_{1})}^{Q_{Y}(\tau_{2})}\partial_{d}\left(\frac{\partial F_Y(y|d,x;\epsilon)}{\partial{\epsilon}}\bigg|_{\epsilon=\epsilon_0}\right)dy\right]dxdd.
\end{aligned}\]

For $\dagger_{4.2.3}$, we get
\[\begin{aligned}
	\dagger_{4.2.3}=&E\left[\left(1\{Y\leq Q_Y(\tau_1)\}-\tau_1\right)\bigg(\check{f}_Y(Y|D,X;\epsilon_{0})+\check{f}_X(X|D;\epsilon_{0})+\check{f}_D(D;\epsilon_{0})\bigg)\right]\\
	=&E\left[1\{Y\leq Q_Y(\tau_1)\}\bigg(\check{f}_Y(Y|D,X;\epsilon_{0})+\check{f}_X(X|D;\epsilon_{0})+\check{f}_D(D;\epsilon_{0})\bigg)\right]\\
	=&E\left[1\{Y\leq Q_Y(\tau_1)\}\bigg(\check{f}_D(D|X,Y;\epsilon_{0})+\check{f}_X(X|Y;\epsilon_{0})+\check{f}_{Y}(Y;\epsilon_{0})\bigg)\right] \\
	=&E\left[1\{Y\leq Q_Y(\tau_1)\}\cdot\check{f}_{Y}(Y;\epsilon_{0})\right] \\
	=&\int_{-\infty}^{Q_Y(\tau_1)}\frac{\partial f_Y(y;\epsilon)}{\partial\epsilon}\bigg|_{\epsilon=\epsilon_0}dy. \\
\end{aligned}\]

Similarly, we get
\[
\dagger_{4.2.4}=\int_{-\infty}^{Q_Y(\tau_2)}\frac{\partial f_Y(y;\epsilon)}{\partial\epsilon}\bigg|_{\epsilon=\epsilon_0}dy, \\
\]
which verifies condition (i). Thus, the proof is completed. $\blacksquare$\\

	\noindent\textbf{Proof of Theorem 4.3.} Let $P^{*}=P\times{P}_{\xi}$. Then the operator $E_{P^{*}}$ then denotes the expectation with respect to $P^{*}=P\times{P_{\xi}}$ and $\mathbb{G}_{n}$ denotes the corresponding empirical process, that is
\[
\mathbb{G}_{n}B(\xi,W)=\frac{1}{\sqrt{n}}\sum_{i=1}^{n}\bigg[B(\xi_{i},W_{i})-E_{P^{*}}B(\xi,W)\bigg].
\]

Recall that we define the bootstrap draw as
\[
\widehat{Z}_{n}^{*}(u)=\sqrt{n}\left(\widehat{\theta}^{*}(u)-\widehat{\theta}(u)\right)=\frac{1}{\sqrt{n}}\sum_{i=1}^{n}\xi_{i}\psi\bigg(W_{i},\widehat{\theta},\widehat{\eta};u\bigg)=\mathbb{G}_{n}\xi\psi\bigg(W,\widehat{\theta},\widehat{\eta};u\bigg),
\]
since $E_{P^{*}}\left[\xi\psi\bigg(W,\widehat{\theta},\widehat{\eta};u\bigg)\right]=0$ because $\xi$ is independent of $W$ and has zero mean. The proof also consists of two steps.

\noindent\textbf{Step 1.} In this step, we establish that
\[
\widehat{Z}_{n}^{*}(u)={Z}_{n}^{*}(u)+o_{p^{*}}(1), \quad in \quad \mathbb{D}=\ell^{\infty}\left(\mathcal{U}\right),
\]
where $Z_{n}^{*}(u)=\mathbb{G}_{n}\xi\psi(W,\theta,\eta;u)$.

Recall that by definition, $\mathbb{G}_{n}\xi\psi\left(W,\theta,{\eta};u\right)=\mathbb{G}_{n}\xi\bar{\psi}\left(W,{\eta};u\right)$. We then have the representation
\[\begin{aligned}
	&\sqrt{n}\left(\widehat{\theta}^{*}(u)-\widehat{\theta}(u)\right)\\
	=&\mathbb{G}_{n}\xi\psi(W,\theta,\eta;u)-\left(\widehat{\theta}(u)-{\theta}(u)\right)\mathbb{G}_{n}\xi+\underbrace{\mathbb{G}_{n}\bigg[\xi\bar{\psi}\left(W,\widehat{\eta};u\right)-\xi\bar{\psi}\left(W,\eta;u\right)\bigg]}_{\dagger_{4.3.1}}. \\
\end{aligned}\]
According to Theorem 4.1, $\left(\widehat{\theta}(u)-{\theta}(u)\right)\mathbb{G}_{n}\xi=O_{p^{*}}\left(n^{-1/2}\right)=o_{p^{*}}(1)$. For $\dagger_{4.3.1}$, applying similar arguments, we have 
\[
\left|\dagger_{4.3.1}\right|\leq\sup_{f\in\xi\mathcal{J}_{1}}\left|\mathbb{G}_{n}f\right|+\sup_{\widetilde{\eta}\in\mathcal{R},t\in\{1,2\}}\left|\mathbb{G}_{n}\bigg[\xi\bar{\psi}_{2}^{t}\left(W,\widetilde{\eta};u\right)-\xi\bar{\psi}_{2}^{t}(W,\eta;u)\bigg]\right|.
\]
By Lemma L.1 of Belloni et al. (2017), multiplication of class $\mathcal{J}_{1}$ by $\xi$ does not change the entropy bound modulo an absolute constant. Thus, we have 
\[
\sup_{Q}\log N\left(\epsilon\left\Vert{T}_{\xi\mathcal{J}_{1}}\right\Vert_{Q,2},\xi\mathcal{J}_{1},\Vert\cdot\Vert_{Q,2}\right)\lesssim{(s_{\beta}\vee s_{\gamma})}\log\left(p_{b}\vee{p_{h}}\right)+(s_{\beta}\vee s_{\gamma})\log(e/\epsilon).
\]
Similarly, as per Lemma C.1 of Belloni et al. (2017), we can show that
\[
\sup_{f\in\xi\mathcal{J}_{1}}\left|\mathbb{G}_{n}f\right|=o_{p^*}(1).
\]
Then uniformly over $\widetilde{\eta}\in\mathcal{R}$ and $t\in\{1,2\}$, we can also deduce that
\[\begin{aligned}
	&\mathbb{G}_{n}\bigg[\xi\bar{\psi}_{2}^{t}\left(W,\widetilde{\eta};u\right)-\xi\bar{\psi}_{2}^{t}(W,\eta;u)\bigg]\\ 
	=&\Big(\zeta^{t}\left(\widetilde{\eta};u\right)-\zeta^{t}\left({\eta};u\right)\Big)\mathbb{G}_{n}\xi\Big(1\left\{Y<Q_{Y}(\tau_{t})\right\}-\tau_{t}\Big)\\
	&+\zeta^{t}\left({\eta};u\right)\mathbb{G}_{n}\bigg[\xi1\left\{Y<\widetilde{Q}_{Y}(\tau_{t})\right\}-\xi1\left\{Y<Q_{Y}(\tau_{t})\right\}\bigg] 
	+o_{p^*}(1). \\
\end{aligned}\]
Similarly, we have
\[
\mathbb{G}_{n}\xi\Big(1\left\{Y<Q_{Y}(\tau_{t})\right\}-\tau_{t}\Big)=O_{p^*}(1),
\]
for $t\in\{1,2\}$, and
\[
\sup_{\widetilde{\eta}\in\mathcal{R},t\in\{1,2\}}\left|\mathbb{G}_{n}\bigg[\xi1\left\{Y<\widetilde{Q}_{Y}(\tau_{t})\right\}-\xi1\left\{Y<Q_{Y}(\tau_{t})\right\}\bigg]\right|=o_{p^*}(1).
\]
Thus,
\[
\sup_{\widetilde{\eta}\in\mathcal{R},t\in\{1,2\}}\left|\mathbb{G}_{n}\bigg[\xi\bar{\psi}_{2}^{t}\left(W,\widetilde{\eta};u\right)-\xi\bar{\psi}_{2}^{t}(W,\eta;u)\bigg]\right|={o}_{p^*}(1).
\]
Combining the results above, we can conclude that
\[
\left|\dagger_{4.3.1}\right|\lesssim {o}_{p^*}(1)+o_{p^*}(1)={o}_{p^*}(1).
\]

\noindent\textbf{Step 2.} Here we claim that
\[
\widehat{Z}_{n}^{*}(u)\leadsto_{B} Z(u) \quad in \quad \mathbb{D}=\ell^{\infty}\left(\mathcal{U}\right).
\]
Applying Theorem B.2 in Belloni et al. (2017) or equivalently, Theorem 2 in Kosorok (2003), we have ${Z}_{n}^{*}(u)\leadsto_{B} Z(u)$ in $\mathbb{D}=\ell^{\infty}\left(\mathcal{U}\right)$. Then by Lemma 2 in Chiang et al. (2019) and the result in Step 1, we have $\widehat{Z}_{n}^{*}(u)\leadsto_{B} Z(u)$ in $\mathbb{D}=\ell^{\infty}\left(\mathcal{U}\right)$. $\blacksquare$
\\

	\noindent\textbf{Proof of Theorem 4.4.} According to Theorem 6.2 in Belloni et al. (2017), we have
\[
\sup_{y\in\mathcal{H}}\left\Vert b(D,X)^{\prime}\left(\widehat{\beta}(y)-\beta(y)\right)\right\Vert_{\mathbb{P}_{n},2}=O_{p}\left(\sqrt{\frac{s_{\beta}\log(p_{b}\vee{n})}{n}}\right),
\]
\[
K_{nb}\sup_{y\in\mathcal{H}}\left\Vert \widehat{\beta}(y)-\beta(y)\right\Vert_{1}=O_{p}\left(\sqrt{\frac{K_{nb}^{2}s_{\beta}^{2}\log(p_{b}\vee{n})}{n}}\right)
\]
and
\[
\sup_{y\in\mathcal{H}}\left\|\widehat{\beta}(y)\right\|_{0}\leq{C}s_{\beta}.
\]
Then by Lemma D.2 in Sasaki et al. (2022), we have
\[
\sup_{y\in\mathcal{H}}\left\Vert \partial_{D}\Lambda\left(b(D,X)^{\prime}\widehat{\beta}(y)\right)-\partial_{D}F_{Y}(y|D,X)\right\Vert_{\mathbb{P}_{n},2}=O_{p}\left(\sqrt{\frac{s_{\beta}\log(p_{b}\vee{n})}{n}}\right).
\]
Notice that
\[\begin{aligned}
	&\partial_{D}\Lambda\left(b(D,X)^{\prime}\widehat{\beta}(y)\right)-\partial_{D}F_{Y}(y|D,X)\\
	=&\partial_{D}\left[\Lambda\left(b(D,X)^{\prime}\widehat{\beta}(y)\right)-\Lambda\left(b(D,X)^{\prime}{\beta}(y)\right)\right]-\partial_{D}r_{F}(D,X,y) \\
	=&\partial{\Lambda}\left(b(D,X)^{\prime}\widehat{\beta}(y)\right)\partial_{D}b(D,X)^{\prime}\left(\widehat{\beta}(y)-\beta(y)\right)\\
	&+\left(\partial{\Lambda}\left(b(D,X)^{\prime}\widehat{\beta}(y)\right)-\partial{\Lambda}\left(b(D,X)^{\prime}{\beta}(y)\right)\right)\partial_{D}b(D,X)^{\prime}\beta(y)-\partial_{D}r_{F}(D,X,y), \\
\end{aligned}\]
where $\partial{\Lambda}$ denotes the first derivative of $\Lambda$. By triangle inequality, Assumptions 4.3(iii) and 4.5-4.6, we have that there exists a positive constant $C$ such that
\[\begin{aligned}
	\sup_{y\in\mathcal{H}}\left\Vert \partial_{D}b(D,X)^{\prime}\left(\widehat{\beta}(y)-\beta(y)\right)\right\Vert_{\mathbb{P}_{n},2}\leq&{C}\sup_{y\in\mathcal{H}}\left\Vert b(D,X)^{\prime}\left(\widehat{\beta}(y)-\beta(y)\right)\right\Vert_{\mathbb{P}_{n},2}\\
	&+C\sup_{y\in\mathcal{H}}\left\Vert\partial_{D}r_{F}(D,X,y)\right\Vert_{\mathbb{P}_{n},2} \\
	&+C\sup_{y\in\mathcal{H}}\left\Vert \partial_{D}\Lambda\left(b(D,X)^{\prime}\widehat{\beta}(y)\right)-\partial_{D}F_{Y}(y|D,X)\right\Vert_{\mathbb{P}_{n},2} \\
	=&O_{p}\left(\sqrt{\frac{s_{\beta}\log(p_{b}\vee{n})}{n}}\right).
\end{aligned}\]
Thus, the proof is completed. $\blacksquare$ \\

Before providing the proof of Theorem 4.5, we first introduce some useful notations. In what follows for a vector of $\phi\in\mathbb{R}^{p_{h}}$ and a set of indices $T\subseteq\{1,2,\dots,p_{h}\}$, we denote $\phi_{T}\in\mathbb{R}^{p_{h}}$ as the vector such that $\left(\phi_{T}\right)_{j}=\phi_{j}$ if $j\in{T}$ and $\left(\phi_{T}\right)_{j}=0$ if $j\notin{T}$. For a set $T$, $|T|$ denotes the cardinality of $T$. Moreover, for a given positive constant $\widetilde{C}$, let
\[
\Delta_{\widetilde{C},\sigma}=\Big\{\phi\in\mathbb{R}^{p_{h}}:\left\Vert\phi_{T_{\sigma}^{c}}\right\Vert_{1}\leq{\widetilde{C}}\left\Vert\phi_{T_{\sigma}}\right\Vert_{1}\Big\}.
\]
The analysis relies on $T_{\sigma}=\text{supp}(\gamma(\sigma))$, $s(\sigma)=\Vert\gamma(\sigma)\Vert_{0}\leq{s_{\gamma}}$, and on the restricted eigenvalues
\[
\varphi_{\widetilde{C}}=\inf_{\sigma\in\mathcal{S}}\min_{\phi\in\Delta_{\widetilde{C},\sigma}}\frac{\Big\Vert{h(D,X)'\phi}\Big\Vert_{\mathbb{P}_{n},2}}{\Vert\phi_{T_{\sigma}}\Vert}
\]
and maximum and minimum sparse eigenvalues
\[\begin{aligned}
	&\varphi_{min}(s)=\min_{1\leq\Vert\phi\Vert_{0}\leq{s}}\frac{\Big\Vert{h(D,X)'\phi}\Big\Vert_{\mathbb{P}_{n},2}^{2}}{\Vert\phi\Vert^{2}}, \\
	&\varphi_{max}(s)=\max_{1\leq\Vert\phi\Vert_{0}\leq{s}}\frac{\Big\Vert{h(D,X)'\phi}\Big\Vert_{\mathbb{P}_{n},2}^{2}}{\Vert\phi\Vert^{2}}. \\
\end{aligned}\]
\\

\noindent\textbf{Proof of Theorem 4.5.} The following results hold uniformly over $\sigma\in\mathcal{S}$. So we suppress the claim ``uniformly over $\sigma\in\mathcal{S}$" throughout the proof. We denote $C$ as the generic positive constant which may varies case by case. Let $\widehat{\phi}(\sigma)=\widehat{\gamma}(\sigma)-\gamma(\sigma)$ and $\varepsilon_{nh}=\sqrt{\log(p_{h}\vee{n})/n}$. Recall that
\[
\widehat{M}(\sigma)=\mathbb{E}_{n}{M}_{i}(\sigma), \quad \widehat{G}=\mathbb{E}_{n}G_{i},
\]
where
\[
{M}_{i}(\sigma)=-\partial_{D_i}h(D_i,X_i)\vartheta\left(D_i;\sigma\right), \quad G_{i}=h(D_{i},X_{i})h(D_{i},X_{i})'.
\]
By definition of $\widehat{\gamma}(\sigma)$, 
\[
\widehat{\gamma}(\sigma)=\arg\min_{\gamma}-2\widehat{M}(\sigma)'\gamma+\gamma'\widehat{G}\gamma+2\lambda_{\gamma}\Vert\gamma\Vert_{1},
\]
we have
\[\begin{aligned}
	&\mathbb{E}_{n}\Big[h'(D_{i},X_{i})\widehat{\phi}(\sigma)\Big]^{2}-2\Big(\widehat{M}(\sigma)-\widehat{G}\gamma(\sigma)\Big)'\widehat{\phi}(\sigma) \\
	=&\Big(-2\widehat{M}(\sigma)'\widehat{\gamma}(\sigma)+\widehat{\gamma}(\sigma)'\widehat{G}\widehat{\gamma}(\sigma)\Big)-\Big(-2\widehat{M}(\sigma)'\gamma(\sigma)+\gamma(\sigma)'\widehat{G}\gamma(\sigma)\Big) \\
	\leq&2\lambda_{\gamma}\Big\Vert\gamma(\sigma)\Big\Vert_{1}-2\lambda_{\gamma}\Big\Vert\widehat{\gamma}(\sigma)\Big\Vert_{1} \\
	\leq&2\lambda_{\gamma}\Big\Vert\widehat{\phi}_{T_{\sigma}}(\sigma)\Big\Vert_{1}-2\lambda_{\gamma}\Big\Vert\widehat{\phi}_{T_{\sigma}^{c}}(\sigma)\Big\Vert_{1}, \\
\end{aligned}\]
according to the fact that $\Big\Vert\widehat{\gamma}(\sigma)\Big\Vert_{1}=\Big\Vert\widehat{\gamma}_{T_{\sigma}}(\sigma)\Big\Vert_{1}+\Big\Vert\widehat{\gamma}_{T_{\sigma}^{c}}(\sigma)\Big\Vert_{1}$, $\gamma(\sigma)=\gamma_{T_{\sigma}}(\sigma)$, $\widehat{\gamma}_{T_{\sigma}^{c}}(\sigma)=\widehat{\phi}_{T_{\sigma}^{c}}(\sigma)$ and triangle inequality. Notice that
\[\begin{aligned}
	&\Big(\widehat{M}(\sigma)-\widehat{G}\gamma(\sigma)\Big)'\widehat{\phi}(\sigma)
	=\mathbb{E}_{n}\Big[M_{i}(\sigma)-G_{i}\gamma(\sigma)\Big]'\widehat{\phi}(\sigma) \\
	=&\left(\mathbb{E}_{n}\Big[M_{i}(\sigma)-L(D_{i},X_{i};\sigma)h(D_{i},X_{i})\Big]+\mathbb{E}_{n}\Big[r_{L}(D_{i},X_{i},\sigma)h(D_{i},X_{i})\Big]\right)'\widehat{\phi}(\sigma) \\
	\leq&\left\Vert\mathbb{E}_{n}\Big[M_{i}(\sigma)-L(D_{i},X_{i};\sigma)h(D_{i},X_{i})\Big]\right\Vert_{\infty}\Big\Vert\widehat{\phi}(\sigma)\Big\Vert_{1} +\mathbb{E}_{n}\Big[r_{L}(D_{i},X_{i},\sigma)h(D_{i},X_{i})'\widehat{\phi}(\sigma)\Big] \\
	\leq&\left\Vert\mathbb{E}_{n}\Big[M_{i}(\sigma)-L(D_{i},X_{i};\sigma)h(D_{i},X_{i})\Big]\right\Vert_{\infty}\Big\Vert\widehat{\phi}(\sigma)\Big\Vert_{1} +\|r_{L}(D_{i},X_{i},\sigma)\|_{\mathbb{P}_{n},2}\left\{\mathbb{E}_{n}\Big[h(D_{i},X_{i})'\widehat{\phi}(\sigma)\Big]^{2}\right\}^{1/2}. \\
\end{aligned}\]
By Assumption 4.9(i)-(ii), we have
\[
\|r_{L}(D_{i},X_{i},\sigma)\|_{\mathbb{P}_{n},2}\leq{C\varepsilon_{nh}^{\frac{2\xi}{1+2\xi}}}
\]
with probability approaching to 1. Since $E\Big[M_{i}(\sigma)-L(D_{i},X_{i};\sigma)h(D_{i},X_{i})\Big]=0$. Applying similar arguments as the proof of Lemma J.1 in Belloni et al. (2017), we have
\[
\left\Vert\mathbb{E}_{n}\Big[M_{i}(\sigma)-L(D_{i},X_{i};\sigma)h(D_{i},X_{i})\Big]\right\Vert_{\infty}\leq{C\varepsilon_{nh}}
\]
with probability approaching to 1. Recall that $\lambda_{\gamma}=\kappa_{n}\varepsilon_{nh}$ with $\kappa_{n}\to\infty$. Combining the results above, we obtain that
\[\begin{aligned}
	&\mathbb{E}_{n}\Big[h(D_{i},X_{i})'\widehat{\phi}(\sigma)\Big]^{2} \\
	\leq&2C\varepsilon_{nh}^{\frac{2\xi}{1+2\xi}}\left\{\mathbb{E}_{n}\Big[h(D_{i},X_{i})'\widehat{\phi}(\sigma)\Big]^{2}\right\}^{1/2}+2\lambda_{\gamma}\left(1+C\frac{\varepsilon_{nh}}{\lambda_{\gamma}}\right)\Big\Vert\widehat{\phi}_{T_{\sigma}}(\sigma)\Big\Vert_{1}-2\lambda_{\gamma}\left(1-C\frac{\varepsilon_{nh}}{\lambda_{\gamma}}\right)\Big\Vert\widehat{\phi}_{T_{\sigma}^{c}}(\sigma)\Big\Vert_{1} \\
	\leq&2C\varepsilon_{nh}^{\frac{2\xi}{1+2\xi}}\left\{\mathbb{E}_{n}\Big[h(D_{i},X_{i})'\widehat{\phi}(\sigma)\Big]^{2}\right\}^{1/2}+3\lambda_{\gamma}\Big\Vert\widehat{\phi}_{T_{\sigma}}(\sigma)\Big\Vert_{1}-\lambda_{\gamma}\Big\Vert\widehat{\phi}_{T_{\sigma}^{c}}(\sigma)\Big\Vert_{1}, \\
\end{aligned}\eqno{(A.5)}\]
with probability approaching to 1, where $\varepsilon_{nh}/\lambda_{\gamma}=1/\kappa_{n}\leq{1/(2C)}$. Let $\widetilde{C}=3$. Suppose $\widehat{\phi}(\sigma)\notin\Delta_{\widetilde{C},\sigma}=\Big\{\phi\in\mathbb{R}^{p_{h}}:\left\Vert\phi_{T_{\sigma}^{c}}\right\Vert_{1}\leq{\widetilde{C}}\left\Vert\phi_{T_{\sigma}}\right\Vert_{1}\Big\}$. Then we have that
\[
3\Big\Vert\widehat{\phi}_{T_{\sigma}}(\sigma)\Big\Vert_{1}\leq\Big\Vert\widehat{\phi}_{T_{\sigma}^{c}}(\sigma)\Big\Vert_{1},
\]
which yields that
\[
\left\{\mathbb{E}_{n}\Big[h(D_{i},X_{i})'\widehat{\phi}(\sigma)\Big]^{2}\right\}^{1/2}\leq2C\varepsilon_{nh}^{\frac{2\xi}{1+2\xi}}
\]
with probability approaching to 1. Otherwise, suppose $\widehat{\phi}(\sigma)\in\Delta_{\widetilde{C},\sigma}$. Notice that $\widetilde{C}$ is uniformly bounded in our setting. Thus, under Assumption 4.10, $\varphi_{\widetilde{C}}$ is bounded away from zero with probability approaching to 1 (See proof of Theorem 6.1 in Belloni et al. (2017) or Lemma 4.1 in Bickel et al. (2009)).  Then by the fact that $\Big\Vert\widehat{\phi}_{T_{\sigma}}(\sigma)\Big\Vert_{1}\leq\sqrt{s_{\gamma}}\Big\Vert\widehat{\phi}_{T_{\sigma}}(\sigma)\Big\Vert$, we have
\[\begin{aligned}
	\mathbb{E}_{n}\Big[h(D_{i},X_{i})'\widehat{\phi}(\sigma)\Big]^{2}&\leq2C\varepsilon_{nh}^{\frac{2\xi}{1+2\xi}}\left\{\mathbb{E}_{n}\Big[h(D_{i},X_{i})'\widehat{\phi}(\sigma)\Big]^{2}\right\}^{1/2}+3\lambda_{\gamma}\sqrt{s_{\gamma}}\Big\Vert\widehat{\phi}_{T_{\sigma}}(\sigma)\Big\Vert \\
	&\leq2C\varepsilon_{nh}^{\frac{2\xi}{1+2\xi}}\left\{\mathbb{E}_{n}\Big[h(D_{i},X_{i})'\widehat{\phi}(\sigma)\Big]^{2}\right\}^{1/2}+\frac{3}{\varphi_{\widetilde{C}}}\lambda_{\gamma}\sqrt{s_{\gamma}}\left\{\mathbb{E}_{n}\Big[h(D_{i},X_{i})'\widehat{\phi}(\sigma)\Big]^{2}\right\}^{1/2}, \\
\end{aligned}\]
which implies that
\[
\left\{\mathbb{E}_{n}\Big[h(D_{i},X_{i})'\widehat{\phi}(\sigma)\Big]^{2}\right\}^{1/2}\leq2C\varepsilon_{nh}^{\frac{2\xi}{1+2\xi}}+\frac{3}{\varphi_{\widetilde{C}}}\lambda_{\gamma}\sqrt{s_{\gamma}}
\]
with probability approaching to 1. By definition, we have $\varepsilon_{nh}^{\frac{2\xi}{1+2\xi}}/(\lambda_{\gamma}\sqrt{s_{\gamma}})=O(1/\kappa_{n})=o(1)$. Combining the results above, we can conclude that
\[
\left\{\mathbb{E}_{n}\Big[h(D_{i},X_{i})'\widehat{\phi}(\sigma)\Big]^{2}\right\}^{1/2}\leq\frac{4}{\varphi_{\widetilde{C}}}\lambda_{\gamma}\sqrt{s_{\gamma}} \eqno{(A.6)}
\]
with probability approaching to 1. Recall that $\lambda_{\gamma}\sqrt{s_{\gamma}}=O\left(\kappa_{n}\varepsilon_{nh}^{\frac{2\xi}{1+2\xi}}\right)=O\left(\kappa_{n}\left(\frac{\log(p_{h}\vee{n})}{n}\right)^{\frac{\xi}{1+2\xi}}\right)$. The proof of the first part is completed.

Then we consider the second part. Suppose $\widehat{\phi}(\sigma)\notin\Delta_{2\widetilde{C},\sigma}$, which implies that
\[
6\Big\Vert\widehat{\phi}_{T_{\sigma}}(\sigma)\Big\Vert_{1}\leq\Big\Vert\widehat{\phi}_{T_{\sigma}^{c}}(\sigma)\Big\Vert_{1}.
\]
Equations (A.5)-(A.6) yield
\[
\frac{1}{2}\lambda_{\gamma}\Big\Vert\widehat{\phi}_{T_{\sigma}^{c}}(\sigma)\Big\Vert_{1}\leq2C\varepsilon_{nh}^{\frac{2\xi}{1+2\xi}}\left\{\mathbb{E}_{n}\Big[h(D_{i},X_{i})'\widehat{\phi}(\sigma)\Big]^{2}\right\}^{1/2}-\mathbb{E}_{n}\Big[h(D_{i},X_{i})'\widehat{\phi}(\sigma)\Big]^{2}\leq\frac{8C}{\varphi_{\widetilde{C}}}\kappa_{n}\varepsilon_{nh}^{\frac{4\xi}{1+2\xi}}
\]
with probability approaching to 1. Thus, 
\[
\Big\Vert\widehat{\phi}(\sigma)\Big\Vert_{1}\leq\frac{7}{6}\Big\Vert\widehat{\phi}_{T_{\sigma}^{c}}(\sigma)\Big\Vert_{1}\leq\frac{56C}{3\varphi_{\widetilde{C}}}\frac{\kappa_{n}\varepsilon_{nh}^{\frac{4\xi}{1+2\xi}}}{\lambda_{\gamma}}\leq\frac{56C}{3\varphi_{\widetilde{C}}}\varepsilon_{nh}^{\frac{2\xi-1}{1+2\xi}}
\]
with probability approaching to 1. Otherwise, suppose $\widehat{\phi}(\sigma)\in\Delta_{2\widetilde{C},\sigma}$. Then it is straightforward to show that
\[\begin{aligned}
	\Big\Vert\widehat{\phi}(\sigma)\Big\Vert_{1}&\leq7\Big\Vert\widehat{\phi}_{T_{\sigma}}(\sigma)\Big\Vert_{1}\leq\frac{7}{\varphi_{2\widetilde{C}}}\sqrt{s_{\gamma}}\left\{\mathbb{E}_{n}\Big[h(D_{i},X_{i})'\widehat{\phi}(\sigma)\Big]^{2}\right\}^{1/2} \\
	&\leq\frac{28}{\varphi_{\widetilde{C}}\varphi_{2\widetilde{C}}}\lambda_{\gamma}s_{\gamma}\leq\frac{28}{\varphi_{\widetilde{C}}\varphi_{2\widetilde{C}}}\kappa_{n}\varepsilon_{nh}^{\frac{2\xi-1}{1+2\xi}}
\end{aligned}\]
with probability approaching to 1. Combining the results above, we can conclude that
\[
\Big\Vert\widehat{\phi}(\sigma)\Big\Vert_{1}\leq\frac{28}{\varphi_{\widetilde{C}}\varphi_{2\widetilde{C}}}\kappa_{n}\varepsilon_{nh}^{\frac{2\xi-1}{1+2\xi}}\leq\frac{28}{\varphi_{\widetilde{C}}\varphi_{2\widetilde{C}}}\kappa_{n}\left(\frac{\log(p_{h}\vee{n})}{n}\right)^{\frac{2\xi-1}{2(1+2\xi)}}
\]
with probability approaching to 1, which completes the second part.

Finally,  we prove the uniform sparsity. Define $\widehat{T}_{\sigma}=\text{supp}\left(\widehat{\gamma}(\sigma)\right)$. By the first-order condition, we have
\[
\left\Vert\Big\{\widehat{M}(\sigma)'-\widehat{G}\widehat{\gamma}(\sigma)\Big\}_{\widehat{T}(\sigma)}\right\Vert=\lambda_{\gamma}\Big\Vert\widehat{\gamma}(\sigma)\Big\Vert_{0}^{1/2}.
\]
Thus, by triangle inequality,
\[\begin{aligned}
	\lambda_{\gamma}\Big\Vert\widehat{\gamma}(\sigma)\Big\Vert_{0}^{1/2}=&\left\Vert\Big\{\widehat{M}(\sigma)'-\widehat{G}\widehat{\gamma}(\sigma)\Big\}_{\widehat{T}(\sigma)}\right\Vert \\
	\leq&\left\Vert\Big\{\mathbb{E}_{n}\Big[M_{i}(\sigma)-L(D_{i},X_{i};\sigma)h(D_{i},X_{i})\Big]\Big\}_{\widehat{T}(\sigma)}\right\Vert \\
	&+\left\Vert\Big\{\mathbb{E}_{n}\Big[r_{L}(D_{i},X_{i},\sigma)h(D_{i},X_{i})\Big]\Big\}_{\widehat{T}(\sigma)}\right\Vert+\left\Vert\Big\{\mathbb{E}_{n}h(D_{i},X_{i})h(D_{i},X_{i})'\widehat{\phi}(\sigma)\Big\}_{\widehat{T}(\sigma)}\right\Vert. \\
\end{aligned}\]
Notice that
\[\begin{aligned}
	&\left\Vert\Big\{\mathbb{E}_{n}\Big[M_{i}(\sigma)-L(D_{i},X_{i};\sigma)h(D_{i},X_{i})\Big]\Big\}_{\widehat{T}(\sigma)}\right\Vert \\
	\leq&\left\Vert\Big\{\mathbb{E}_{n}\Big[M_{i}(\sigma)-L(D_{i},X_{i};\sigma)h(D_{i},X_{i})\Big]\Big\}_{\widehat{T}(\sigma)}\right\Vert^{1/2}_{0}\left\Vert\Big\{\mathbb{E}_{n}\Big[M_{i}(\sigma)-L(D_{i},X_{i};\sigma)h(D_{i},X_{i})\Big]\Big\}_{\widehat{T}(\sigma)}\right\Vert_{\infty} \\
	\leq&\Big\Vert\widehat{\gamma}(\sigma)\Big\Vert_{0}^{1/2}\left\Vert\mathbb{E}_{n}\Big[M_{i}(\sigma)-L(D_{i},X_{i};\sigma)h(D_{i},X_{i})\Big]\right\Vert_{\infty}\\
	\leq&C\Big\Vert\widehat{\gamma}(\sigma)\Big\Vert_{0}^{1/2}\varepsilon_{nh}
\end{aligned}\]
with probability approaching to 1. Notice that
\[\begin{aligned}
	&\left\Vert\Big\{\mathbb{E}_{n}\Big[r_{L}(D_{i},X_{i},\sigma)h(D_{i},X_{i})\Big]\Big\}_{\widehat{T}(\sigma)}\right\Vert\\
	\leq&\sup_{\Vert{a}\Vert_{0}\leq\Vert\widehat{\gamma}(\sigma)\Vert_{0},\Vert{a}\Vert=1}\Big|a'\mathbb{E}_{n}\Big[r_{L}(D_{i},X_{i},\sigma)h(D_{i},X_{i})\Big]\Big| \\
	\leq&\left\{\mathbb{E}_{n}\Big[{r}_{L}^{2}(D_{i},X_{i},\sigma)\Big]\right\}^{1/2}\sup_{\Vert{a}\Vert_{0}\leq\Vert\widehat{\gamma}(\sigma)\Vert_{0},\Vert{a}\Vert=1}\left\{\mathbb{E}_{n}\Big[a'h(D_{i},X_{i})h(D_{i},X_{i})'a\Big]\right\}^{1/2} \\
	\leq&C\varepsilon_{nh}^{\frac{2\xi}{1+2\xi}}\sqrt{\varphi_{max}\Big(\Vert\widehat{\gamma}(\sigma)\Vert_{0}\Big)}
\end{aligned}\]
with probability approaching to 1. For the last term, we have
\[\begin{aligned}
	&\left\Vert\Big\{\mathbb{E}_{n}h(D_{i},X_{i})h(D_{i},X_{i})'\widehat{\phi}(\sigma)\Big\}_{\widehat{T}(\sigma)}\right\Vert \\
	\leq&\sup_{\Vert{a}\Vert_{0}\leq\Vert\widehat{\gamma}(\sigma)\Vert_{0},\Vert{a}\Vert=1}\Big|a'\mathbb{E}_{n}h(D_{i},X_{i})h(D_{i},X_{i})'\widehat{\phi}(\sigma)\Big| \\
	\leq&\left\{\mathbb{E}_{n}\Big[h(D_{i},X_{i})'\widehat{\phi}(\sigma)\Big]^{2}\right\}^{1/2}\sup_{\Vert{a}\Vert_{0}\leq\Vert\widehat{\gamma}(\sigma)\Vert_{0},\Vert{a}\Vert=1}\left\{\mathbb{E}_{n}\Big[a'h(D_{i},X_{i})h(D_{i},X_{i})'a\Big]\right\}^{1/2} \\
	\leq&\frac{4}{\varphi_{\widetilde{C}}}\lambda_{\gamma}\sqrt{s_{\gamma}}\sqrt{\varphi_{max}\Big(\Vert\widehat{\gamma}(\sigma)\Vert_{0}\Big)}
\end{aligned}\]
with probability approaching to 1. Combining the results above, we have
\[\begin{aligned}
	\Big\Vert\widehat{\gamma}(\sigma)\Big\Vert_{0}^{1/2}\leq&C\frac{\varepsilon_{nh}}{\lambda_{\gamma}}\Big\Vert\widehat{\gamma}(\sigma)\Big\Vert_{0}^{1/2}+\left(C\frac{\varepsilon_{nh}^{\frac{2\xi}{1+2\xi}}}{\lambda_{\gamma}}+\frac{4}{\varphi_{\widetilde{C}}}\sqrt{s_{\gamma}}\right)\sqrt{\varphi_{max}\Big(\Vert\widehat{\gamma}(\sigma)\Vert_{0}\Big)} \\
	\leq&\frac{1}{2}\Big\Vert\widehat{\gamma}(\sigma)\Big\Vert_{0}^{1/2}+\left(C\frac{\sqrt{s_{\gamma}}}{\kappa_{n}}+\frac{4}{\varphi_{\widetilde{C}}}\sqrt{s_{\gamma}}\right)\sqrt{\varphi_{max}\Big(\Vert\widehat{\gamma}(\sigma)\Vert_{0}\Big)}, \\
\end{aligned}\]
which leads to
\[
\Big\Vert\widehat{\gamma}(\sigma)\Big\Vert_{0}\leq\frac{25}{\varphi_{\widetilde{C}}^{2}}s_{\gamma}\varphi_{max}\Big(\Vert\widehat{\gamma}(\sigma)\Vert_{0}\Big)
\]
with probability approaching to 1, where $C/\kappa_{n}\leq1/\varphi_{\widetilde{C}}$. Let $C^{*}=25/\varphi_{\widetilde{C}}^{2}$. Suppose $\Big\Vert\widehat{\gamma}(\sigma)\Big\Vert_{0}>3C^{*}\overline{c}s_{\gamma}$. Then
\[
s_{\gamma}\varphi_{max}\Big(\Vert\widehat{\gamma}(\sigma)\Vert_{0}\Big)\leq s_{\gamma}\left\lceil\frac{\Vert\widehat{\gamma}(\sigma)\Vert_{0}}{3C^{*}\overline{c}s_{\gamma}}\right\rceil\varphi_{max}\Big(3C^{*}\overline{c}s_{\gamma}\Big),
\]
where the inequality holds by Lemma 3 in Belloni and Chernozhukov (2013). Since $\lceil{k}\rceil\leq{2k}$ for any $k\geq{1}$ and $\varphi_{max}\Big(3C^{*}\overline{c}s_{\gamma}\Big)\leq\overline{c}$ by Assumption 6.6, we have 
\[
\Big\Vert\widehat{\gamma}(\sigma)\Big\Vert_{0}\leq C^{*}s_{\gamma}\left\lceil\frac{\Vert\widehat{\gamma}(\sigma)\Vert_{0}}{3C^{*}\overline{c}s_{\gamma}}\right\rceil\varphi_{max}\Big(3C^{*}\overline{c}s_{\gamma}\Big)\leq\frac{2}{3}\Big\Vert\widehat{\gamma}(\sigma)\Big\Vert_{0},
\]
which leads to contradiction. Therefore, we have
\[
\Big\Vert\widehat{\gamma}(\sigma)\Big\Vert_{0}\leq3C^{*}\overline{c}s_{\gamma}
\]
with probability approaching to 1. Thus, the proof is completed. $\blacksquare$\\
	
	\subsection*{A.4. Proofs for Section 5}
	\noindent\textbf{Proof of Proposition 5.1.} Applying similar arguments as the proof in Proposition 2.1, we have
	\[\begin{aligned}
		\theta^{D}(\tau_1,\tau_2,\Psi)=&E\left[\lim_{\delta\rightarrow 0}\frac{m\left(\Psi_{\delta}^{-1}(R_{D}),X,U\right)-m\left(D,X,U\right)}{\delta}\bigg|Y\in\left(Q_Y(\tau_1),Q_Y(\tau_2)\right)\right] \\
		=&E\left[\frac{\partial}{\partial \delta}\Psi_{\delta}^{-1}\circ{F}_{D}(D)\Big|_{\delta=0}\cdot\partial_D m(D,X,U)\bigg|Y\in\left(Q_Y(\tau_1),Q_Y(\tau_2)\right)\right] \\
		=&\frac{1}{\tau_2-\tau_1}\int_{Q_Y(\tau_1)}^{Q_Y(\tau_2)}E\left[\frac{F_{D}(D)-G_{0}(D)}{f_{D}(D)}\partial_D m(D,X,U)\bigg|Y=y\right]f_Y(y)dy \\
		=&-\frac{1}{\tau_2-\tau_1}E\left[\frac{F_{D}(D)-G_{0}(D)}{f_{D}(D)}\int_{Q_Y(\tau_1)}^{Q_Y(\tau_2)}\partial_DF_Y(y|D,X)dy\right], \\
	\end{aligned}\]
	where the last equality holds by the fact that
	\[
	E\left[\partial_D m(D,X,U)\bigg|Y=y,D=d,X=x\right]=-\frac{\partial_dF_Y(y|d,x)}{f_Y(y|d,x)}.
	\]
	$\blacksquare$
	\\

\noindent\textbf{Proof of Theorem 5.1.} Recall that $\vartheta\Big(F_{D}(D);\Psi\Big)=\Big(F_{D}(D)-G_{0}(D)\Big)\Big/f_{D}(D)$. By a slight abuse of notation, denote nuisance parameters in Theorem 5.1 as $\eta_{1}(\cdot)$, $\eta_{2}(D,X;\cdot)$, $\eta_{3}(D)$, $\eta_{4}(D,X;\cdot)$, $\eta_{5}(\cdot)$ and $\eta_{6}(D;\cdot)$, respectively, namely, 
\[\begin{aligned}
	&\eta^{D}(W;\cdot)\\
	=&\Big(Q_{Y}(\cdot),F_Y(\cdot|D,X),F_{D}(D),L(D,X;\cdot),f_{Y}(\cdot),\alpha(D;\cdot),E\Big[\vartheta\Big(F_{D}(D);\cdot\Big)\partial_DF_Y\left(Q_Y(\cdot)\big|D,X\right)\Big]\Big) \\
	\equiv&\left(\eta_{1}(\cdot),\eta_{2}(D,X;\cdot),\eta_{3}(D),\eta_{4}(D,X;\cdot),\eta_{5}(\cdot),\eta_{6}(D;\cdot),E\Big[\vartheta\Big(\eta_{3}(D);\cdot\Big)\partial_D\eta_2\Big(D,X;\eta_1(\cdot)\Big)\Big]\right),
\end{aligned}\]
where
\[
L(D,X;\Psi)=\frac{\partial_D\left(\vartheta\Big(F_{D}(D);\Psi\Big)f(D,X)\right)}{f(D,X)}.
\]
For any real function $\iota(\cdot)$, $\vartheta_{F_{D}}'\Big(\iota(d);\Psi\Big)$ denotes the Gateaux derivative uniformly in $\iota(\cdot)$. Specifically,
\[
\vartheta_{F_{D}}'\Big(\iota(d);\Psi\Big)=\frac{\iota(d)f_D(d)+G(d)\partial_d\iota(d)-F_D(d)\partial_d\iota(d)}{ f_D^2(d)}.
\]
We can easily check that $\alpha(D;\tau_{1},\tau_{2},\Psi)$ satisfies
\[
E\left[\vartheta_{F_{D}}'\Big(\iota(D);\Psi\Big)\int_{Q_Y(\tau_1)}^{Q_Y(\tau_2)}\partial_DF_Y(y|D,X)dy+\alpha(D;\tau_{1},\tau_{2},\Psi)\iota(D)\right]=0.
\]

We rewrite the score function $\psi^{D}$ in terms of $\eta^{D}$ as
\[\begin{aligned}
	&\psi^{D}\bigg(W,\theta,\eta^{D};\tau_1,\tau_2,\Psi\bigg)\\
	=&\frac{-1}{\tau_2-\tau_1}\vartheta\Big(\eta_{3}(D);\Psi\Big)\int_{\eta_1(\tau_1)}^{\eta_1(\tau_2)}\partial_D\eta_2(D,X;y)dy-\theta\\
	&+\frac{1}{\tau_2-\tau_1}\eta_4(D,X;\Psi)\left(\int_{\eta_1(\tau_1)}^{\eta_1(\tau_2)}\eta_2(D,X;y)dy-\int_{\eta_1(\tau_1)}^{\eta_1(\tau_2)}1\{Y<y\}dy\right)\\
	&+\frac{1}{\tau_2-\tau_1}\int \eta_{6}\left(\widetilde{d};\tau_{1},\tau_{2},\Psi\right)\Big(1\left\{D\le \widetilde{d}\right\}-\eta_{3}\left(\widetilde{d}\right)\Big)d\eta_{3}\left(\widetilde{d}\right)\\
	&-\frac{1}{\tau_2-\tau_1}\frac{E\left[\vartheta\Big(\eta_{3}(D);\Psi\Big)\partial_D\eta_2\Big(D,X;\eta_1(\tau_1)\Big)\right]}{\eta_5\Big(\eta_1(\tau_1)\Big)}\left(1\Big\{Y\leq \eta_1(\tau_1)\Big\}-\tau_1\right)\notag\\
	&+\frac{1}{\tau_2-\tau_1}\frac{E\left[\vartheta\Big(\eta_{3}(D);\Psi\Big)\partial_D\eta_2\Big(D,X;\eta_1(\tau_2)\Big)\right]}{\eta_5\Big(\eta_1(\tau_2)\Big)}\left(1\Big\{Y\leq\eta_1(\tau_2)\Big\}-\tau_2\right).\\
\end{aligned}\]
It is straightforward to show that $E\psi^{D}\bigg(W,\theta^{D}(\tau_1,\tau_2,\Psi),\eta^{D};\tau_1,\tau_2,\Psi\bigg)=0$, and then part (i) is proved.

Applying similar arguments as the proceeding one, part (ii) can be proved by the fact that $E\left[\eta_2(D,X;y)-1\Big\{Y<y\Big\}\Big|D,X\right]=0$, $E\left[1\Big\{Y\leq \eta_1(\tau_j)\Big\}-\tau_j\right]=0$ for $j\in\{1,2\}$, and applying integration by part. 

The proof of part (iii) is similar to which in Proposition 2.2(iii) and is omitted. 

Finally, we turn to prove part (iv). Notice that
\[\begin{aligned}
	\theta^{D}(\tau_1,\tau_2,\Psi)=&\frac{-1}{\tau_2-\tau_1}\int\vartheta\left(\int 1\{t\leq d\}f_D(t)dt;\Psi\right)\\
	&\times\left(\int1\{Q_Y(\tau_1)<y<Q_Y(\tau_2)\}1\{t\leq y\}\partial_df_Y(t|d,x)dtdy\right)f_{X}(x|d)f_{D}(d)dddx. \\
\end{aligned}\]
Therefore, the parameter $\theta(\tau_1,\tau_2,\Psi;\epsilon)$ induced by the submodel $f(y,d,x;\epsilon)$ satisfies
\[\begin{aligned}
	&\theta^{D}(\tau_1,\tau_2,\Psi;\epsilon)\\
	=&\frac{-1}{\tau_2-\tau_1}\int\vartheta\left(\int 1\{t\leq d\}f_D(t;\epsilon)dt;\Psi\right)\\
	&\times\left(\int1\{Q_Y(\tau_1;\epsilon)<y<Q_Y(\tau_2;\epsilon)\}1\{t\leq y\}\partial_df_Y(t|d,x;\epsilon)dtdy\right)f_{X}(x|d;\epsilon)f_{D}(d;\epsilon)dddx \\
	=&\frac{-1}{\tau_2-\tau_1}\displaystyle\int\vartheta\Big(F_{D}(D;\epsilon);\Psi\Big)
	\left(\int_{Q_{Y}(\tau_{1};\epsilon)}^{Q_{Y}(\tau_{2};\epsilon)}\partial_{d}F_Y(y|d,x;\epsilon)dy\right)f_{X}(x|d;\epsilon)f_{D}(d;\epsilon)dxdd. \\
\end{aligned}\]

Recall that the tangent space of the model is the set of functions that are mean zero and satisfy the additive structure of the score:
\begin{eqnarray}
	\Im=\{s_y(y|d,x)+s_x(x|d)+s_d(d)\}\notag
\end{eqnarray}
for any functions $s_y$, $s_x$ and $s_d$ satisfying the mean zero property
\[
E[s_y(Y|D,X)|D,X]=E[s_x(X|D)|D]=Es_d(D)=0.
\]
To justify our theorem, it suffice to show that (i)
\[
\dfrac{\partial\theta^{D}(\tau_1,\tau_2,\Psi;\epsilon)}{\partial\epsilon}\bigg|_{\epsilon=\epsilon_0}=E\Big[\psi^{D}\left(W,\theta^{D}(\tau_1,\tau_2,\Psi),\eta^{D};\tau_1,\tau_2,\Psi\right)\cdot s(W;\epsilon_0)\Big].
\]
and (ii) $\psi^{D}\left(W,\theta^{D}(\tau_1,\tau_2,\Psi),\eta^{D};\tau_1,\tau_2,\Psi\right)$ lies in the tangent space $\Im$.

Part (ii) can be easily verified. For part (i), we can similarly derive that
\[\begin{aligned}
	&\dfrac{\partial\theta^{D}(\tau_1,\tau_2,\Psi;\epsilon)}{\partial\epsilon}\bigg|_{\epsilon=\epsilon_0} \\
		=&-\frac{1}{\tau_2-\tau_1}\int\vartheta_{F_{D}}'\left(\frac{\partial{F_{D}(d;\epsilon)}}{\partial\epsilon}\bigg|_{\epsilon=\epsilon_0};\Psi\right)\left(\int_{Q_{Y}(\tau_{1})}^{Q_{Y}(\tau_{2})}\partial_{d}F_Y(y|d,x)dy\right)f_{X}(x|d)f_{D}(d)dxdd \\
		&-\frac{1}{\tau_2-\tau_1}\frac{\displaystyle\int_{-\infty}^{Q_Y(\tau_1)}\dfrac{\partial f_Y(y;\epsilon)}{\partial\epsilon}\big|_{\epsilon=\epsilon_0}dy}{f_Y\left(Q_Y(\tau_1)\right)}E\Big[\vartheta\Big(F_{D}(D);\Psi\Big)\partial_DF_Y\left(Q_Y(\tau_1)|D,X\right)\Big]\\
	&+\frac{1}{\tau_2-\tau_1}\frac{\displaystyle\int_{-\infty}^{Q_Y(\tau_2)}\dfrac{\partial f_Y(y;\epsilon)}{\partial\epsilon}\big|_{\epsilon=\epsilon_0}dy}{f_Y\left(Q_Y(\tau_2)\right)}E\Big[\vartheta\Big(F_{D}(D);\Psi\Big)\partial_DF_Y\left(Q_Y(\tau_2)|D,X\right)\Big]\\
	&-\frac{1}{\tau_2-\tau_1}\int\vartheta\Big(F_{D}(D);\Psi\Big)\left[\int_{Q_{Y}(\tau_{1})}^{Q_{Y}(\tau_{2})}\partial_{d}\left(\frac{\partial F_Y(y|d,x;\epsilon)}{\partial{\epsilon}}\bigg|_{\epsilon=\epsilon_0}\right)dy\right]f_{X}(x|d)f_D(d)dxdd \\
	&-\frac{1}{\tau_2-\tau_1}\int\vartheta\Big(F_{D}(D);\Psi\Big)\left(\int_{Q_Y(\tau_1)}^{Q_Y(\tau_2)}\partial_dF_Y(y|d,x)dy\right)\frac{\partial f_X(x|d;\epsilon)}{\partial\epsilon}\bigg|_{\epsilon=\epsilon_0}f_D(d)dxdd\\
	&-\frac{1}{\tau_2-\tau_1}\int\vartheta\Big(F_{D}(D);\Psi\Big)\left(\int_{Q_Y(\tau_1)}^{Q_Y(\tau_2)}\partial_dF_Y(y|d,x)dy\right)f_X(x|d)\frac{\partial f_D(d;\epsilon)}{\partial\epsilon}\bigg|_{\epsilon=\epsilon_0}dxdd. \\
\end{aligned}\]
After applying similar arguments as the proof of Theorem 4.2 and the fact that
\[
E\left[\vartheta_{F_{D}}'\left(\frac{\partial{F_{D}(D;\epsilon)}}{\partial\epsilon}\bigg|_{\epsilon=\epsilon_0};\Psi\right)\int_{Q_Y(\tau_1)}^{Q_Y(\tau_2)}\partial_DF_Y(y|D,X)dy+\alpha(D;\tau_{1},\tau_{2},\Psi)\frac{\partial{F_{D}(D;\epsilon)}}{\partial\epsilon}\bigg|_{\epsilon=\epsilon_0}\right]=0,
\]
we can show that
\[
\dfrac{\partial\theta^{D}(\tau_1,\tau_2,\Psi;\epsilon)}{\partial\epsilon}\bigg|_{\epsilon=\epsilon_0}=E\Big[\psi^{D}\left(W,\theta^{D}(\tau_1,\tau_2,\Psi),\eta^{D};\tau_1,\tau_2,\Psi\right)\cdot s(W;\epsilon_0)\Big],
\]
which completes the proof. $\blacksquare$

		\section*{Appendix B: Simulation}
	
	In this section, we study the finite-sample performance of the naive estimator, using the moment condition derived in Proposition 2.1, and the DML estimator, using the orthogonal score derived in Proposition 2.2. We consider the estimation of nine quantities: $\theta(0.1,0.2,\mathcal{G})$, $\theta(0.2,0.3,\mathcal{G})$, ..., $\theta(0.8,0.9,\mathcal{G})$.
	
	The data generating process is defined as
	\begin{eqnarray*}
		&&Y=D+X'\left(c_y\delta_0\right)+DX_1+ U, \\
		&&D=X'(c_d\delta_0)+V,
	\end{eqnarray*}
	where $U$ and $V$ are independently distributed as $N(0,1)$, and $X = (X_1, \dots, X_{p_x})' \sim N(0,\Sigma)$ with $\Sigma_{kj} = (0.5)^{|j-k|}$. The vector $\delta_0$ is of dimension $p_x \times 1$ with elements defined as $\delta_{0,j} = (1/j)^2$ for $j \in {1, 2, \dots, p_x}$.
	$c_d$ and $c_y$ are scalars to control the level of dependence between $X$ and $D$, as well as between $X$ and $Y$.
	We set $c_d=\sqrt{\frac{(\pi^2/3)R_d^2}{(1-R_d^2)\delta_0'\Sigma\delta_0}}$ and $c_y=\sqrt{\frac{R_y^2}{(1-R_y^2)\delta_0'\Sigma\delta_0}}$ and consider nine combinations of $c_{d}$ and $c_{y}$ by choosing $R_d^2\in\{0.2,0.3,0.4\}$ and $R_y^2\in\{0.2,0.3,0.4\}$. Note that smaller $R_d^2$ values reflect higher sparsity levels of the effects of $X$ on $D$, and similar reasoning applies to $R_y^2$. We consider the following policy intervention
	\[
	\mathcal{G}_\delta(D)=(D+3\delta)(1+\delta),
	\]
	which corresponds to a general location-scale transformation. We choose $n=500$, $p_x=30$ throughout the simulation. We construct the basis functions $b(d,x) = h(d,x)$ by including all first-order, second-order, and interaction terms among $(D,X)$. Thus, the dimension of the basis functions is 527, which exceeds the sample size $n$.

	For each design, we calculate the naive and DML estimators.
	We perform 500 iterations to compute the bias ratio, standard errors (Std), mean square error (MSE), and the probability that the 500 estimators lie within the nominal 95\% confidence interval (Cvg). The estimation results are reported in Tables B.1-B.3. Our results indicate that DML estimators outperform naive estimators in all cases examined.

	\begin{table}[H]
		\centering
		\begin{threeparttable}
			\begin{tabular}{llccccccccccc}	\multicolumn{13}{l}{\small{\textbf{Table B.1 }}\small{Sparsity Design for $R_d^2=0.2$}}\\
				\toprule
				&       & \multicolumn{2}{c}{Bias Ratio} &       & \multicolumn{2}{c}{Std} &       & \multicolumn{2}{c}{MSE} &       & \multicolumn{2}{c}{Cvg} \\
				\cmidrule{3-4}\cmidrule{6-7}\cmidrule{9-10}\cmidrule{12-13}    $R_y^2$ & Quantile & Naive & DML  &       & Naive & DML  &       & Naive & DML  &       & Naive & DML \\
				\midrule
				\multirow{8}[1]{*}{0.2} & 0.1-0.2 & .201  & \textbf{.069} &       & .317  & \textbf{.286} &       & .184  & \textbf{.091} &       & .854  & \textbf{.934} \\
				& 0.2-0.3 & .242  & \textbf{.075} &       & .351  & \textbf{.309} &       & .268  & \textbf{.109} &       & .822  & \textbf{.942} \\
				& 0.3-0.4 & .153  & \textbf{.046} &       & .373  & \textbf{.324} &       & .213  & \textbf{.111} &       & .888  & \textbf{.930} \\
				& 0.4-0.5 & .029  & \textbf{.000} &       & .396  & \textbf{.363} &       & .160  & \textbf{.132} &       & .944  & \textbf{.940} \\
				& 0.5-0.6 & .019  & \textbf{.005} &       & .465  & \textbf{.453} &       & .218  & \textbf{.205} &       & \textbf{.946} & .940 \\
				& 0.6-0.7 & .006  & \textbf{-.003} &       & .585  & \textbf{.581} &       & .342  & \textbf{.337} &       & \textbf{.964} & \textbf{.964} \\
				& 0.7-0.8 & .021  & \textbf{.011} &       & .841  & \textbf{.836} &       & .715  & \textbf{.701} &       & \textbf{.948} & .954 \\
				& 0.8-0.9 & .015  & \textbf{.006} &       & \textbf{1.059} & \textbf{1.059} &       & 1.131 & \textbf{1.121} &       & \textbf{.962} & .964 \\
				\midrule
				&       & \multicolumn{2}{c}{Bias Ratio} &       & \multicolumn{2}{c}{Std} &       & \multicolumn{2}{c}{MSE} &       & \multicolumn{2}{c}{Cvg} \\
				\cmidrule{3-4}\cmidrule{6-7}\cmidrule{9-10}\cmidrule{12-13}    $R_y^2$ & Quantile & Naive & DML  &       & Naive & DML  &       & Naive & DML  &       & Naive & DML \\
				\midrule
				\multirow{8}[1]{*}{0.3} & 0.1-0.2 & .265  & \textbf{.140} &       & .298  & \textbf{.262} &       & .204  & \textbf{.101} &       & .806  & \textbf{.894} \\
				& 0.2-0.3 & .252  & \textbf{.074} &       & .358  & \textbf{.297} &       & .264  & \textbf{.100} &       & .814  & \textbf{.924} \\
				& 0.3-0.4 & .108  & \textbf{.038} &       & .374  & \textbf{.324} &       & .173  & \textbf{.109} &       & .922  & \textbf{.940} \\
				& 0.4-0.5 & \textbf{-.025} & -.031 &       & .354  & \textbf{.347} &       & .128  & \textbf{.124} &       & \textbf{.948} & \textbf{.948} \\
				& 0.5-0.6 & \textbf{-.001} & -.009 &       & .441  & \textbf{.440} &       & .195  & \textbf{.194} &       & \textbf{.948} & .946 \\
				& 0.6-0.7 & .018  & \textbf{.009} &       & .586  & \textbf{.580} &       & .347  & \textbf{.336} &       & \textbf{.948} & .954 \\
				& 0.7-0.8 & .013  & \textbf{.003} &       & .834  & \textbf{.828} &       & .697  & \textbf{.685} &       & .952  & \textbf{.950} \\
				& 0.8-0.9 & .007  & \textbf{-.003} &       & 1.049 & \textbf{1.047} &       & 1.100 & \textbf{1.094} &       & \textbf{.954} & .958 \\
				\midrule
				&       & \multicolumn{2}{c}{Bias Ratio} &       & \multicolumn{2}{c}{Std} &       & \multicolumn{2}{c}{MSE} &       & \multicolumn{2}{c}{Cvg} \\
				\cmidrule{3-4}\cmidrule{6-7}\cmidrule{9-10}\cmidrule{12-13}    $R_y^2$ & Quantile & Naive & DML  &       & Naive & DML  &       & Naive & DML  &       & Naive & DML \\
				\midrule
				\multirow{8}[1]{*}{0.4} & 0.1-0.2 & .255  & \textbf{.149} &       & .315  & \textbf{.262} &       & \textbf{.185} & \textbf{.098} &       & .836  & \textbf{.900} \\
				& 0.2-0.3 & .225  & \textbf{.074} &       & .360  & \textbf{.287} &       & .222  & \textbf{.092} &       & .876  & \textbf{.936} \\
				& 0.3-0.4 & .040  & \textbf{.006} &       & .339  & \textbf{.308} &       & .119  & \textbf{.095} &       & .942  & \textbf{.954} \\
				& 0.4-0.5 & \textbf{-.006} & -.009 &       & .374  & \textbf{.369} &       & .140  & \textbf{.136} &       & \textbf{.958} & .960 \\
				& 0.5-0.6 & \textbf{.004} & \textbf{-.004} &       & .469  & \textbf{.470} &       & \textbf{.220} & .221  &       & \textbf{.948} & .942 \\
				& 0.6-0.7 & .018  & \textbf{.008} &       & .600  & \textbf{.594} &       & .363  & \textbf{.352} &       & .940  & \textbf{.946} \\
				& 0.7-0.8 & .007  & \textbf{-.004} &       & .862  & \textbf{.854} &       & .743  & \textbf{.728} &       & \textbf{.946} & \textbf{.946} \\
				& 0.8-0.9 & .022  & \textbf{.011} &       & 1.057 & \textbf{1.048} &       & 1.141 & \textbf{1.104} &       & .940  & \textbf{.942} \\
				\bottomrule
			\end{tabular}%
		\end{threeparttable}
		\label{tab:addlabel}%
	\end{table}%
	\begin{table}[H]
		\centering
		\begin{threeparttable}
			\begin{tabular}{llccccccccccc}	\multicolumn{13}{l}{\small{\textbf{Table B.2 }}\small{Sparsity Design for $R_d^2=0.3$}}\\
				\toprule
				&       & \multicolumn{2}{c}{Bias Ratio} &       & \multicolumn{2}{c}{Std} &       & \multicolumn{2}{c}{MSE} &       & \multicolumn{2}{c}{Cvg} \\
				\cmidrule{3-4}\cmidrule{6-7}\cmidrule{9-10}\cmidrule{12-13}    $R_y^2$ & Quantile & Naive & DML  &       & Naive & DML  &       & Naive & DML  &       & Naive & DML \\
				\midrule
				\multirow{8}[1]{*}{0.2} & 0.1-0.2 & .147  & \textbf{.056} &       & .306  & \textbf{.274} &       & .136  & \textbf{.081} &       & .904  & \textbf{.934} \\
				& 0.2-0.3 & .234  & \textbf{.072} &       & .336  & \textbf{.307} &       & .246  & \textbf{.107} &       & .806  & \textbf{.930} \\
				& 0.3-0.4 & .177  & \textbf{.051} &       & .344  & \textbf{.310} &       & .219  & \textbf{.104} &       & .850  & \textbf{.936} \\
				& 0.4-0.5 & .082  & \textbf{.022} &       & .401  & \textbf{.359} &       & .191  & \textbf{.131} &       & .924  & \textbf{.954} \\
				& 0.5-0.6 & .052  & \textbf{.026} &       & .486  & \textbf{.465} &       & .255  & \textbf{.220} &       & .930  & \textbf{.942} \\
				& 0.6-0.7 & .013  & \textbf{-.001} &       & .627  & \textbf{.621} &       & .395  & \textbf{.385} &       & .958  & \textbf{.954} \\
				& 0.7-0.8 & .027  & \textbf{.015} &       & .863  & \textbf{.861} &       & .763  & \textbf{.745} &       & .958  & \textbf{.952} \\
				& 0.8-0.9 & .018  & \textbf{.008} &       & 1.131 & \textbf{1.130} &       & 1.297 & \textbf{1.277} &       & \textbf{.964} & .966 \\
				\midrule
				&       & \multicolumn{2}{c}{Bias Ratio} &       & \multicolumn{2}{c}{Std} &       & \multicolumn{2}{c}{MSE} &       & \multicolumn{2}{c}{Cvg} \\
				\cmidrule{3-4}\cmidrule{6-7}\cmidrule{9-10}\cmidrule{12-13}    $R_y^2$ & Quantile & Naive & DML  &       & Naive & DML  &       & Naive & DML  &       & Naive & DML \\
				\midrule
				\multirow{8}[1]{*}{0.3} & 0.1-0.2 & .192  & \textbf{.112} &       & .283  & \textbf{.245} &       & .140  & \textbf{.080} &       & .866  & \textbf{.906} \\
				& 0.2-0.3 & .275  & \textbf{.083} &       & .324  & \textbf{.281} &       & .265  & \textbf{.094} &       & .766  & \textbf{.918} \\
				& 0.3-0.4 & .180  & \textbf{.057} &       & .392  & \textbf{.327} &       & .248  & \textbf{.116} &       & .884  & \textbf{.938} \\
				& 0.4-0.5 & .031  & \textbf{-.001} &       & .402  & \textbf{.370} &       & .166  & \textbf{.136} &       & .938  & \textbf{.940} \\
				& 0.5-0.6 & \textbf{.004} & -.010 &       & .481  & \textbf{.476} &       & .231  & \textbf{.227} &       & .960  & \textbf{.954} \\
				& 0.6-0.7 & .024  & \textbf{.011} &       & .632  & \textbf{.627} &       & .406  & \textbf{.394} &       & \textbf{.950} & .952 \\
				& 0.7-0.8 & .022  & \textbf{.009} &       & .903  & \textbf{.892} &       & .827  & \textbf{.796} &       & \textbf{.948} & \textbf{.952} \\
				& 0.8-0.9 & .015  & \textbf{.002} &       & 1.128 & \textbf{1.124} &       & 1.282 & \textbf{1.262} &       & \textbf{.950} & .956 \\
				\midrule
				&       & \multicolumn{2}{c}{Bias Ratio} &       & \multicolumn{2}{c}{Std} &       & \multicolumn{2}{c}{MSE} &       & \multicolumn{2}{c}{Cvg} \\
				\cmidrule{3-4}\cmidrule{6-7}\cmidrule{9-10}\cmidrule{12-13}    $R_y^2$ & Quantile & Naive & DML  &       & Naive & DML  &       & Naive & DML  &       & Naive & DML \\
				\midrule
				\multirow{8}[2]{*}{0.4} & 0.1-0.2 & .201  & \textbf{.130} &       & .300  & \textbf{.253} &       & .144  & \textbf{.087} &       & .866  & \textbf{.910} \\
				& 0.2-0.3 & .324  & \textbf{.106} &       & .336  & \textbf{.282} &       & .306  & \textbf{.100} &       & .744  & \textbf{.910} \\
				& 0.3-0.4 & .127  & \textbf{.036} &       & .397  & \textbf{.323} &       & .200  & \textbf{.108} &       & .906  & \textbf{.958} \\
				& 0.4-0.5 & .015  & \textbf{.001} &       & .407  & \textbf{.389} &       & .167  & \textbf{.151} &       & .966  & \textbf{.956} \\
				& 0.5-0.6 & .011  & \textbf{-.002} &       & .499  & \textbf{.498} &       & .249  & \textbf{.248} &       & \textbf{.952} & .964 \\
				& 0.6-0.7 & .027  & \textbf{.013} &       & .654  & \textbf{.643} &       & .437  & \textbf{.415} &       & \textbf{.948}  & .942 \\
				& 0.7-0.8 & .020  & \textbf{.005} &       & .919  & \textbf{.908} &       & .855  & \textbf{.824} &       &\textbf{ .956} &.960 \\
				& 0.8-0.9 & .026  & \textbf{.012} &       & 1.133 & \textbf{1.126} &       & 1.324 & \textbf{1.275} &       & \textbf{.948} & \textbf{.948} \\
				\bottomrule
			\end{tabular}%
		\end{threeparttable}
		\label{tab:addlabel}%
	\end{table}%
	\begin{table}[H]
		\centering
		\begin{threeparttable}
			\begin{tabular}{llccccccccccc}	\multicolumn{13}{l}{\small{\textbf{Table B.3 }}\small{Sparsity Design for $R_d^2=0.4$}}\\
				\toprule
				&       & \multicolumn{2}{c}{Bias Ratio} &       & \multicolumn{2}{c}{Std} &       & \multicolumn{2}{c}{MSE} &       & \multicolumn{2}{c}{Cvg} \\
				\cmidrule{3-4}\cmidrule{6-7}\cmidrule{9-10}\cmidrule{12-13}    $R_y^2$ & Quantile & Naive & DML  &       & Naive & DML  &       & Naive & DML  &       & Naive & DML \\
				\midrule
				\multirow{8}[1]{*}{0.2} & 0.1-0.2 & .081  & \textbf{.038} &       & .291  & \textbf{.263} &       & .097  & \textbf{.072} &       & .942  & \textbf{.946} \\
				& 0.2-0.3 & .214  & \textbf{.069} &       & .305  & \textbf{.288} &       & .201  & \textbf{.094} &       & .816  & \textbf{.932} \\
				& 0.3-0.4 & .162  & \textbf{.042} &       & .336  & \textbf{.310} &       & .195  & \textbf{.102} &       & .854  & \textbf{.934} \\
				& 0.4-0.5 & .119  & \textbf{.042} &       & .409  & \textbf{.370} &       & .232  & \textbf{.145} &       & .916  & \textbf{.946} \\
				& 0.5-0.6 & .071  & \textbf{.033} &       & .533  & \textbf{.502} &       & .322  & \textbf{.260} &       & .938  & \textbf{.952} \\
				& 0.6-0.7 & .044  & \textbf{.023} &       & .688  & \textbf{.679} &       & .500  & \textbf{.467} &       & \textbf{.948} & .958 \\
				& 0.7-0.8 & .025  & \textbf{.011} &       & .904  & \textbf{.901} &       & .837  & \textbf{.813} &       & \textbf{.948} & \textbf{.952} \\
				& 0.8-0.9 & .020  & \textbf{.008} &       & 1.194 & \textbf{1.196} &       & 1.452 & \textbf{1.431} &       & \textbf{.952} & \textbf{.952} \\
				\midrule
				&       & \multicolumn{2}{c}{Bias Ratio} &       & \multicolumn{2}{c}{Std} &       & \multicolumn{2}{c}{MSE} &       & \multicolumn{2}{c}{Cvg} \\
				\cmidrule{3-4}\cmidrule{6-7}\cmidrule{9-10}\cmidrule{12-13}    $R_y^2$ & Quantile & Naive & DML  &       & Naive & DML  &       & Naive & DML  &       & Naive & DML \\
				\midrule
				\multirow{8}[1]{*}{0.3} & 0.1-0.2 & .121  & \textbf{.083} &       & .270  & \textbf{.229} &       & .096  & \textbf{.063} &       & .894  & \textbf{.928} \\
				& 0.2-0.3 & .245  & \textbf{.070} &       & .300  & \textbf{.266} &       & .216  & \textbf{.081} &       & .770  & \textbf{.938} \\
				& 0.3-0.4 & .212  & \textbf{.072} &       & .357  & \textbf{.309} &       & .258  & \textbf{.110} &       & .834  & \textbf{.936} \\
				& 0.4-0.5 & .101  & \textbf{.034} &       & .439  & \textbf{.388} &       & .238  & \textbf{.155} &       & .920  & \textbf{.948} \\
				& 0.5-0.6 & .024  & \textbf{-.002} &       & .515  & \textbf{.498} &       & .269  & \textbf{.248} &       & .938  & \textbf{.950} \\
				& 0.6-0.7 & .032  & \textbf{.015} &       & .673  & \textbf{.668} &       & .468  & \textbf{.449} &       & \textbf{.950} & .956 \\
				& 0.7-0.8 & .031  & \textbf{.014} &       & .916  & \textbf{.908} &       & .870  & \textbf{.829} &       & .946  & \textbf{.952} \\
				& 0.8-0.9 & .018  & \textbf{.004} &       & 1.221 & \textbf{1.215} &       & 1.513 & \textbf{1.475} &       & .938  & \textbf{.944} \\
				\midrule
				&       & \multicolumn{2}{c}{Bias Ratio} &       & \multicolumn{2}{c}{Std} &       & \multicolumn{2}{c}{MSE} &       & \multicolumn{2}{c}{Cvg} \\
				\cmidrule{3-4}\cmidrule{6-7}\cmidrule{9-10}\cmidrule{12-13}    $R_y^2$ & Quantile & Naive & DML  &       & Naive & DML  &       & Naive & DML  &       & Naive & DML \\
				\midrule
				\multirow{8}[2]{*}{0.4} & 0.1-0.2 & .136  & \textbf{.102} &       & .285  & \textbf{.243} &       & .106  & \textbf{.073} &       & .918  & \textbf{.930} \\
				& 0.2-0.3 & .333  & \textbf{.115} &       & .300  & \textbf{.264} &       & .294  & \textbf{.094} &       & .704  & \textbf{.908} \\
				& 0.3-0.4 & .211  & \textbf{.070} &       & .395  & \textbf{.328} &       & .275  & \textbf{.121} &       & .852  & \textbf{.928} \\
				& 0.4-0.5 & .064  & \textbf{.021} &       & .450  & \textbf{.402} &       & .220  & \textbf{.163} &       & \textbf{.950} & .956 \\
				& 0.5-0.6 & .024  & \textbf{.003} &       & .520  & \textbf{.512} &       & .274  & \textbf{.262} &       & \textbf{.954} & .956 \\
				& 0.6-0.7 & .039  & \textbf{.020} &       & .738  & \textbf{.722} &       & .569  & \textbf{.526} &       & .944  & \textbf{.952} \\
				& 0.7-0.8 & .029  & \textbf{.010} &       & .976  & \textbf{.964} &       & .979  & \textbf{.932} &       & .938  & \textbf{.950} \\
				& 0.8-0.9 & .035  & \textbf{.018} &       & 1.187 & \textbf{1.178} &       & 1.501 & \textbf{1.410} &       & .932  & \textbf{.944} \\
				\bottomrule
			\end{tabular}%
		\end{threeparttable}
		\label{tab:addlabel}%
	\end{table}%
	
		\section*{Appendix C: Additional Results for the Empirical Example}
		Figure C.1 shows a concerning heavy tail of low birthweights for black mothers. Therefore, we focus on the sample of black mothers.
		\begin{figure}[H]
		\centering
		\includegraphics[width=6.2in]{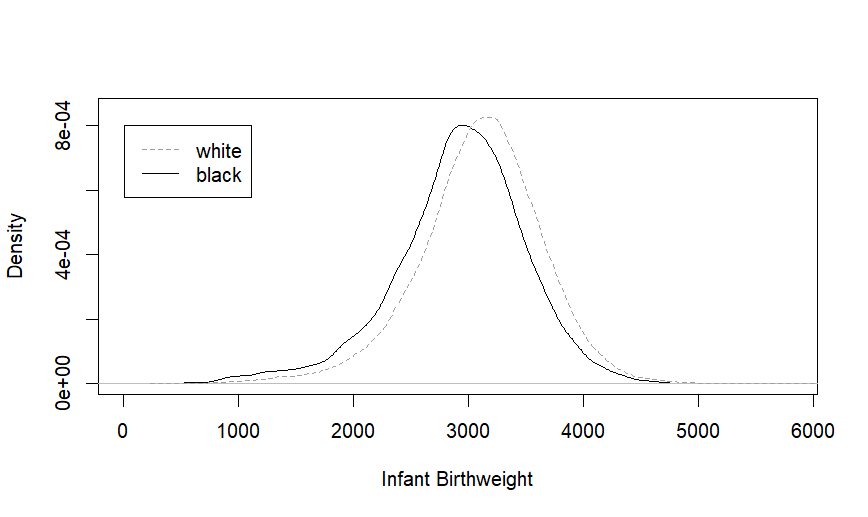}
		\caption*{\textbf{Figure C.1.} Birthweight densities for black and white mothers}
	\end{figure}
	The estimation results of standard quantile regression for the extremal low tails, such that $\tau\in\{0.1,0.15,0.2,0.25\}$, are presented in Table C.1.
	We find that the variables Prenatal Times and Mother's BMI are significant at the 1\% level across all the quantiles considered. Therefore, we consider conditioning treatment assignment on two dummy covariates: (i) whether a mother paid a prenatal visit and (ii) whether a mother's BMI is higher than the median of all mothers' BMI in the sample.
	

	\begin{table}[H]\centering\small
		\begin{threeparttable}
			\begin{tabular}{lllllllll}
				\multicolumn{9}{l}{\small{\textbf{Table C.1: Quantile regression at $\tau\in\{0.1,0.15,0.2,0.25\}$}}}\\
				\toprule
				& \multicolumn{2}{c}{0.1} & \multicolumn{2}{c}{0.15} & \multicolumn{2}{c}{0.2} & \multicolumn{2}{c}{0.25} \\
				& Estimate & p-value & Estimate & p-value & Estimate & p-value & Estimate & p-value \\
				\midrule
				Cigarettes & -1.210 & .270  & -1.176 & .338  & -.437 & .639  & -.479 & .509 \\
				Married & 30.149* & .099  & 8.771 & .593  & 10.297 & .538  & 22.102* & .082 \\
				Mother's Age  & -2.174 & .205  & -2.685* & .095  & -3.235** & .019  & -2.952** & .011 \\
				Mother's Education & 13.100* & .063  & 9.568 & .136  & 5.967 & .277  & 9.298** & .038 \\
				Father's Age  & -2.307*** & .009  & -2.813*** & .009  & -2.156** & .013  & -2.588*** & .000 \\
				Father's Education & -5.030 & .608  & -5.847 & .497  & -6.391 & .365  & -6.597 & .304 \\
				Prenatal Times & 8.754*** & .000  & 7.234*** & .000  & 8.309*** & .000  & 6.678*** & .000 \\
				Prenatal Second & 55.257*** & .000  & 27.675* & .075  & 41.237*** & .006  & 38.862*** & .003 \\
				Prenatal Third & 106.595*** & .007  & 89.075** & .013  & 116.262*** & .000  & 93.661*** & .000 \\
				Mother's BMI  & 7.740*** & .000  & 8.396*** & .000  & 9.178*** & .000  & 9.148*** & .000 \\
				Mother's Height & 13.053*** & .000  & 15.629*** & .000  & 17.564*** & .000  & 17.478*** & .000 \\
				Mother's Weight Gain & 3.643*** & .000  & 3.741*** & .000  & 4.195*** & .000  & 4.405*** & .000 \\
				WIC   & 1.508 & .921  & 5.120 & .739  & 2.380 & .850  & -2.375 & .832 \\
				Gestation & 114.478*** & .000  & 118.541*** & .000  & 118.200*** & .000  & 118.463*** & .000 \\
				Boy   & 89.836*** & .000  & 104.082*** & .000  & 112.836*** & .000  & 116.952*** & .000 \\
				\bottomrule
			\end{tabular}%
			\begin{tablenotes}
				\footnotesize
				\item Notes: *, ** and *** respectively indicate the significance at 10, 5 and 1 percent level.
			\end{tablenotes}
		\end{threeparttable}
		\label{tab:addlabel}%
	\end{table}%

	We now demonstrate the identification of $V\left(\tau_1,\tau_2,\pi,\mathcal{G}\right)$. By definition,
	\[\begin{aligned}
				&V\left(\tau_1,\tau_2,\pi,\mathcal{G}\right)\notag\\
		=&E\left[\pi(X)\cdot\lim_{\delta\rightarrow 0}\frac{m\left(\mathcal{G}_{\delta}(D),X,U\right)-m\left(D,X,U\right)}{\delta}\bigg|Y\in\left(Q_Y(\tau_1),Q_Y(\tau_2)\right)\right]\notag\\
		=&E\left[\pi(X)\vartheta\left(D;\mathcal{G}\right)\partial_D m\left(D,X,U\right)\bigg|Y\in\left(Q_Y(\tau_1),Q_Y(\tau_2)\right)\right]\notag\\
		=&\frac{-1}{\tau_2-\tau_1}E\left[\pi(X)\vartheta\left(D;\mathcal{G}\right)\int_{Q_Y(\tau_1)}^{Q_Y(\tau_2)}\partial_DF_Y(y|D,X)dy\right],
	\end{aligned}\]
which is identifiable for any given $\pi$. Consequently, the (infeasible) optimal policy assignment rule can be equivalently expressed as
\[
\pi^*\left(\tau_1,\tau_2,\mathcal{G}\right)=\arg\max_{\pi\in\Pi}E\left[\Big(2\pi(X)-1\Big)\cdot\frac{-\vartheta\left(D;\mathcal{G}\right)}{\tau_2-\tau_1}\int_{Q_Y(\tau_1)}^{Q_Y(\tau_2)}\partial_DF_Y(y|D,X)dy\right].
\]

	Finally, the estimation results for reducing the number of cigarettes by a fixed amount,  i.e., $\mathcal{G}_\delta(D)=D-\delta$, are presented in Table C.2. The baseline welfare gain reported in the first item of Table C.2 is -0.882, which is not statistically significant at the 10\% level. For the policy assigned to mothers who did not have any prenatal visits during pregnancy, corresponding to three assignment rules $(0,0,1,0)$, $(0,0,0,1)$ and $(0,0,1,1)$, the welfare gains are 0.065, 0.089, and -0.030, respectively. The first two values are significant at the 5\% level, while the last one is not statistically significant. Although the welfare gain for the assignment rule $(0,0,1,1)$ is negative, it significantly improves welfare gain compared to other assignment rules. This result supports the finding in the main text that the optimal intervention to achieve the highest welfare gain for low birth weight infants is highly influenced by the mother's prenatal visits during pregnancy.
	
	\begin{table}[H]\centering
	\begin{threeparttable}
		\begin{tabular}{cccccccccc}
			\multicolumn{10}{l}{\small{\textbf{Table C.2: Estimated welfare gain for each assignment rule for $\mathcal{G}_\delta(D)=D-\delta$}}}\\
\toprule
\multicolumn{10}{c}{$\mathcal{G}_\delta(D)=D-\delta$} \\
\multicolumn{1}{l}{$\pi$} &       & \multicolumn{1}{l}{$\pi$} &       & \multicolumn{1}{l}{$\pi$} &       & \multicolumn{1}{l}{$\pi$} &       &  
\multicolumn{1}{l}{$\pi$}     &  \\
\midrule
\multicolumn{1}{l}{(1,1,1,1)} & -.882 & \multicolumn{1}{l}{(0,0,1,0)} & \textbf{.065**} & \multicolumn{1}{l}{(1,0,1,0)} & -.524 & \multicolumn{1}{l}{(0,1,0,1)} & -.315 & \multicolumn{1}{l}{(1,1,0,1)} & -.932 \\
& (1.198) &       & \textbf{(.025)} &       & (.729) &       & (.568) &       & (1.147) \\
& \{1.210\} &       & \textbf{\{.026\}} &       & \{.708\} &       & \{.565\} &       & \{1.170\} \\
\multicolumn{1}{l}{(1,0,0,0)} & -.549 & \multicolumn{1}{l}{(0,0,0,1)} & \textbf{.089**} & \multicolumn{1}{l}{(1,0,0,1)} & -.530 & \multicolumn{1}{l}{(0,0,1,1)} & -.030 & \multicolumn{1}{l}{(1,0,1,1)} & -.540 \\
& (.691) &       & \textbf{(.045)} &       & (.686) &       & (.055) &       & (.722) \\
& \{.675\} &       & \textbf{\{.045\}} &       & \{.709\} &       & \{.055\} &       & \{.734\} \\
\multicolumn{1}{l}{(0,1,0,0)} & -.428 & \multicolumn{1}{l}{(1,1,0,0)} & -1.067 & \multicolumn{1}{l}{(0,1,1,0)} & -.413 & \multicolumn{1}{l}{(1,1,1,0)} & -1.023 & \multicolumn{1}{l}{(0,1,1,1)} & -.322 \\
& (.548) &       & (1.136) &       & (.548) &       & (1.187) &       & (.593) \\
& \{.551\} &       & \{1.136\} &       & \{.547\} &       & \{1.174\} &       & \{.582\} \\
\bottomrule

		\end{tabular}%
		\begin{tablenotes}
			\footnotesize
			\item Notes:
			Analytical standard errors are provided in parentheses.
			Bootstrap standard errors based on 1000 repetitions
			with standard normal variables are provided in braces.
			*, ** and *** respectively indicate the significance at 10, 5 and 1 percent level.
		\end{tablenotes}
	\end{threeparttable}
	\label{tab:addlabel}%
\end{table}%
\end{document}